\documentclass[]{Liebert_Author}

\usepackage{amssymb, amsmath,framed}
\usepackage{color} 

\title{Waste Heat and Habitability: Constraints from Technological Energy Consumption}

\author{Amedeo Balbi,$^{1\ast}$ Manasvi Lingam,$^{2,3}$\\
{$^{1}$Dipartimento di Fisica,}\\
{Universit\`a di Roma Tor Vergata, Roma, I-00133, Italy}\\
{$^{2}$Department of Aerospace, Physics and Space Sciences,}\\ 
{Florida Institute of Technology, Melbourne, FL 32901, USA}\\
{$^{3}$Department of Chemistry and Chemical Engineering,}\\ 
{Florida Institute of Technology, Melbourne, FL 32901, USA}\\
{$^\ast$To whom correspondence should be addressed;}\\
{E-mail:  amedeo.balbi@roma2.infn.it}
}

\begin{document}
\maketitle

\begin{abstract}
Waste heat production represents an inevitable consequence of energy conversion as per the laws of thermodynamics. Based on this fact, by using simple theoretical models, we analyze constraints on the habitability of Earth-like terrestrial planets hosting putative technological species and technospheres characterized by persistent exponential growth of energy consumption and waste heat generation. In particular, we quantify the deleterious effects of rising surface temperature on biospheric processes and the eventual loss of liquid water. Irrespective of whether these sources of energy are ultimately stellar or planetary (e.g., nuclear, fossil fuels) in nature, we demonstrate that the loss of habitable conditions on such terrestrial planets may be expected to occur on timescales of $\lesssim 1000$ years, as measured from the start of the exponential phase, provided that the annual growth rate of energy consumption is of order $1\%$. We conclude with a discussion of the types of evolutionary trajectories that might be feasible for industrialized technological species, and we sketch the ensuing implications for technosignature searches.
\end{abstract}

\keywords{astrobiology  -- Earth (planet) -- habitable planets -- habitable zone -- planetary climates -- technosignatures}
 
\section{Introduction}\label{sec:intro}

It is now firmly established that the upsurge in atmospheric CO$_2$ concentration caused by human technological activities has been elevating Earth's average surface temperature \citep[e.g.,][]{DA16,DH16,AED21,MPD21,AQS22,JJR22,ET22}. Nevertheless, it is worth appreciating that ongoing anthropogenic warming is not exclusively a consequence of the intensified greenhouse effect stemming from fossil fuel utilization (though the latter phenomenon is unequivocally proven to be the chief driver). Instead, as elucidated in a number of analyses and overviews \citep[e.g.,][]{Chaisson2008,MGF09,Kleidon2012,MLC14,LWW17,KMB23}, this facet of anthropogenic warming embodies a broader outcome of energy production and conversion that underpins  modern (viz., globalized and industrialized) human societies. The pioneering 1969 publication by the Soviet scientist Mikhail I. Budyko explicitly underscored this theme \citep[pg. 618]{Budyko1969}:
\begin{quotation}
All the energy used by man is transformed into heat, the main portion of this energy being an additional source of heat as compared to the present radiation gain. Simple calculations show \citep{Budyko1961} that with the present rate of growth of using energy the heat produced by man in less than two hundred years will be comparable with the energy coming from the sun. 
\end{quotation}

This feature arises from the fundamental fact that energy transduction inevitably generates waste heat \citep[e.g.,][]{Budyko1969,CMS71,VS08}\footnote{Interestingly, around the same time as Budyko's paper, Larry Niven's classic science fiction novel \textit{Ringworld} (1970) also acknowledged the importance of waste heat as an evolutionary factor for technological species \citep{LN70}.}. Although the current contribution of this factor to global warming on Earth is negligible compared to the impact of carbon emissions \citep{MLC14,KMB23},\footnote{In localized regions endowed with high industrialization (e.g., large and dense cities), however, it is already a notable contributor to warming \citep{ISH99,BKS04,OGK07,ALG11,ZCH13,BHD14,SGM14,SK17,SWC18,RPC20,MKG20}, and this effect is anticipated to increase in the future.} it is expected to amplify quite substantially in the next century in the event that the current growth rate of global energy consumption persists \citep{LWW17}. The consequent rise in global temperature (just from this process) derived from assorted climate models may reach up to $\sim 1$ K in this duration \citep{MGF09,AN09,SMF19}. 

Cruder estimates, when extrapolated over longer intervals, project a temperature increase of $\sim 10$ K over a period of $350$ years, with the assumption of energy trends similar to those of today \citep{Chaisson2008}. Likewise, a cognate analysis by \citet{MHM19} (see also \citealt{SVH75,DD00}) concluded that direct heating engendered by progressively higher energy consumption consistent with current rates could dramatically impede human habitability on timescales of merely hundreds of years in the future. It is crucial to realize that this contribution to global warming would persist even in the scenario of achieving a complete transition to zero CO$_2$-emissions energy production.

While this effect is of profound importance for the long-term future and fate of human civilization, it also has broader significance in the general context of planetary evolution and the potential for harboring technologically capable life in the universe -- this latter domain is closely linked with the detection of  technosignatures, a burgeoning area of astrobiology \citep{MC12,ML21,SHW21}, with roots traceable to the 1960s \citep{CS63,Drake1965,Sagan1966,JT01}. To be specific, we are free to inquire whether energetic growth, often innately presumed to be associated with technological activity \citep{NSK85}, imposes a universal fundamental limit on the habitability of a planetary environment, arising simply from thermodynamic considerations, as indicated in \citet{SVH75}.

The notion that technological advancement is correlated with the amount of energy consumed was initially articulated vis-\`a-vis extraterrestrial life by \cite{Kardashev1964}; early studies implicitly adopted the premise that technological species would pursue and evince growth of their (energetic) activities \citep{FJD60,Sagan1966,NSK85}. Kardashev's idea has been predominantly employed to categorize hypothetical extraterrestrial technological entities,\footnote{We remark, however, that a few proposed alternatives to the Kardashev scale exist \citep{CS73,RZ00,ZG05,FKA17,IBC20,LFB23}.} and gauge their detectability through the types of technosignatures they might produce \citep{MMC15,Gray2020,ML21}, but studies that strive to understand how heightened energy consumption by these entities impacts a planetary environment and whether this trend actually aligns with long-term habitability are comparatively limited in scope and in number.

Thus, the \emph{summum bonum} of this paper is to perform a preliminary investigation of how the habitability of Earth-like planets would be affected by the presence of life specifically capable of engaging in progressively energy-demanding technological activities. One of our primary objectives is to gauge the maximum lifespan of habitable planets that harbor an evolving ``technosphere'' \citep{DLM17,ZWW17,CHP18}, by making certain assumptions regarding the growth of energy consumption by an industrialized technological species. This analysis parallels the canonical approach used to estimate the lifetime of continuously habitable zones via incorporation of the increase in stellar luminosity, which exclusively accounts for abiotic factors \citep[see, e.g.,][]{KWR93,Rushby2013,GCA17,TW23}, and not biological/technological feedbacks.

The outline of the current paper is as follows. We present the methods and accompanying results that pertain to the impacts of global heating (from energy production) on putative biospheres (and technospheres) and climates in Section \ref{SecMethod}, where it is shown that the loss of habitable conditions might occur in $\sim 10^3$ yr. Next, we elucidate their ramifications for the potential trajectories of technological species (e.g., collapse; slowdown or cessation of growth; expansion into space) and the detection of technosignatures in Section \ref{SecConc}.

\section{Methods and Results}\label{SecMethod}
In this section, we describe some of the thermodynamic principles underlying our modeling, and delineate the results that ensue in our framework.

\subsection{Thermodynamics of energy production}\label{SSecThermEP}

The Second Law of thermodynamics dictates that whenever energy $E$ is converted into useful work $W$, waste heat $\mathcal{Q}$ is produced and dissipated in the environment \citep[e.g.,][]{CMS71,LR98,BB10,HS24}. If the conversion mechanism exhibits an efficiency of $\eta = W/E$, the amount of energy that goes into heating is expressible as
\begin{equation}
    \mathcal{Q} =\left(1-\eta \right)E.
\end{equation}
Carnot's theorem canonically sets a theoretical limit on the maximum efficiency attainable by a cyclical process operating between temperatures $T_C$ and $T_H$ \citep[e.g.,][]{DVS20,HS24}, as given by the famous formula:
\begin{equation}
    \eta=1-\frac{T_C}{T_H},
\end{equation}
with the accompanying condition that $T_C\le T_H$. 

It is, therefore, unavoidable that energy conversion (to thermodynamic work) that enables technological activity produces a finite amount of waste heat, $\mathcal{Q}_{\rm tech}$ \citep{Budyko1969,Budyko72}. When energy is extracted from sources that would otherwise be rendered ``inactive'' (such as burning fossil fuels or exploiting nuclear reactions), $\mathcal{Q}_{\rm tech}$ will constitute an extra heat source that is fed into the planetary system, in addition to the ``natural'' heat deposited by stellar irradiation or by the dissipation of ``intrinsic'' planetary energy sources (e.g., radioactivity). It is worth recognizing that the warming connected to $\mathcal{Q}_{\rm tech}$ is conceptually different from that caused by the enhanced (viz., anthropogenic) greenhouse effect that results from alterations of atmospheric chemistry. While in principle a ``zero emissions'' energy conversion strategy ought to be viable, the production of waste heat can only be minimized, but never completely eliminated. 

Although the amount of waste heat dissipated by technological activity is evidently not $100\%$ of the energy input, in most practical applications of current technology, it is comparable to this upper bound. Moreover, waste heat does not seem reducible to a negligible fraction, although it might be curtailed drastically in theory \citep[cf.][]{LEB08,CA10,CAB11,JKA18,GBC20,GLP21}. The energy conversion efficiency attributed to fossil fuels, nuclear and even geothermal sources is distinctly lower than the theoretical Carnot limit. In global society, it has consequently been estimated that $\mathcal{Q}/E$ is typically $\gtrsim 50\%$ \citep{GVW07,VS08,Smil2017,CA10,ZM14,FMP16,FZY19,GBC20,OT21}. Renewable energy sources such as wind, hydroelectric, and tidal power may not entail direct additional heating (although they could impact the climate indirectly), but whether their exclusive usage could meet the demands of a global (industrialized) technological society embarking on sustained exponential growth of the type studied herein remains an open question \citep[cf.][]{DM09,VS15,VS20}. Finally, the conversion of abundant stellar energy warrants a separate treatment that is tackled in Section \ref{SSecSolEn}.

As underscored in the preceding exposition, the waste heat is comparable to the input energy for known sources. Therefore, we will hereafter adopt the ostensibly reasonable premise that the waste heat generated is roughly equal to the energy consumption associated with technological activity, unless indicated otherwise. If we postulate that the magnitude of waste heat generated is diminished by about an order of magnitude, this does not alter our qualitative findings, and it modifies the quantitative results by merely a modest degree; this is implicit in the paragraph below (\ref{TimeScale}).

\subsection{Global heating from technological activity}

At equilibrium, the incoming stellar energy that reaches a planetary surface has to equal the outgoing flux emitted by the planet itself. The basic (zero-dimensional and equilibrium) energy balance equation can be expressed as (e.g., \citealt[Chapter 7]{DJJ99}, \citealt[Chapter 1]{GVV12}, and \citealt[Chapter 2]{ET22}):
\begin{equation}\label{eq:EBMv1}
    \frac{S}{4}(1-A)+\frac{1}{2}\varepsilon\sigma T^4+Q_{\rm tech}=\sigma T^4,
\end{equation}
where $S$ is the stellar flux incident at the top of the atmosphere, $A$ is the planetary albedo, $\varepsilon$ is the fraction of terrestrial infrared radiation absorbed by the atmosphere (a parameter that can be calibrated to model greenhouse warming), $T$ is the planetary surface temperature (which is not the same as the planet's equilibrium blackbody temperature), and $Q_{\rm tech}$ is the waste heat flux (units of power per unit area) that ensues from technological activity. Where needed (i.e., not for every calculation), we will henceforth adopt an average albedo of $A \approx 0.3$ for modern Earth \citep{TFK09,SOW15}, as well as for ``Earth-like'' habitable worlds (i.e., Earth-analogs) \citep[cf.][]{DKW19,RSJ19,MK20}.\footnote{However, as long as $A$ is not very close to unity (i.e., $A \rightarrow 1$), our forthcoming quantitative results are not particularly sensitive to this parameter.} 

It should, however, be noted that (\ref{eq:EBMv1}) neglects the supplemental role of convection \citep{DH16,ET22}, which may be of importance for certain temperate planets orbiting M-dwarfs \citep{SBJ16}. Likewise, we neglect internal heat sources in (\ref{eq:EBMv1}), which would enter as additional terms on the left-hand side (LHS), and would serve to elevate the temperature further, and might possibly diminish the timescales computed in this paper. The possible sources are radioactivity and tidal heating, both of which are relatively modest on Earth compared to $S$; the former translates to a heat flux of $\sim 7.4 \times 10^{-2}$ Wm$^{-2}$ \citep{AAA20}. On some habitable worlds, however, it is plausible that tidal heating \citep{JGB08,BJG09,DB15,DBK19} and radiogenic heating \citep{FMM14,LL20,NPF20} could be higher and thus necessitate inclusion in the global energy budget equation.

The energy balance in (\ref{eq:EBMv1}) is applied at the planetary surface. The first and the second terms on the LHS account for incoming stellar radiation and the thermal radiation emitted by the atmosphere (toward the surface), whereas the right-hand side (RHS) is the outgoing thermal radiation from the planetary surface. The last term on the LHS is the extra contribution arising from technological activity. For the sake of the subsequent exposition, we rewrite (\ref{eq:EBMv1}) in terms of the parameter $\tau \equiv 1 - \varepsilon/2$, which has been interpreted as an effective ``transmittance'' in some publications \citep[e.g.,][]{Steiglechner2021}, thereupon simplifying to yield
\begin{equation}\label{eq:EB}
    \frac{S}{4}(1-A) + Q_{\rm tech} = \tau \sigma T^4,
\end{equation}
where it should be noted that $\tau$ is innately dependent on atmospheric properties, such as atmospheric composition, and encapsulates a fair amount of complexity.

All values should be inherently perceived as global averages. As mentioned in Section \ref{SSecThermEP}, $Q_{\rm tech}$ is anticipated to be comparable to the total power converted to support technological activity. As a reference, adopting a global primary energy consumption of $1.8\times 10^5$ TWh in 2022 \citep{OWID20},\footnote{\url{https://ourworldindata.org/energy-production-consumption}} and averaging over the entire surface of Earth ($5.1\times 10^{14}\ \rm m^2$) yields an estimate of $Q_{\rm tech} \sim 0.04$ Wm$^{-2}$, which matches the calculation reported in \citet{KMB23}, acting as a consistency check. This is currently a negligible fraction of the radiative forcing attributable to the anthropogenic greenhouse effect (computed to be $2.7$ Wm$^{-2}$) responsible for global warming \citep{KMB23}. 

The planetary temperature can be further decomposed as $T=T_0+\Delta T$, where we have defined
\begin{equation}\label{T0def}
T_0=\left[\frac{S(1-A)}{4\tau\sigma}\right]^{1/4}
\end{equation} 
as the temperature when $Q_{\rm tech}\approx 0$, and $\Delta T$ is the ensuing warming when $Q_{\rm tech} \ne 0$. On solving (\ref{eq:EB}), we find
\begin{equation}\label{eq:deltat}
    \Delta T=\left( T_0^4 +\frac{Q_{\rm tech}}{\tau\sigma}\right)^{1/4}-T_0.
\end{equation}
This expression can also be recast in terms of a fractional temperature increase by dividing throughout with $T_0$ as follows:
\begin{equation}\label{eq:fracdeltat}
    \frac{\Delta T}{T_0}=\left(1 + \frac{4}{1-A}\frac{Q_{\rm tech}}{S}\right)^{1/4}-1,
\end{equation}
where we have substituted (\ref{T0def}) into (\ref{eq:deltat}). An interesting feature of the above equation is that the lower bound on $\Delta T/T_0$ is attained in the low-albedo limit of $A \rightarrow 0$ and is given by
\begin{equation}
    \frac{\Delta T}{T_0} \geq \left(1 + 4 \frac{Q_{\rm tech}}{S}\right)^{1/4}-1,
\end{equation}
which is independent of the atmospheric parameter $\tau$, just like the parent equation (\ref{eq:fracdeltat}) from which it was derived.

Next, if we suppose that $Q_{\rm tech}/S \ll 1$ in (\ref{eq:fracdeltat}), we obtain a simple relation through a Taylor series expansion,
\begin{equation}
     \frac{\Delta T}{T_0} \sim \frac{1}{1-A}\frac{Q_{\rm tech}}{S},
\end{equation}
which implies that the increase in global surface temperature is linearly proportional to the extent of waste heat generation (encapsulated in $Q_{\rm tech}$); to put it differently, the surface temperature responds in a roughly linear fashion to an increase in energy consumption. The exact same scaling law was predicted by \citet{Budyko72}, thus acting as a consistency check, and this trend remains approximately valid $\sim 50$ years after the paper was published \citep{AL20}.

Lastly, let us apply (\ref{eq:fracdeltat}) to a future projection with the goal of loosely gauging the validity of our simple global model by comparing it against climate simulations undertaken by \citet{MGF09}; the latter forecast the temperature increase caused exclusively by anthropogenic heat emissions in 2100. We adopt a magnitude of $Q_{\rm tech} \approx 1.2$ Wm$^{-2}$ for select industrialized large-scale regions (continental USA and western Europe) \citep[Section 3]{MGF09}, and we hold Earth's albedo fixed (at $A=0.3$) owing to the multitude of unknowns concerning the evolution of $A$ at play.\footnote{The albedo of urban environs is slightly different than the surrounding milieu \citep{JDZ05}, so this premise seems tenable.} On plugging these values in (\ref{eq:fracdeltat}), we end up with $\Delta T \approx 0.36$ K, which displays good agreement with the simulated scenario(s) of $\Delta T \approx 0.44$--$0.54$ K for these regions \citep[Table 2]{MGF09}.

It must be recalled that our goal in this work is not to perform state-of-the-art climate modeling but to formulate a heuristic treatment that captures the salient trends (in consideration of the many variables at play). Furthermore, as intimated in Section \ref{SSecThermEP} and particularly Section \ref{SSecKScale}, the supposition of exponential growth in power consumption and waste heat generation ensures that most of our predictions pertaining to the timescales (of various variables) are quite robust, owing to a weak (i.e., logarithmic) dependence.

\subsection{Global heating from stellar energy conversion}\label{SSecSolEn}

Hitherto, $Q_{\rm tech}$ has been interpreted as extra heating, in addition to that naturally produced by stellar radiation, from a source that would otherwise be sequestered within the planet (e.g. by fossil fuel burning or nuclear reactions). We will now analyze the case wherein energy for technological activity is entirely produced via suitable (e.g., photovoltaic) conversion of the stellar flux that reaches the planetary surface (refer to \citealt{ADV08,WLB15,KA16,KK18,GLW19,HAM19}). 

Ultimately, an appropriate fraction of such energy converted into useful work (for instance, through the production of electricity) is dissipated as heat. At first, it may appear that this energy usage ought not to translate to an elevation of the planetary surface temperature, since all the incoming and outgoing energy is already seemingly included in (\ref{eq:EB}) with $Q_{\rm tech}= 0$. However, the efficient conversion of stellar energy necessitates the collection of incident flux that would otherwise be reflected to space by tapping into methods such as surface texturing \citep{NC13,KLK20} and anti-reflective coatings \citep{RGN11,SES20,SPM20}, which thereby can induce a net decrease of the planetary albedo \citep{Berg2015,BMA16,BKG19,ZSI19,SAK21,XLQ24}.\footnote{We caution, however, that solar energy conversion by means of photovoltaic cells engenders an intricate cascade of climatic effects on Earth currently, which may contribute to either regional heating \citep{BMA16,BKG19,SAK21,SZQ22,KS23} or cooling \citep{MBS14,HLM16,ZX20,XLQ24}; by the same token, future extrapolations to extensive deployment of these cells could have major attendant uncertainties.} If $f_s$ is the fractional area covered by solar panels and $A_s$ is the panel albedo, the balance equation becomes:
\begin{equation}
    \frac{S}{4}\left(1-(1-f_s)A-f_sA_s\right)-\tau\sigma T^4=0,
\end{equation}
which reduces to (\ref{eq:EB}) with $Q_{\rm tech}= 0$ in the limit of $f_s \rightarrow 0$ (i.e., when solar panels and other energy sources like fossil fuels are absent), along expected lines. The above expression is therefore rewritten to highlight this aspect:
\begin{equation}\label{StellEBM}
    \frac{S}{4}(1-A) + Q_s = \tau\sigma T^4,
\end{equation}
where $Q_s$ constitutes an additional heating term stemming from the decrease in albedo attributable to the collecting area covered by photovoltaic panels:
\begin{equation}
    Q_s\equiv\frac{S}{4}f_s\left(A-A_s\right).
\end{equation}
The above result may be viewed as a lower bound because it only comprises the heating from change(s) in albedo and does not incorporate the finite efficiency of photovoltaics \citep{KR18,KLJ20,GDY23}\footnote{For a single junction photovoltaic cell, \citet{SQ61} derived a theoretical efficiency limit, which has been subsequently modified both empirically and theoretically \citep{JN03,KJH13,SR16,EAT20,TM22}.} and further losses during energy conversion. The fraction of stellar flux converted into usable energy flux $\mathcal{P}$ (e.g., by way of photovoltaics) is 
\begin{equation}\label{PowStell}
    \mathcal{P} = \frac{S}{4}f_s\left(1-A_s\right).
\end{equation}
The corresponding temperature increase can be computed by substituting $Q_s$ in (\ref{eq:deltat}). By combining the preceding two equations, we find that 
\begin{equation}\label{QsPrel}
    Q_s = \left(\frac{A-A_s}{1-A_s}\right) \mathcal{P}.
\end{equation} 
Hence, the power derived from stellar radiation conversion is responsible for reduced global heating compared to the previous scenario of $Q_{\rm tech} \sim \mathcal{P}$ motivated in Section \ref{SSecThermEP} for the category of intrinsic technological power sources (e.g., fossil fuels and nuclear reactions). This statement is a consequence of the fact that the term preceding $\mathcal{P}$ in (\ref{QsPrel}) is smaller than unity.

In the idealized limit of $A_s=0$, the contribution to heating can decrease by a factor of $A$ for the same energy consumption (i.e., a fixed value of $\mathcal{P}$), as revealed from inspecting (\ref{QsPrel}); this case is consistent with the simplified results presented in \citet[Section 4]{Berg2015}. The maximum power attainable via photovoltaic cells (assuming an unrealistic 100\% efficiency in stellar energy conversion) corresponds to the case with $f_s=1$ and $A_s=0$ in (\ref{PowStell}), to wit, when all the stellar flux that reaches the entire planetary surface ($S/4$) is collected by panels, without any reflection to space or miscellaneous losses. In this situation, the planet is anticipated to experience heating by an extra $A\times (S/4)$, as revealed by (\ref{QsPrel}). On plugging these values in (\ref{StellEBM}), it can be shown that $T$ for this planet corresponds to that of a perfect absorber, as expected.

To exceed the aforementioned power limit exclusively by means of stellar power, additional stellar flux would need to be redirected towards the surface, or the collecting area should be increased; potential avenues include the deployment of mirrors or photovoltaic panels in space and the construction of Stapledon-Dyson spheres \citep[e.g.,][]{Stap37,FJD60,KSL15}. In such scenarios, the outcome is predicted to yield a further temperature increase, which would be effectively regulated by the analog of (\ref{eq:deltat}) that is pertinent when external energy is fed into the system. 

\subsection{Connection to the Kardashev scale}\label{SSecKScale}

\begin{figure*}
    \centering
\includegraphics[width=1.0\textwidth]{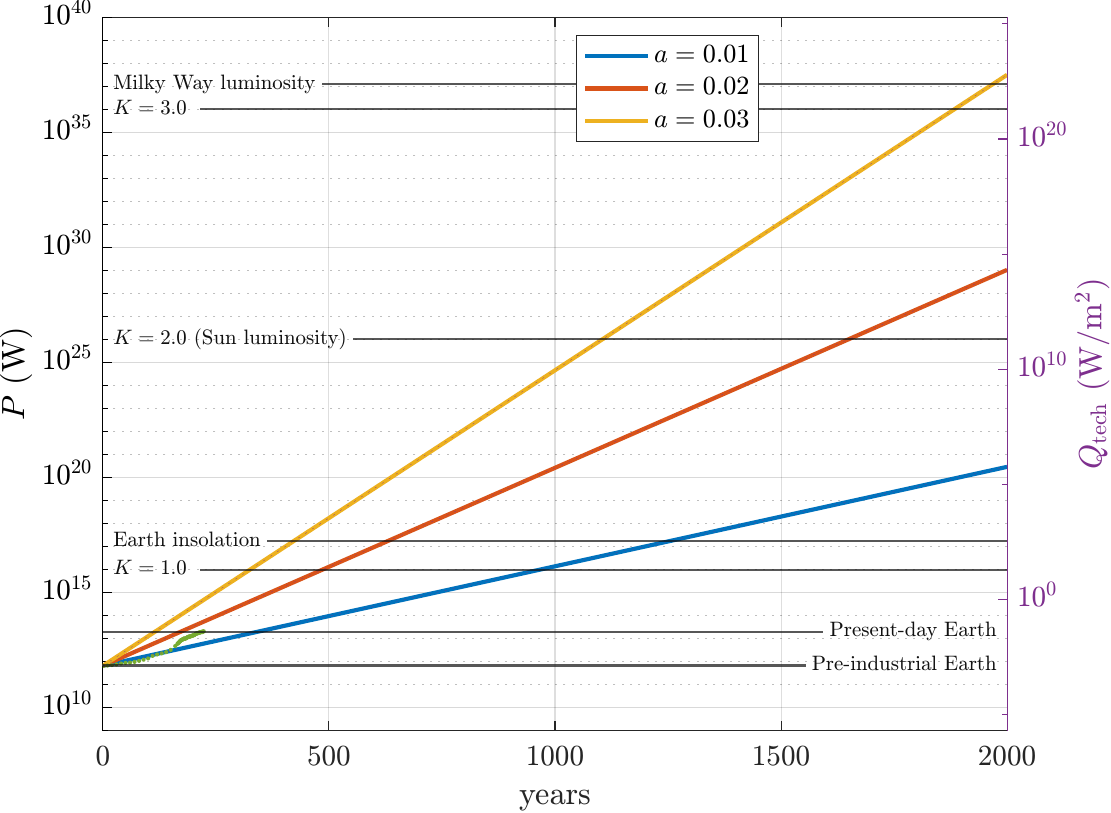}
    \caption{Growth of power consumption $P(t)$ of technological species as per (\ref{eq:heatgrowth}), upon assuming a constant annual growth rate $a$ of 0.01 (blue), 0.02 (red) and 0.03 (yellow). The start time corresponds to that of humankind circa 1800 (i.e., cessation of the preindustrial phase), with an associated power consumption of $P_0$. The green curve represent the actual data for humanity's global power consumption (from \citep{OWID20}). The right axis roughly shows the ensuing global heating flux $Q_{\rm tech}$, obtained from dividing the global power consumption by the total surface area of Earth. 
    }
    \label{fig:growth}
\end{figure*}

Circling back to the discussion in Section \ref{sec:intro}, in his classic paper, \cite{Kardashev1964} proposed that energy consumption can be used to broadly categorize the technological level of extraterrestrial intelligences. Quantitatively, this classification is expressed using the so-called \emph{Kardashev scale}, which is mathematically represented as \citep{Gray2020}:
\begin{equation}\label{Kscale}
    K=\frac{1}{10}\log_{10}\left(\frac{P}{10^6\,\mathrm{W}}\right)
\end{equation}
where $P$ is the power available to the technological species and $K$ is its corresponding level in this scale. 


A frequently adopted prescription for estimating the time required for an advancing technological species to ascend the rungs of the Kardashev scale is to employ a constant exponential growth rate $a$ of energy utilization,\footnote{It is known, however, that presupposing any particular growth pattern involves attendant uncertainty \citep[e.g.,][]{VS00} and may not be tenable, as partly elaborated in Section \ref{SecConc}.} reflecting the observed trend on Earth during the industrial era \citep{SVH75}. This assumption can be quantitatively encapsulated as follows:
\begin{equation}\label{eq:heatgrowth}
    P(t) \approx P_0(1+a)^{t/{\rm 1 yr}}
\end{equation}
where $P_0$ represents the energy employed by the technological species at some initial time: a suitable choice might be the epoch of transition from pre-industrial to industrial societies, although demarcating such an epoch naturally entails a bevy of accompanying subtleties. If we take the logarithm of (\ref{eq:heatgrowth}) and postulate that $a \ll 1$ is valid, we arrive at
\begin{equation}\label{TimeScale}
    t\,({\rm in\, yr}) \approx \frac{1}{a} \ln\left(\frac{P(t)}{P_0}\right),
\end{equation}
on simplification, which reveals that $t$ has a modest dependence on $a$ (inverse proportionality) and a very weak (i.e., logarithmic) dependence on $P_0$. Even if the latter were to vary by many orders of magnitude, it ought to alter the predicted value of $t$ by a relatively moderate amount (e.g., an order of magnitude), thus conferring some degree of robustness. These two vital points should be borne in mind with respect to upcoming results. For instance, as remarked toward the end of Section \ref{SSecThermEP}, reducing the waste heat production -- which is crudely tantamount to diminishing the magnitude of $P$ -- ought not modify our core conclusions substantively.

For reference, global primary energy consumption at the onset of industrial revolution on Earth (circa 1800) was $P_0 \approx 6000$ TWh \citep{Smil2017},\footnote{\scriptsize{\url{https://ourworldindata.org/grapher/global-primary-energy}}} and the average growth rate has been $a \sim 0.02$ during the industrial epoch \citep{OWID20}\footnote{\scriptsize{\url{https://www.oecd-ilibrary.org/energy/primary-energy-supply/indicator/english_1b33c15a-en}}}. Figure~\ref{fig:growth} depicts the evolution of $P$ from (\ref{eq:heatgrowth}), with various assumptions for the growth rate and some useful values for reference. A consequence of the postulated exponential growth is that the relevant parameter (energy consumption) grows by orders of magnitude on timescales of merely hundreds of years, which is consistent with prior publications centered on this theme \citep{SVH75,DD00,HMB09,MHM19,KKH24}.

While the adoption of a constant growth rate may create the impression that a swift transition to ascending levels of the Kardashev scale is bound to happen, our preceding discussion warrants caution. Specifically, we must take into account the inevitable global heating associated with energy consumption (represented by $P$). If, as assumed implicitly in this paper (see Section \ref{SSecThermEP}), the waste heat flux $Q_{\rm tech}$ broadly obeys the relation
\begin{equation}\label{QPrel}
    Q_{\rm tech} \sim \frac{P}{4\pi R_\oplus^2},
\end{equation}
for Earth-sized worlds, it is anticipated that the concomitant warming would escalate as rapidly as $P$. If $P$ is instead generated through the conversion of stellar flux, the accompanying heating only decreases by a fairly modest factor of $A$ at most, as explained in Section \ref{SSecSolEn}. However, the maximum power attainable in this scenario is $P=\left(4\pi R_\oplus^2\right) \times \left(S/4\right)$ (refer to Section \ref{SSecSolEn}). On substituting $S = S_\oplus \equiv 1360$ Wm$^{-2}$ in this expression for $P$, and using (\ref{Kscale}), we end up with $K \approx 1.12$. As a result, even the application of perfectly efficient conversion of incident stellar power cannot enable a putative technological species to exceed $K = 1$ by a sizable margin.

In the forthcoming Section \ref{SSecHeatHab}, we will further delve into the implications for planetary habitability, and we will follow this up by calculating the maximal lifetime and Kardashev scale of technological species in Section \ref{SSecMaxTechno}.

\subsection{Implications of heating for habitability}\label{SSecHeatHab}

\begin{figure*}
    \centering
\includegraphics[width=1.0\textwidth]{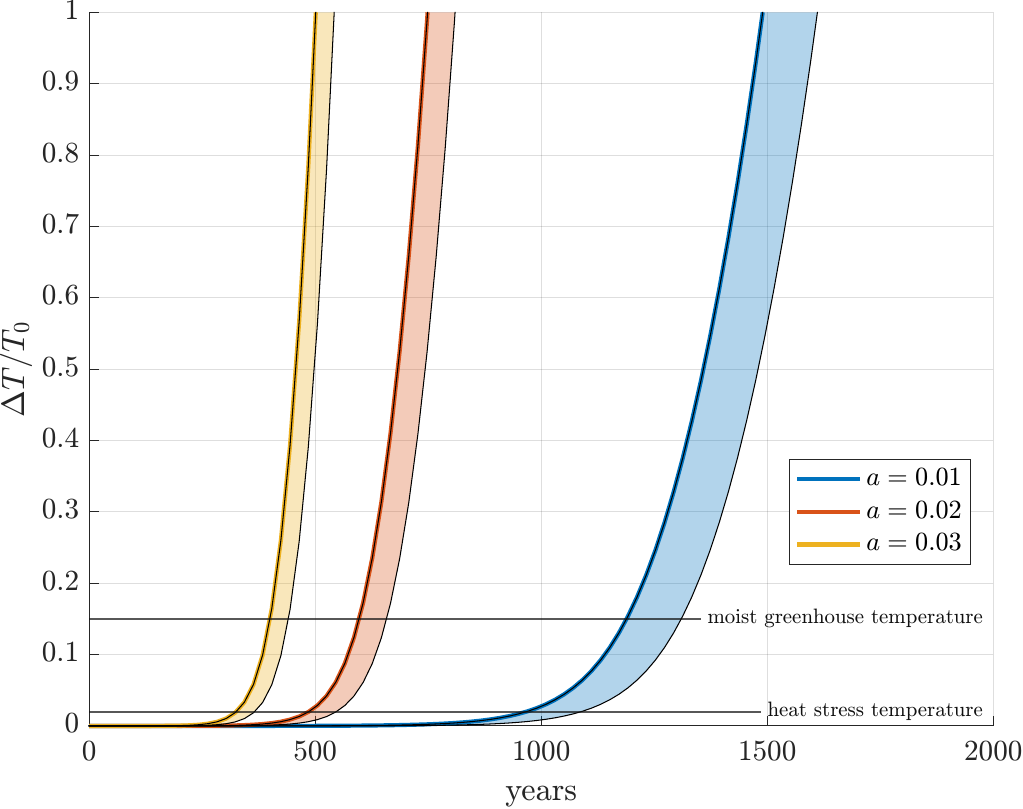}
    \caption{The fractional increase in planetary temperature when power consumption $P$ grows at a constant annual rate $a$ of 0.01 (blue), 0.02 (red) and 0.03 (yellow). Timescale shown on the horizontal axis corresponds to that of Figure~\ref{fig:growth}. The shaded regions encompass the upper and lower limits for $\Delta T/T_0$, with optimistic estimates corresponding to the production of all power via conversion of stellar flux. In plotting the curves, we have assumed a fiducial planetary albedo of $A=0.3$ for Earth-like worlds.}
    \label{fig:heating}
\end{figure*}

We can apply the previous considerations to estimate the ramifications for planetary habitability that ensue from the production of waste heat associated with energy consumption by technological species. In essence, we are interested in forecasting the time it would take for a planet that hosts a technological species -- which pursues an exponential growth of energy consumption -- to experience a rise in temperature so significant as to jeopardize its habitability. This matter was studied by the likes of \citet{SVH75} and \citet{MHM19}.

In Figure \ref{fig:heating}, the fractional temperature increase $\Delta T/T_0$ is plotted as a function of time lapsed (since the commencement of exponential growth) for different annual growth rates of energy assumption by invoking (\ref{eq:fracdeltat}), where the heating term is specified appropriately based on the power source (viz., non-stellar vs stellar energy). The two horizontal lines of $\Delta T/T_0 = 0.02$ and $\Delta T/T_0 = 0.15$ represent the crossing of conceivably vital ``planetary thresholds'' for habitability justified below.\footnote{These thresholds are partially reminiscent of the central ``planetary boundaries'' and ``planetary climate tipping points'' concepts in climate science \citep{LHK08,RSN09,BHB12,SRR15,KR17,OFL18,RCC21,PCC22,RSL23,RGQ23,RDF24,WVA24}, as well as cognate paradigms, but are not equivalent to them.} We notice that a major elevation of surface temperature is expected on short timescales of $\lesssim 1000$ years.

First, let us consider Earth's biosphere as a reference point. It has been known for well over a century that heat stress poses severe impediments to human activities \citep{BH20}, owing to which formulations like the wet-bulb temperature have been introduced for quantifying such effects \citep{JSH05}. Based on heat stress considerations, \citet{SH10} proposed that $\Delta T = 12$ K could render large swathes of Earth uninhabitable for \emph{Placentalia} (a taxon which includes humans), which is also consistent with the study by \citet{ASH21}, who concluded that $\Delta T > 10$ K would induce heat stress across many species. Thus, this numerical value was consequently employed by \citet{MHM19} in their analysis.

However, it is plausible that $\Delta T = 12$ K may not represent an accurate upper bound because a potentially crucial factor could be missing, namely, the intricate interplay of ecological interactions. By incorporating such kinds of dependencies into a suite of numerical models and modeling possible trajectories, \citet{SB18} concluded that a swift temperature increase of merely $\Delta T \approx 6$ K might suffice to drive global biodiversity collapse (see also \citealt{SKT21,MFL24} for elaborations of this Earth system picture). State-of-the-art evidence garnered from mass extinctions in the Phanerozoic indicates that a couple of them were characterized by temperature changes of $\sim 10$ K \citep[e.g.,][]{DSC22,MBB22}. Therefore, when we adopt a fiducial temperature of $T_0 = T_\oplus \equiv 288$ K for the planet and heating of $\Delta T \approx 6$ K, we duly obtain $\Delta T/T_0 \approx 0.02$. 

Second, we pivot to a relatively general phenomenon grounded in physics, since it may be argued that the temperature increases delineated in the preceding paragraphs are innately parochial, to wit, overly reliant on Earth's biomes and biota. A number of numerical studies have demonstrated that the ``moist greenhouse effect'' can be instantiated at a typical temperature of around $330 \pm 10$ K \citep{KWR93,LFC13,GKL18,RMR20}, although the precise value is predicted to be modulated by a variety of stellar and planetary parameters \citep[e.g.,][]{WT15,PSM16,FDA17,KWA17}. 

As a moist greenhouse effect could result in the depletion of substantial inventories of liquid water, it represents a plausible candidate for stymieing planetary habitability for life-as-we-know-it. For the fiducial temperature of $T_0 = 288$ K chosen earlier, a moist greenhouse threshold of approximately $330$ K amounts to setting $\Delta T \approx 42$ K. In turn, the fractional temperature increase in this scenario translates to $\Delta T/T_0 \approx 0.15$, which was introduced previously.

One general point that should be recognized is that $\Delta T/T_0$ is exclusively a function of the ratio of the heating $Q_{\rm tech}$ and the stellar flux at the surface $S\left(1-A\right)/4$, as seen from (\ref{eq:fracdeltat}). Thus, it does not incorporate additional heating that might arise from climate changes (e.g., atmospheric composition) during the growth in energy consumption. Hence, it seems prudent to regard $\Delta T/T_0$ calculated from (\ref{eq:fracdeltat}) as an underestimate of the actual increase in the surface temperature, because of which the attendant timescale gauged herein to reach a given $\Delta T/T_0$ may represent an overestimate (i.e., upper bound of sorts).

\subsection{Maximum lifespan of technospheres}\label{SSecMaxTechno}

\begin{figure*}
    \centering
\includegraphics[width=1.0\textwidth]{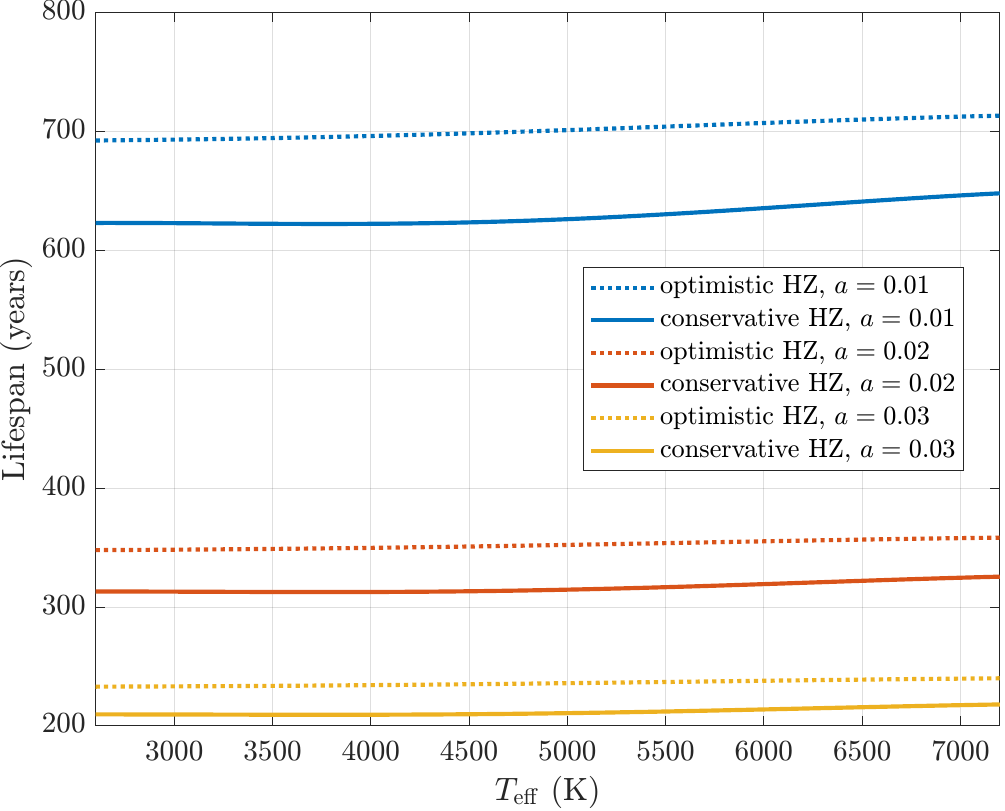}
    \caption{The upper bound on the lifespan of technospheres with increasing energy usage, computed as the maximum time the planet spends within the habitable zone of a star of effective temperature $T_{\rm eff}$; it is assumed that the power consumption of technospheres displays exponential growth at a constant annual rate of $a$. The continuous (dotted) lines are based on the conservative (optimistic) HZ boundaries computed in \citet{Kopparapu2013}.}
    \label{fig:lifespan}
\end{figure*}

The circumstellar habitable zone (HZ) is a cornerstone of astrobiology; its roots date back to the 19th century \citep{Man21}. The lifetime of the continuous habitable zone -- which solely involves abiotic (i.e., neither biological nor technological) factors -- was analyzed in several publications \citep{KWR93,Kopparapu2013,Rushby2013,GCA17,BPL21,WYT22,TW23}, based on the consideration of a gradually increasing stellar luminosity. We will now proceed to estimate the upper bound on the lifetime of technospheres, where the key determinant of longevity is the energy consumption growth rate of technological species. 

In classic models of the HZ, the inner and outer limits are expressed in terms of the corresponding top-of-atmosphere stellar flux $S_{\rm max}$ and $S_{\rm min}$, respectively regulated by the maximum and moist greenhouse effect \citep{KWR93,Kopparapu2013,RMR18,RAF19,MA24}. These thresholds can, to a certain degree, be perceived as ``universal'' because they are built from well-defined atmospheric and surface physics and chemistry; hence they may pose fairly stringent constraints on planetary habitability, although exceptions could exist.

At an initial time (e.g., as when $P = P_0$ in Section \ref{SSecKScale}), let us suppose that the planet harboring a technological species -- albeit one that has not yet entered the exponential growth phase -- is at the outer edge of the habitable zone and receives a stellar flux of $S_{\rm min}$. The planet cannot exist any further away from the host star, because liquid water would no longer be viable on the planetary surface (for a given atmospheric composition), and the world may therefore not be capable of sustaining a technological species with an Earth-like biochemistry (viz., with liquid water as the solvent).

With the onset of exponential growth in energy consumption, the surface temperature would rise due to the increasing additional heating resulting from technological activity, as revealed from (\ref{eq:fracdeltat}). It may be valid to suppose that the stellar flux and the planetary albedo both remain constant throughout the period of interest (but we do briefly comment on exceptions to this implicit norm in Section \ref{SecConc}), because this interval is relatively short, as demonstrated hereafter. Consequently, any changes in the net energy input at the surface are attributable to variations in $Q_{\rm tech}$ in our model. 

We will now treat $Q_{\rm tech}$ as though it were an extra energy flux applied at the top of the atmosphere; the benefit will become apparent in the next paragraph. Of this energy input, only a fraction would pass through the atmosphere and reach the surface; if we denote the latter by $\tilde{Q}_{\rm tech}$, it is likely that $\tilde{Q}_{\rm tech} < Q_{\rm tech}$. Another way to frame this statement is that the real warming induced by $Q_{\rm tech}$ would be more severe than the assumption made herein, since this heat flux is actually engendered \emph{within} the system (i.e., as an additional heat layer at the planetary surface). Thus, the lifetime of technospheres derived below constitutes an upper bound because the heating at the surface in our modified framework (which is $\tilde{Q}_{\rm tech}$) is smaller than the true heat flux $Q_{\rm tech}$. 

Given that we have ``lifted'' $Q_{\rm tech}$ to the top of the atmosphere to calculate the upper bound on the technosphere lifetime, we can write the total top-of-the-atmosphere flux at any moment in time as
\begin{equation}\label{Sgrowth}
    S(t) \approx Q_{\rm tech}(t) + S_{\rm min},
\end{equation}
with the initial condition $S(t=0) \approx S_{\rm min}$, as motivated earlier.\footnote{Strictly speaking, the initial condition should be $S(t=0) = S_{\rm min} + Q_{\rm tech}(t=0)$, but the second term on the RHS (i.e., waste heat flux) may be neglected compared to the first term.} As $Q_{\rm tech}(t)$ grows exponentially with time, the same trend is evinced by $S(t)$. Eventually, the magnitude of $S(t)$ will exceed $S_{\rm max}$, at which stage the loss of liquid water and habitability is instantiated. In other words, because we require $S(t) \leq S_{\rm max}$ for maintaining habitable conditions, this relation is expressible as
\begin{equation}\label{eq:cond}
    Q_{\rm tech} \leq S_{\rm max}-S_{\rm min} \equiv \Delta S,
\end{equation}
upon employing (\ref{Sgrowth}). Instead of choosing an initial stellar flux of $S_{\rm min}$, if we selected a higher value, then the upper bound for $Q_{\rm tech}$ analogous to (\ref{eq:cond}) would decrease, as would the timescale to reach the threshold of $S_{\rm max}$. It is plausible, therefore, that the timescale(s) for truly crossing the boundaries of habitability may be lower than the upper bound computed herein.

The values of $S_{\rm max}$ and $S_{\rm min}$ exhibit some variation across publications \citep{SBJ16,LL19}, such as dependence on atmospheric composition \citep{PG11,AZ15,RMR18,MPC21}, planetary mass and rotation rate \citep{KRS14,YBF14,KWH16}, and the properties of biota \citep{SRO19,ML20}, to name a handful. We shall adopt the well-known prescription of \citet{Kopparapu2013} for the HZ boundaries around F-, G-, K- and M-type stars. We will compute the duration $\Delta t$ required for $Q_{\rm tech}$ to surpass $\Delta S$, which represents the maximal lifetime of technospheres, as justified in the prior paragraphs. On invoking (\ref{eq:heatgrowth}) and presuming that (\ref{QPrel}) holds true, we substitute these expressions in (\ref{eq:cond}), thereby yielding
\begin{equation}\label{eq:lifespan}
    \Delta t \approx \frac{\ln\left(\Delta S/Q_{\rm tech,0}\right)}{\ln\left(1+a\right)}\ {\rm years}
\end{equation}
where $Q_{\rm tech,0} \sim P_0/\left(4\pi R_\oplus^2\right)$ is the heating from technological activity at the initial time (i.e., commencement of exponential growth). This lifetime is shown in Figure \ref{fig:lifespan} for various values of $a$, after drawing on both the conservative and optimistic HZ boundaries from \citet{Kopparapu2013}. In all instances, we observe that the maximal lifetime of technospheres is $<1000$ years, provided that $a \gtrsim 0.01$ throughout the epoch of interest.

\begin{figure*}
    \centering
\includegraphics[width=1.0\textwidth]{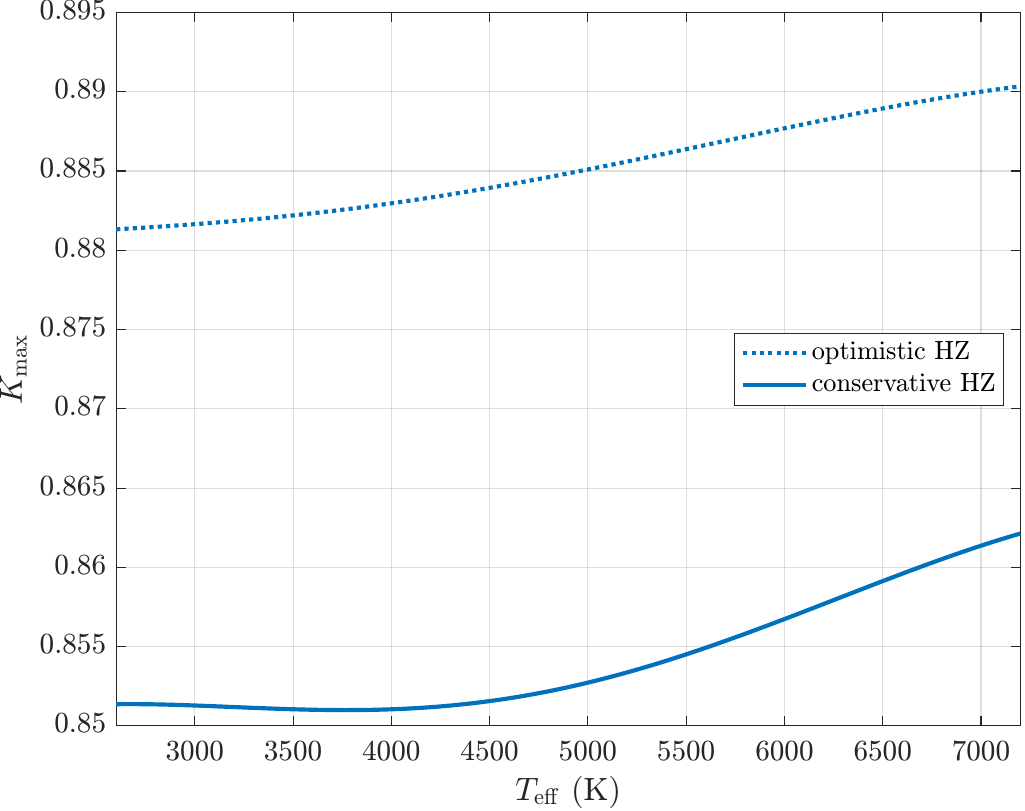}
    \caption{The upper bound on the Kardashev scale level attainable by a technological species inhabiting an Earth-like planet in the habitable zone, before the loss of habitability due to the growth in energy consumption. The two scenarios depicted represent optimistic and conservative limits for staying within the habitable zone.}
    \label{fig:kmax}
\end{figure*}

Lastly, a perusal of (\ref{eq:cond}) reveals that the upper bound on $Q_{\rm tech}$ is set by $\Delta S$, from which we can duly estimate the maximal power $P_{\rm max}$ associated with the loss of habitability on an Earth-like planet containing a growing technological species. If (\ref{QPrel}) is applicable, we end up with $P_{\rm max} \sim 4\pi R_\oplus^2 \Delta S$. Upon plugging this value in (\ref{Kscale}), the upper limit of the Kardashev scale ($K_{\rm max}$) attainable by such an ostensible technological species is found to be
\begin{equation}\label{KscaleMax}
    K_{\rm max} \approx \frac{1}{10}\log_{10}\left(\frac{4\pi R_\oplus^2 \Delta S}{10^6\,\mathrm{W}}\right),
\end{equation}
and the ensuing results are plotted in Figure \ref{fig:kmax}. It is evident that the maximal Kardashev scale $K_{\rm max}$ increases with the stellar temperature, but only by a minimal amount - in all the analyzed cases, $K_{\rm max}$ exists in a narrow range of about $0.85$--$0.9$, slightly lower than unity. A corollary is that, for the simple model(s) in this paper, the maximal Kardashev scale under this particular set of assumptions may be less than unity. A vital facet of (\ref{KscaleMax}) is that this limit is, at first glimpse, independent of both the modality and rate (or equivalently timescale) of growth in energy consumption. Hence, this ceiling theoretically represents the maximal Kardashev level attainable by a technological species within a particular terrestrial planetary environment, irrespective of the specific energy growth model adopted, and might be accordingly construed as a physical (i.e., thermodynamical) limit endowed with a certain degree of generality.

\section{Discussion and Conclusions}\label{SecConc}

As articulated in Section \ref{sec:intro}, waste heat production is an inevitable consequence of thermodynamics. In this paper, we investigated the myriad effects of technological species -- and their accompanying technospheres -- that experience exponential growth with respect to their energy consumption and waste production. 

Our analysis suggests that, if the energy growth rate is of order $1\%$ per year, the maximal lifetime of such putative technospheres is ephemeral compared to stellar evolution. From Figure \ref{fig:heating}, we notice that significant heating -- which can drive the planet beyond potentially dangerous thresholds (consult Section \ref{SSecHeatHab}) -- due to exponential waste heat generation is predicted to occur on a short timescale of $\lesssim 1000$ years; these findings are broadly compatible with prior studies \citep{SVH75,MHM19}. Likewise, a scrutiny of Figure \ref{fig:lifespan} reveals that the upper bound on the lifetime of technospheres is relatively insensitive to stellar spectral type, and is merely hundreds of years in duration. These results stand in stark contrast to Earth's long-lived biosphere that has existed for perhaps $\sim 4$ Gyr, as per the available geological evidence \citep{EJJ19,KL20,WBF23}, phylogenetic inferences \citep{BPC18,MMS23,MAM24}, and simple mathematical models \citep{ML17}.

We may identify three classes of trajectories that seem consistent with our modeling in this work, as listed below. 
\begin{enumerate}
    \item Technological species that pursue relentless exponential growth of energy consumption beyond the planet's safe operating thresholds render themselves extinct on short timescales of typically $\lesssim 1000$ years.
    \item Technological species transition from the phase of exponential growth in energy consumption to either an indefinite period of (near-)zero growth or even intervals of negative growth.
    \item Technological species venture beyond their home planet(s), thereby utilizing space infrastructure to produce and dissipate energy, as well as to perform other technological activities.
\end{enumerate}
We shall now tackle each of these trajectories in more detail and briefly comment on their implications. We emphasize, however, that we do not claim to exhaust all conceivable routes in our pr\'ecis. A comparatively exhaustive treatment of Earth's technosphere $1000$ years from the present day, which draws on the domain of future studies, was recently explored in \citet{HPK24}.

For instance, technological species may deliberately diminish the net incident stellar flux (which we held constant on the timescales considered) to offset increasing waste heat generation.\footnote{On the other hand, such a reduction could translate to less energy accessible for, say, photovoltaic conversion, thus possibly stymieing growth (and dovetailing with Trajectory \#2).} Loosely overlapping with the forthcoming trajectory \#3, this procedure may entail undertaking large-scale (macro)engineering activities near the surface and in the atmosphere -- such as inputting aerosols to raise albedo \citep{RTT08,DK13,MRK18,MTO18,MK19} -- or even in outer space: constructing starshades \citep{JTE89,RA06,EG17,BLS22,KR22} and variants thereof \citep{POL06,BLCS,BH23,BKK23} or nudging the planet into a wider orbit \citep{KLA01,CRM02,DGK04}are a couple of such examples. While a subset of these projects have witnessed extensive theoretical explication and assessments, note that potential lacunae and issues with regard to the engineering, environmental, economic, social, political, and ethical dimensions have been broached \citep[e.g.,][]{VL11,BCV12,GW12,CBC13,IKL16,OM16,PGB20,RWS23}.

With the aforementioned caveat and one credible example of a trajectory not elaborated in this paper out of the way, we will now proceed to elucidate the three categories previously introduced a couple of paragraphs ago. \\

\noindent \underline{Trajectory \#1:} In case the majority of technological species are driven to extinction over the course of their postulated interval of exponential growth on account of violating one or more ``planetary thresholds'' (sketched in Section \ref{SSecHeatHab}), their mean lifetime during the detectable phase might become markedly low, as intimated earlier. In this specific scenario, the estimated number of communicative technological species in the Galaxy would drop commensurately as per the famous Drake equation formulated by Frank \cite{Drake1965}, since one of the constituent factors is the average longevity of such species \citep{MB65,Sagan1966,DS92,MJB06,SS11,Vakoch2015}.

If the number of these technological species is indeed modest -- based on the Drake equation and its probabilistic descendants \citep{GBB12,CM12,MVM23}, which are regulated by the longevity -- this outcome might offer a (partial) ``solution'' to the famous Fermi paradox \citep{MS02,Webb2015,Cirkovic2018,DF19}: we have not encountered technological species because they are rare at any given moment in time. This explanation of the Fermi paradox in terms of long-term sustainability -- or rather, the lack thereof -- was surveyed by the likes of \citet{DD00} and \citet{HMB09} and fleshed out in subsequent publications by \citet{FCA18,MHM19,BA20,ML21,SFC21,WB22}. 

However, even if the majority of technological species are short-lived, in accordance with Lindy's law \citep{NNT12}, this facet does not necessarily preclude us from detecting their attendant technosignatures \citep{BG24}. As this trajectory encompasses energy-intensive species, a variety of detectable technosignatures are conceivable \citep[e.g.,][Chapter 9]{ML21}, such as silicon-based photovoltaic cells and space-based mirrors for implementing stellar energy conversion \citep{KSL15,LL17,KKH24}, nitrogen dioxide from combustion \citep{KAH21}, and urban heat islands \citep{KB15,BKL18}. Moreover, a fraction of these signatures may persist over substantial timescales, even perhaps long after their creators have become extinct or departed the world \citep{DLH91,CVS19,WHF22}. \\

\noindent \underline{Trajectory \#2:} The pivotal outcomes that could stem from reifying protracted exponential growth of any given resource(s) have attracted intense debates since at least the classic treatise by \citet{TRM}. A diverse collection of perspectives advanced by intellectuals spanning multifarious schools of thought -- dating back to \citet{JSM48} at the minimum -- has advocated in favor of humankind ultimately shifting to a paradigm of (near-)zero growth of that particular resource(s) \emph{sensu lato} (i.e., a quasi-steady state in essence) or even embracing suitable intervals of negative growth, which may be interpreted as degrowth \citep{NGR,HED73,MMR,FH78,HED96,PB06,HTO07,SL09,CK10,FCY10,UB11,DF11,JR12,DDK14,CCD15,MS15,SGB15,KKL18,JV19,VS19,JH20,KPD20,MSR20,WLK20,GH21,KL21,MRG21,JBF22,HKJ22,KS22,SVV22,BKN24}. It should be recognized, however, that these paradigms have garnered a plethora of criticisms, critiques, and responses (e.g., highlighting purported lacunae) from multiple standpoints \citep{PE00,SLH05,MPV10,VDB11,VDB17,AM12,MF12,HHF12,DS12,RN13,SQ13,CSO17,KQ17,BO17,HBH18,GL18,SSB18,BK19,CLS20,TT20,DHS,BG21,PHK,HR24}.

For an annual growth rate ($a$) obeying $a \ll 1$, it can be shown that the maximal lifetime of technospheres is proportional to $1/a$ on taking this limit in (\ref{eq:lifespan}). Hence, if a technological species is capable of drastically reducing the growth rate, its longevity will be proportionally enhanced, although the threshold imposed by (\ref{eq:cond}) is still at play. By the same token, shifting from exponential growth to, say, power-law growth can also extend the lifespan of the technosphere because an exponential function increases faster than a power law in due course; this statement is likewise valid for logistic growth, which is tantamount to having the growth rate approach zero asymptotically. Lastly, the pursuit of growth might engender collapse, but not extinction, of technological species, thence serving as a negative feedback that suppresses growth and elevates their lifespan, albeit with a reduced energy footprint.

At this juncture, it is worth underscoring that, as long as some version of energy growth is existent, the magnitude of $Q_{\rm tech}$ would presumably continue to rise until the planetary thresholds from Section \ref{SSecHeatHab} are breached. In other words, these thresholds are typically applicable to generic planet-bound technological species, but the particular timescales for crossing them are contingent on the modalities of growth and the corrective actions taken by technological species. Since the latter are challenging to forecast, this work has focused on a single growth pattern (viz., exponential growth), with different ansatzen (e.g., non-monotonic patterns) representing the basis of desirable future work.

Thus, given that several studies have proposed that humanity is more likely to detect long-lived technological species \citep[cf.][]{KFS20,BC21,BG24}, it is not implausible that these species might have switched to a sustainable state of minimal (perhaps even near-zero or negative) energy growth or to slower modes (e.g., power laws) of growth; of the $10$ future scenarios envisioned for Earth's technosphere in $1000$ years by \citet{HPK24}, $5$ of them evince zero growth. One subtlety that should, however, be borne in mind is that stable technological species may entail modest energy consumption and waste heat dissipation, thereby perhaps lowering their likelihood of detection on the one hand while simultaneously boosting longevity on the other hand \citep[cf.][]{TRS18}.\footnote{\url{https://www.kschroeder.com/weblog/the-deepening-paradox}} \\

\noindent \underline{Trajectory \#3:} Hitherto, we have chronicled either the looming specter of extinction because of the pursuit of untenable growth (Trajectory \#1) or longevity with the possible stipulation of maintaining a restricted energy footprint (Trajectory \#2), the latter of which stems from (\ref{eq:cond}). It is natural to wonder, therefore, whether the existence of alternative pathways that can synthesize the pros of Trajectory \#1 and Trajectory \#2 is feasible from a theoretical viewpoint.

To resolve this matter, let us examine (\ref{eq:cond}), which signifies a rough upper bound on the allowed heat flux. This limit was interpreted as capping the permitted energy consumption (i.e., power) and the Kardashev scale at the end of Section \ref{SSecMaxTechno}: refer to (\ref{KscaleMax}) and the associated exposition. However, this line of reasoning is implicitly valid only if the total surface area is held fixed. In contrast, if technological species were to expand their activities into space -- which has been advocated by many to yield a myriad of benefits to humankind \citep{JDB29,Stap37,FJD60,ONeill1977,FJ85,CS94,NB03,AAH01,DL07,MB09,IAC10,RJR13,GM14,LF15,IAC16,MS17,RZ19,RZ21,MMC22,JR23}, although this stance has concomitantly garnered its share of critiques \citep{LW10,DV12,RES15,SM16,LB19,KS19,JT19,DD20,MK21,GSP21,SWT21} -- these entities could decide to deploy space infrastructure to generate energy and dissipate heat over a much wider area \citep[cf.][]{ZLZ01,KVW17}.\footnote{From the perspective of creating heat sinks, an extremely speculative example by \citet{Opatrny2016} proposed that technological species with $K\gtrsim 1$ might control and utilize small black holes for this purpose (see also \citealt{AL24}).}

By doing so, the aforementioned bound(s) on the power and the Kardashev scale may be circumvented to some degree. A patently similar scenario involves transferring various energy-intensive processes and activities off-world, not to empty space as such, but to uninhabited and uninhabitable objects instead: on these worlds -- which can constitute a subset of the putative so-called ``service worlds'' \citep{WHF22} -- exceeding the prior flux and power thresholds may not pose significant impediments to implementing certain energetic components of a technosphere (e.g., stellar energy transduction) owing to the posited absence of habitable conditions and/or a biosphere to begin with.

Technological species belonging to Trajectory \#3 might, in principle, readily surpass a Kardashev scale of unity -- unlike their planet-based counterparts in Section \ref{SSecMaxTechno} exhibiting $K_{\rm max}\approx 1$ -- without perhaps sacrificing their longevity or sustainability (we address this theme in the next two paragraphs); the scenario S9 (out of $10$) outlined in \citet{HPK24} for Earth's technosphere $1000$ years in the future envisages expansion into space by postbiological agents (artificial intelligences) unfolding at a growth rate of $1\%$. Trajectory \#3, as remarked before in Section \ref{SecConc}, shares select thematic links (albeit not equivalence) with alternative pursuits of large-scale engineering on the planet or in space as potential solutions. In case such technological species are viable and actually exist in the Milky Way, they may present us with a relatively strong opportunity to detect their accompanying technosignatures; the possible salient characteristics (e.g., optimization) of this category of technological species were discussed, \emph{inter alia}, by \citet{CB06}. 

We caution that unceasing exponential growth, even amidst the much broader horizons of space, might be rendered impractical at some stage in particular cases (as touched on shortly): if so, technological species may eventually adopt a variant of Trajectory \#2 and curb their growth. Furthermore, whether Trajectory \#3 can be instantiated while conceivably grappling with the specter of Trajectory \#1 at a certain juncture (e.g., early on in history) could constitute an open question. An interesting feature of partial relevance in this context is the timescales corresponding to various phenomena. We have shown in this paper that the exponential growth of putative technological species on Earth-like planets can become unsustainable in $\sim 10^3$ yr under specific assumptions, whereas the duration for effectuating self-sustaining and stable planetary-scale modification (terraforming) of another world \emph{in toto} might be up to $\lesssim 10^5$ yr with current human capabilities \citep[cf.][]{MM01,MB09}, for reaching neighboring stars with chemical propulsion is conventionally $\sim 10^5$ yr, and for settling the entire Milky Way is estimated to be $\sim 10^7$--$10^9$ yr \citep{EMJ76,EMJ81,CM09,NF13,ML16,CFW19,HMF22}.

We will not delve further into the complex dynamics of space settlement and galactic expansion because this is not the central theme of the paper, and equally crucially, there are a multitude of intricate variables at play, ranging from engineering (e.g., propulsion systems, self-replication) and scientific to sociological and economic considerations. It has been suggested by some authors that exponential growth of technological species on a galactic scale is ultimately not viable, due to bottlenecks on availability of resources or energy \citep{SVH75,NS81,MHM19}. \\

\noindent {\bf Coda:} To round off this discourse, two of the three core themes that underpin astrobiology (arguably comprising its \emph{raison d'{\^e}tre}) are centered on the fundamental questions of (1) the potential existence and distribution of extraterrestrial life and (2) the future of life on Earth and beyond \citep[e.g.,][]{CH05,DNA08,DGW16,MA24}. By employing simple thermodynamic models, we have attempted to gauge the maximal lifetime of putative technological species and their technospheres characterized by exponential growth of energy consumption (a category which crucially encompasses humans in this epoch), since this endeavor has a direct bearing on the preceding two themes in this discipline.

Hence, even while several avenues that evidently warrant further research beyond this preliminary undertaking are apparent -- such as notably harnessing sophisticated climate models to self-consistently calculate the temporal climate (co)evolution of temperate terrestrial planets that host technospheres fuelled by sundry sources of energy or incorporating alternative modes of energy growth (e.g., logistic or power law growth) -- which might consequently refine or modify (or perhaps even contradict) our central findings, we nevertheless hope that this work may serve to capture some of the salient physical details and pave the way for in-depth treatments to chart the trajectories accessible by technological species in the universe, including humans, as well as the ensuing signatures of technological activities (viz., technosignatures).

\section*{Acknowledgments}
We are grateful to the reviewers for their constructive reports, which helped improve the quality and scope of the paper. M.L. acknowledges support from the NASA Exobiology program under grant 80NSSC22K1009.

\bibliographystyle{abbrvnat}
\bibliography{TechnoLifetime}

\begin{thebibliography}{344}
\providecommand{\natexlab}[1]{#1}
\providecommand{\url}[1]{\texttt{#1}}
\expandafter\ifx\csname urlstyle\endcsname\relax
  \providecommand{\doi}[1]{doi: #1}\else
  \providecommand{\doi}{doi: \begingroup \urlstyle{rm}\Url}\fi

\bibitem[{Abbass} et~al.(2022){Abbass}, {Qasim}, {Song}, {Murshed}, {Mahmood}, and {Younis}]{AQS22}
K.~{Abbass}, M.~Z. {Qasim}, H.~{Song}, M.~{Murshed}, H.~{Mahmood}, and I.~{Younis}.
\newblock {A review of the global climate change impacts, adaptation, and sustainable mitigation measures}.
\newblock \emph{Environ. Sci. Pollut. Res.}, 29\penalty0 (28):\penalty0 42539--42559, June 2022.
\newblock \doi{10.1007/s11356-022-19718-6}.

\bibitem[{Allen} et~al.(2011){Allen}, {Lindberg}, and {Grimmond}]{ALG11}
L.~{Allen}, F.~{Lindberg}, and C.~S.~B. {Grimmond}.
\newblock {Global to city scale urban anthropogenic heat flux: model and variability}.
\newblock \emph{Int. J. Climatol.}, 31\penalty0 (13):\penalty0 1990--2005, Nov. 2011.
\newblock \doi{10.1002/joc.2210}.

\bibitem[{Anderson} and {M'Gonigle}(2012)]{AM12}
B.~{Anderson} and M.~{M'Gonigle}.
\newblock {Does ecological economics have a future?: Contradiction and reinvention in the age of climate change}.
\newblock \emph{Ecol. Econ.}, 84:\penalty0 37--48, Dec. 2012.
\newblock \doi{10.1016/j.ecolecon.2012.06.009}.

\bibitem[{Angel}(2006)]{RA06}
R.~{Angel}.
\newblock {Feasibility of cooling the Earth with a cloud of small spacecraft near the inner Lagrange point (L1)}.
\newblock \emph{Proc. Natl. Acad. Sci.}, 103\penalty0 (46):\penalty0 17184--17189, Nov. 2006.
\newblock \doi{10.1073/pnas.0608163103}.

\bibitem[{Archer}(2016)]{DA16}
D.~{Archer}.
\newblock \emph{{The Long Thaw: How Humans Are Changing the Next 100,000 Years of Earth's Climate}}.
\newblock Princeton: Princeton University Press, 2016.

\bibitem[{Asseng} et~al.(2021){Asseng}, {Sp{\"a}nkuch}, {Hernandez-Ochoa}, and {Laporta}]{ASH21}
S.~{Asseng}, D.~{Sp{\"a}nkuch}, I.~M. {Hernandez-Ochoa}, and J.~{Laporta}.
\newblock {The upper temperature thresholds of life}.
\newblock \emph{Lancet Planet. Health}, 5\penalty0 (6):\penalty0 e378--e385, 2021.
\newblock \doi{10.1016/S2542-5196(21)00079-6}.

\bibitem[{Balbi} and {{\'C}irkovi{\'c}}(2021)]{BC21}
A.~{Balbi} and M.~M. {{\'C}irkovi{\'c}}.
\newblock {Longevity Is the Key Factor in the Search for Technosignatures}.
\newblock \emph{Astron. J.}, 161\penalty0 (5):\penalty0 222, May 2021.
\newblock \doi{10.3847/1538-3881/abec48}.

\bibitem[{Balbi} and {Grimaldi}(2024)]{BG24}
A.~{Balbi} and C.~{Grimaldi}.
\newblock {Technosignatures Longevity and Lindy's Law}.
\newblock \emph{Astron. J.}, 167\penalty0 (3):\penalty0 119, Mar. 2024.
\newblock \doi{10.3847/1538-3881/ad217d}.

\bibitem[{Bardi}(2011)]{UB11}
U.~{Bardi}.
\newblock \emph{{The Limits to Growth Revisited}}.
\newblock SpringerBriefs in Energy. New York: Springer, 2011.
\newblock \doi{10.1007/978-1-4419-9416-5}.

\bibitem[{Barnes} et~al.(2009){Barnes}, {Jackson}, {Greenberg}, and {Raymond}]{BJG09}
R.~{Barnes}, B.~{Jackson}, R.~{Greenberg}, and S.~N. {Raymond}.
\newblock {Tidal Limits to Planetary Habitability}.
\newblock \emph{Astrophys. J. Lett.}, 700\penalty0 (1):\penalty0 L30--L33, July 2009.
\newblock \doi{10.1088/0004-637X/700/1/L30}.

\bibitem[{Barnosky} et~al.(2012){Barnosky}, {Hadly}, {Bascompte}, {Berlow}, {Brown}, {Fortelius}, {Getz}, {Harte}, {Hastings}, {Marquet}, {Martinez}, {Mooers}, {Roopnarine}, {Vermeij}, {Williams}, {Gillespie}, {Kitzes}, {Marshall}, {Matzke}, {Mindell}, {Revilla}, and {Smith}]{BHB12}
A.~D. {Barnosky}, E.~A. {Hadly}, J.~{Bascompte}, E.~L. {Berlow}, J.~H. {Brown}, M.~{Fortelius}, W.~M. {Getz}, J.~{Harte}, A.~{Hastings}, P.~A. {Marquet}, N.~D. {Martinez}, A.~{Mooers}, P.~{Roopnarine}, G.~{Vermeij}, J.~W. {Williams}, R.~{Gillespie}, J.~{Kitzes}, C.~{Marshall}, N.~{Matzke}, D.~P. {Mindell}, E.~{Revilla}, and A.~B. {Smith}.
\newblock {Approaching a state shift in Earth's biosphere}.
\newblock \emph{Nature}, 486\penalty0 (7401):\penalty0 52--58, June 2012.
\newblock \doi{10.1038/nature11018}.

\bibitem[{Barron-Gafford} et~al.(2016){Barron-Gafford}, {Minor}, {Allen}, {Cronin}, {Brooks}, and {Pavao-Zuckerman}]{BMA16}
G.~A. {Barron-Gafford}, R.~L. {Minor}, N.~A. {Allen}, A.~D. {Cronin}, A.~E. {Brooks}, and M.~A. {Pavao-Zuckerman}.
\newblock {The Photovoltaic Heat Island Effect: Larger solar power plants increase local temperatures}.
\newblock \emph{Sci. Rep.}, 6:\penalty0 35070, Oct. 2016.
\newblock \doi{10.1038/srep35070}.

\bibitem[{Baum} et~al.(2022){Baum}, {Low}, and {Sovacool}]{BLS22}
C.~M. {Baum}, S.~{Low}, and B.~K. {Sovacool}.
\newblock {Between the sun and us: Expert perceptions on the innovation, policy, and deep uncertainties of space-based solar geoengineering}.
\newblock \emph{Renew. Sustain. Energy Rev.}, 158:\penalty0 112179, Apr. 2022.
\newblock \doi{10.1016/j.rser.2022.112179}.

\bibitem[{Beech}(2009)]{MB09}
M.~{Beech}.
\newblock \emph{{Terraforming: The Creating of Habitable Worlds}}.
\newblock Astronomers' Universe. New York: Springer, 2009.
\newblock \doi{10.1007/978-0-387-09796-1}.

\bibitem[{Bell}(2008)]{LEB08}
L.~E. {Bell}.
\newblock {Cooling, Heating, Generating Power, and Recovering Waste Heat with Thermoelectric Systems}.
\newblock \emph{Science}, 321\penalty0 (5895):\penalty0 1457, Sept. 2008.
\newblock \doi{10.1126/science.1158899}.

\bibitem[{Bellamy} et~al.(2012){Bellamy}, {Chilvers}, {Vaughan}, and {Lenton}]{BCV12}
R.~{Bellamy}, J.~{Chilvers}, N.~E. {Vaughan}, and T.~M. {Lenton}.
\newblock {A review of climate geoengineering appraisals}.
\newblock \emph{Wiley Interdiscip. Rev. Clim. Change}, 3\penalty0 (6):\penalty0 597--615, Nov. 2012.
\newblock \doi{10.1002/wcc.197}.

\bibitem[{Berdyugina} et~al.(2018){Berdyugina}, {Kuhn}, {Langlois}, {Moretto}, {Krissansen-Totton}, {Catling}, {Grenfell}, {Santl-Temkiv}, {Finster}, {Tarter}, {Marchis}, {Hargitai}, and {Apai}]{BKL18}
S.~V. {Berdyugina}, J.~R. {Kuhn}, M.~{Langlois}, G.~{Moretto}, J.~{Krissansen-Totton}, D.~{Catling}, J.~L. {Grenfell}, T.~{Santl-Temkiv}, K.~{Finster}, J.~{Tarter}, F.~{Marchis}, H.~{Hargitai}, and D.~{Apai}.
\newblock {The Exo-Life Finder (ELF) telescope: New strategies for direct detection of exoplanet biosignatures and technosignatures}.
\newblock In H.~K. {Marshall} and J.~{Spyromilio}, editors, \emph{Ground-based and Airborne Telescopes VII}, volume 10700 of \emph{Society of Photo-Optical Instrumentation Engineers (SPIE) Conference Series}, page 107004I, July 2018.
\newblock \doi{10.1117/12.2313781}.

\bibitem[{Berg} et~al.(2015){Berg}, {Hartley}, and {Richters}]{Berg2015}
M.~{Berg}, B.~{Hartley}, and O.~{Richters}.
\newblock {A stock-flow consistent input-output model with applications to energy price shocks, interest rates, and heat emissions}.
\newblock \emph{New J. Phys.}, 17\penalty0 (1):\penalty0 015011, Jan. 2015.
\newblock \doi{10.1088/1367-2630/17/1/015011}.

\bibitem[{Bernal}(1929)]{JDB29}
J.~D. {Bernal}.
\newblock \emph{{The World, the Flesh and the Devil: An Inquiry into the Future of the Three Enemies of the Rational Soul}}.
\newblock London: Kegan Paul, Trench, Trubner \& Co., 1929.

\bibitem[{Betts} et~al.(2018){Betts}, {Puttick}, {Clark}, {Williams}, {Donoghue}, and {Pisani}]{BPC18}
H.~C. {Betts}, M.~N. {Puttick}, J.~W. {Clark}, T.~A. {Williams}, P.~C.~J. {Donoghue}, and D.~{Pisani}.
\newblock {Integrated genomic and fossil evidence illuminates life’s early evolution and eukaryote origin}.
\newblock \emph{Nat. Ecol. Evol.}, 2\penalty0 (10):\penalty0 1556--1562, 2018.
\newblock \doi{10.1038/s41559-018-0644-x}.

\bibitem[{Bewick} et~al.(2013){Bewick}, {L{\"u}cking}, {Colombo}, {Sanchez}, and {McInnes}]{BLCS}
R.~{Bewick}, C.~{L{\"u}cking}, C.~{Colombo}, J.~P. {Sanchez}, and C.~R. {McInnes}.
\newblock {Heliotropic dust rings for Earth climate engineering}.
\newblock \emph{Adv. Space Res.}, 51\penalty0 (7):\penalty0 1132--1144, Apr. 2013.
\newblock \doi{10.1016/j.asr.2012.10.024}.

\bibitem[{Billings}(2019)]{LB19}
L.~{Billings}.
\newblock {Colonizing other planets is a bad idea}.
\newblock \emph{Futures}, 110:\penalty0 44--46, 2019.
\newblock \doi{10.1016/j.futures.2019.02.020}.

\bibitem[{Block} et~al.(2004){Block}, {Keuler}, and {Schaller}]{BKS04}
A.~{Block}, K.~{Keuler}, and E.~{Schaller}.
\newblock {Impacts of anthropogenic heat on regional climate patterns}.
\newblock \emph{Geophys. Res. Lett.}, 31\penalty0 (12):\penalty0 L12211, June 2004.
\newblock \doi{10.1029/2004GL019852}.

\bibitem[{Blundell} and {Blundell}(2010)]{BB10}
S.~J. {Blundell} and K.~M. {Blundell}.
\newblock \emph{{Concepts in Thermal Physics}}.
\newblock Oxford: Oxford University Press, 2nd edition, 2010.

\bibitem[{Bohnenstengel} et~al.(2014){Bohnenstengel}, {Hamilton}, {Davies}, and {Belcher}]{BHD14}
S.~I. {Bohnenstengel}, I.~{Hamilton}, M.~{Davies}, and S.~E. {Belcher}.
\newblock {Impact of anthropogenic heat emissions on London's temperatures}.
\newblock \emph{Q. J. R. Meteorol. Soc.}, 140\penalty0 (679):\penalty0 687--698, Jan. 2014.
\newblock \doi{10.1002/qj.2144}.

\bibitem[{Bologna} and {Aquino}(2020)]{BA20}
M.~{Bologna} and G.~{Aquino}.
\newblock {Deforestation and world population sustainability: a quantitative analysis}.
\newblock \emph{Sci. Rep.}, 10:\penalty0 7631, May 2020.
\newblock \doi{10.1038/s41598-020-63657-6}.

\bibitem[{Borgue} and {Hein}(2023)]{BH23}
O.~{Borgue} and A.~M. {Hein}.
\newblock {Transparent occulters: A nearly zero-radiation pressure sunshade to support climate change mitigation}.
\newblock \emph{Acta Astronaut.}, 203:\penalty0 308--318, Feb. 2023.
\newblock \doi{10.1016/j.actaastro.2022.12.006}.

\bibitem[{Bostrom}(2003)]{NB03}
N.~{Bostrom}.
\newblock {Astronomical Waste: The Opportunity Cost of Delayed Technological Development}.
\newblock \emph{Utilitas}, 15\penalty0 (3):\penalty0 308--314, 2003.
\newblock \doi{10.1017/S0953820800004076}.

\bibitem[{Broadbent} et~al.(2019){Broadbent}, {Krayenhoff}, {Georgescu}, and {Sailor}]{BKG19}
A.~M. {Broadbent}, E.~S. {Krayenhoff}, M.~{Georgescu}, and D.~J. {Sailor}.
\newblock {The Observed Effects of Utility-Scale Photovoltaics on Near-Surface Air Temperature and Energy Balance}.
\newblock \emph{J. Appl. Meteorol. Climatol.}, 58\penalty0 (5):\penalty0 989--1006, May 2019.
\newblock \doi{10.1175/JAMC-D-18-0271.1}.

\bibitem[{Bromley} et~al.(2023){Bromley}, {Khan}, and {Kenyon}]{BKK23}
B.~C. {Bromley}, S.~H. {Khan}, and S.~J. {Kenyon}.
\newblock {Dust as a solar shield}.
\newblock \emph{PLOS Clim.}, 2\penalty0 (2):\penalty0 e0000133, 2023.
\newblock \doi{10.1371/journal.pclm.0000133}.

\bibitem[{Buch-Hansen}(2018)]{HBH18}
H.~{Buch-Hansen}.
\newblock {The Prerequisites for a Degrowth Paradigm Shift: Insights from Critical Political Economy}.
\newblock \emph{Ecol. Econ.}, 146:\penalty0 157--163, Apr. 2018.
\newblock \doi{10.1016/j.ecolecon.2017.10.021}.

\bibitem[{Buch-Hansen} et~al.(2024){Buch-Hansen}, {Koch}, and {Nesterova}]{BKN24}
H.~{Buch-Hansen}, M.~{Koch}, and I.~{Nesterova}.
\newblock \emph{{Deep transformations: A theory of degrowth}}.
\newblock Manchester: Manchester University Press, 2024.

\bibitem[{B{\"u}chs} and {Koch}(2019)]{BK19}
M.~{B{\"u}chs} and M.~{Koch}.
\newblock {Challenges for the degrowth transition: The debate about wellbeing}.
\newblock \emph{Futures}, 105:\penalty0 155--165, 2019.
\newblock \doi{10.1016/j.futures.2018.09.002}.

\bibitem[{Budyko}(1961)]{Budyko1961}
M.~I. {Budyko}.
\newblock {On the thermal zones of the Earth.}
\newblock \emph{Meteorology and Hydrology}, 11, 1961.

\bibitem[{Budyko}(1969)]{Budyko1969}
M.~I. {Budyko}.
\newblock {The effect of solar radiation variations on the climate of the earth.}
\newblock \emph{Tellus}, 21:\penalty0 611--619, Jan. 1969.
\newblock \doi{10.3402/tellusa.v21i5.10109}.

\bibitem[{Budyko}(1972)]{Budyko72}
M.~I. {Budyko}.
\newblock {The future climate}.
\newblock \emph{EOS Transactions}, 53\penalty0 (10):\penalty0 868--874, Jan. 1972.
\newblock \doi{10.1029/EO053i010p00868}.

\bibitem[{Burchell}(2006)]{MJB06}
M.~J. {Burchell}.
\newblock {W(h)ither the Drake equation?}
\newblock \emph{Int. J. Astrobiol.}, 5\penalty0 (3):\penalty0 243--250, Dec. 2006.
\newblock \doi{10.1017/S1473550406003107}.

\bibitem[{Burkett}(2006)]{PB06}
P.~{Burkett}.
\newblock \emph{{Marxism and Ecological Economics: Toward a Red and Green Political Economy}}, volume~11 of \emph{Historical Materialism}.
\newblock Leiden: Brill, 2006.

\bibitem[{Buzan} and {Huber}(2020)]{BH20}
J.~R. {Buzan} and M.~{Huber}.
\newblock {Moist Heat Stress on a Hotter Earth}.
\newblock \emph{Annu. Rev. Earth Planet. Sci.}, 48:\penalty0 623--655, May 2020.
\newblock \doi{10.1146/annurev-earth-053018-060100}.

\bibitem[{Caldeira} et~al.(2013){Caldeira}, {Bala}, and {Cao}]{CBC13}
K.~{Caldeira}, G.~{Bala}, and L.~{Cao}.
\newblock {The Science of Geoengineering}.
\newblock \emph{Annu. Rev. Earth Planet. Sci.}, 41:\penalty0 231--256, May 2013.
\newblock \doi{10.1146/annurev-earth-042711-105548}.

\bibitem[{Carroll-Nellenback} et~al.(2019){Carroll-Nellenback}, {Frank}, {Wright}, and {Scharf}]{CFW19}
J.~{Carroll-Nellenback}, A.~{Frank}, J.~{Wright}, and C.~{Scharf}.
\newblock {The Fermi Paradox and the Aurora Effect: Exo-civilization Settlement, Expansion, and Steady States}.
\newblock \emph{Astron. J.}, 158\penalty0 (3):\penalty0 117, Sept. 2019.
\newblock \doi{10.3847/1538-3881/ab31a3}.

\bibitem[{Chaisson}(2008)]{Chaisson2008}
E.~J. {Chaisson}.
\newblock {Long-Term Global Heating From Energy Usage}.
\newblock \emph{EOS Transactions}, 89\penalty0 (28):\penalty0 253--254, July 2008.
\newblock \doi{10.1029/2008EO280001}.

\bibitem[{Chyba} and {Hand}(2005)]{CH05}
C.~F. {Chyba} and K.~P. {Hand}.
\newblock {ASTROBIOLOGY: The Study of the Living Universe}.
\newblock \emph{Annu. Rev. Astron. Astrophys.}, 43\penalty0 (1):\penalty0 31--74, Sept. 2005.
\newblock \doi{10.1146/annurev.astro.43.051804.102202}.

\bibitem[{{\'C}irkovi{\'c}}(2012)]{MC12}
M.~M. {{\'C}irkovi{\'c}}.
\newblock \emph{{The Astrobiological Landscape}}.
\newblock Cambridge Astrobiology. Cambridge: Cambridge University Press, 2012.
\newblock \doi{10.1017/CBO9780511667404}.

\bibitem[{{\'C}irkovi{\'c}}(2015)]{MMC15}
M.~M. {{\'C}irkovi{\'c}}.
\newblock {Kardashev's Classification at 50+: A Fine Vehicle With Room for Improvement}.
\newblock \emph{Serb. Astron. J.}, 191:\penalty0 1--15, Dec. 2015.
\newblock \doi{10.2298/SAJ1591001C}.

\bibitem[{Cirkovic}(2018)]{Cirkovic2018}
M.~M. {Cirkovic}.
\newblock \emph{{The Great Silence: Science and Philosophy of Fermi's Paradox}}.
\newblock Oxford: Oxford University Press, 2018.

\bibitem[{{\'C}irkovi{\'c}}(2022)]{MMC22}
M.~M. {{\'C}irkovi{\'c}}.
\newblock {The nutshell kings: Why is human space settlement controversial in the first place?}
\newblock \emph{Futures}, 143:\penalty0 103023, 2022.
\newblock \doi{10.1016/j.futures.2022.103023}.

\bibitem[{{\'C}irkovi{\'c}} and {Bradbury}(2006)]{CB06}
M.~M. {{\'C}irkovi{\'c}} and R.~J. {Bradbury}.
\newblock {Galactic gradients, postbiological evolution and the apparent failure of SETI}.
\newblock \emph{New Astron.}, 11\penalty0 (8):\penalty0 628--639, July 2006.
\newblock \doi{10.1016/j.newast.2006.04.003}.

\bibitem[{{\'C}irkovi{\'c}} et~al.(2019){{\'C}irkovi{\'c}}, {Vukoti{\'c}}, and {Stojanovi{\'c}}]{CVS19}
M.~M. {{\'C}irkovi{\'c}}, B.~{Vukoti{\'c}}, and M.~{Stojanovi{\'c}}.
\newblock {Persistence of Technosignatures: A Comment on Lingam and Loeb}.
\newblock \emph{Astrobiology}, 19\penalty0 (10):\penalty0 1300--1302, Oct. 2019.
\newblock \doi{10.1089/ast.2019.2052}.

\bibitem[{Cosme} et~al.(2017){Cosme}, {Santos}, and {O'Neill}]{CSO17}
I.~{Cosme}, R.~{Santos}, and D.~W. {O'Neill}.
\newblock {Assessing the degrowth discourse: A review and analysis of academic degrowth policy proposals}.
\newblock \emph{J. Clean. Prod.}, 149:\penalty0 321--334, Apr. 2017.
\newblock \doi{10.1016/j.jclepro.2017.02.016}.

\bibitem[{Costanza} et~al.(2015){Costanza}, {Cumberland}, {Daly}, {Goodland}, {Norgaard}, {Kubiszewski}, and {Franco}]{CCD15}
R.~{Costanza}, J.~H. {Cumberland}, H.~{Daly}, R.~{Goodland}, R.~B. {Norgaard}, I.~{Kubiszewski}, and C.~{Franco}.
\newblock \emph{{An Introduction to Ecological Economics}}.
\newblock Boca Raton: CRC Press, 2nd edition, 2015.

\bibitem[{Cotta} and {Morales}(2009)]{CM09}
C.~{Cotta} and A.~{Morales}.
\newblock {A Computational Analysis of Galactic Exploration with Space Probes - Implications for the Fermi Paradox}.
\newblock \emph{J. Br. Interplanet. Soc.}, 62:\penalty0 82--88, Jan. 2009.
\newblock \doi{10.48550/arXiv.0907.0345}.

\bibitem[{Crawford}(2010)]{IAC10}
I.~A. {Crawford}.
\newblock {Astrobiological Benefits of Human Space Exploration}.
\newblock \emph{Astrobiology}, 10\penalty0 (6):\penalty0 577--587, Aug. 2010.
\newblock \doi{10.1089/ast.2010.0476}.

\bibitem[{Crawford}(2016)]{IAC16}
I.~A. {Crawford}.
\newblock {The long-term scientific benefits of a space economy}.
\newblock \emph{Space Policy}, 37:\penalty0 58--61, Aug. 2016.
\newblock \doi{10.1016/j.spacepol.2016.07.003}.

\bibitem[{Cullen} and {Allwood}(2010)]{CA10}
J.~M. {Cullen} and J.~M. {Allwood}.
\newblock {Theoretical efficiency limits for energy conversion devices}.
\newblock \emph{Energy}, 35\penalty0 (5):\penalty0 2059--2069, 2010.
\newblock \doi{10.1016/j.energy.2010.01.024}.

\bibitem[{Cullen} et~al.(2011){Cullen}, {Allwood}, and {Borgstein}]{CAB11}
J.~M. {Cullen}, J.~M. {Allwood}, and E.~H. {Borgstein}.
\newblock {Reducing Energy Demand: What Are the Practical Limits?}
\newblock \emph{Environ. Sci. Technol.}, 45\penalty0 (4):\penalty0 1711--1718, Feb. 2011.
\newblock \doi{10.1021/es102641n}.

\bibitem[{Dal Corso} et~al.(2022){Dal Corso}, {Song}, {Callegaro}, {Chu}, {Sun}, {Hilton}, {Grasby}, {Joachimski}, and {Wignall}]{DSC22}
J.~{Dal Corso}, H.~{Song}, S.~{Callegaro}, D.~{Chu}, Y.~{Sun}, J.~{Hilton}, S.~E. {Grasby}, M.~M. {Joachimski}, and P.~B. {Wignall}.
\newblock {Environmental crises at the Permian-Triassic mass extinction}.
\newblock \emph{Nat. Rev. Earth Environ.}, 3\penalty0 (3):\penalty0 197--214, Mar. 2022.
\newblock \doi{10.1038/s43017-021-00259-4}.

\bibitem[{D'Alisa} et~al.(2014){D'Alisa}, {Demaria}, and {Kallis}]{DDK14}
G.~{D'Alisa}, F.~{Demaria}, and G.~{Kallis}, editors.
\newblock \emph{{Degrowth: A Vocabulary for a New Era}}.
\newblock New York: Routledge, 2014.

\bibitem[{Daly}(1973)]{HED73}
H.~E. {Daly}.
\newblock \emph{{Towards a Steady-state Economy}}.
\newblock San Francisco: Freeman, 1973.

\bibitem[{Daly}(1996)]{HED96}
H.~E. {Daly}.
\newblock \emph{{Beyond Growth: The Economics of Sustainable Development}}.
\newblock Boston: Beacon Press, 1996.

\bibitem[{Daly} and {Farley}(2011)]{DF11}
H.~E. {Daly} and J.~{Farley}.
\newblock \emph{{Ecological Economics: Principles and Applications}}.
\newblock Washington, D.C.: Island Press, 2nd edition, 2011.

\bibitem[{De Vos}(2008)]{ADV08}
A.~{De Vos}.
\newblock \emph{{Thermodynamics of Solar Energy Conversion}}.
\newblock Weinheim: Wiley-VCH, 2008.

\bibitem[{Del Genio} et~al.(2019){Del Genio}, {Kiang}, {Way}, {Amundsen}, {Sohl}, {Fujii}, {Chandler}, {Aleinov}, {Colose}, {Guzewich}, and {Kelley}]{DKW19}
A.~D. {Del Genio}, N.~Y. {Kiang}, M.~J. {Way}, D.~S. {Amundsen}, L.~E. {Sohl}, Y.~{Fujii}, M.~{Chandler}, I.~{Aleinov}, C.~M. {Colose}, S.~D. {Guzewich}, and M.~{Kelley}.
\newblock {Albedos, Equilibrium Temperatures, and Surface Temperatures of Habitable Planets}.
\newblock \emph{Astrophys. J.}, 884\penalty0 (1):\penalty0 75, Oct. 2019.
\newblock \doi{10.3847/1538-4357/ab3be8}.

\bibitem[{Des Marais} et~al.(2008){Des Marais}, {Nuth}, {Allamandola}, {Boss}, {Farmer}, {Hoehler}, {Jakosky}, {Meadows}, {Pohorille}, {Runnegar}, and {Spormann}]{DNA08}
D.~J. {Des Marais}, I.~{Nuth}, Joseph~A., L.~J. {Allamandola}, A.~P. {Boss}, J.~D. {Farmer}, T.~M. {Hoehler}, B.~M. {Jakosky}, V.~S. {Meadows}, A.~{Pohorille}, B.~{Runnegar}, and A.~M. {Spormann}.
\newblock {The NASA Astrobiology Roadmap}.
\newblock \emph{Astrobiology}, 8\penalty0 (4):\penalty0 715--730, Aug. 2008.
\newblock \doi{10.1089/ast.2008.0819}.

\bibitem[{Dessler}(2021)]{AED21}
A.~E. {Dessler}.
\newblock \emph{{Introduction to Modern Climate Change}}.
\newblock Cambridge: Cambridge University Press, 3rd edition, 2021.
\newblock \doi{10.1017/9781108879125}.

\bibitem[{Deudney}(2020)]{DD20}
D.~{Deudney}.
\newblock \emph{{Dark Skies: Space Expansionism, Planetary Geopolitics and the Ends of Humanity}}.
\newblock Oxford: Oxford University Press, 2020.

\bibitem[{Dick} and {Launius}(2007)]{DL07}
S.~J. {Dick} and R.~D. {Launius}, editors.
\newblock \emph{{Societal Impact of Spaceflight}}.
\newblock Washington, D.C.: National Aeronautics \& Space Administration, 2007.

\bibitem[{Dobos} et~al.(2019){Dobos}, {Barr}, and {Kiss}]{DBK19}
V.~{Dobos}, A.~C. {Barr}, and L.~L. {Kiss}.
\newblock {Tidal heating and the habitability of the TRAPPIST-1 exoplanets}.
\newblock \emph{Astron. Astrophys.}, 624:\penalty0 A2, Apr. 2019.
\newblock \doi{10.1051/0004-6361/201834254}.

\bibitem[{Domagal-Goldman} et~al.(2016){Domagal-Goldman}, {Wright}, {Adamala}, {Arina de la Rubia}, {Bond}, {Dartnell}, {Goldman}, {Lynch}, {Naud}, {Paulino-Lima}, {Singer}, {Walter-Antonio}, {Abrevaya}, {Anderson}, {Arney}, {Atri}, {Az{\'u}a-Bustos}, {Bowman}, {Brazelton}, {Brennecka}, {Carns}, {Chopra}, {Colangelo-Lillis}, {Crockett}, {DeMarines}, {Frank}, {Frantz}, {de la Fuente}, {Galante}, {Glass}, {Gleeson}, {Glein}, {Goldblatt}, {Horak}, {Horodyskyj}, {Ka{\c{c}}ar}, {Kereszturi}, {Knowles}, {Mayeur}, {McGlynn}, {Miguel}, {Montgomery}, {Neish}, {Noack}, {Rugheimer}, {St{\"u}eken}, {Tamez-Hidalgo}, {Walker}, and {Wong}]{DGW16}
S.~D. {Domagal-Goldman}, K.~E. {Wright}, K.~{Adamala}, L.~{Arina de la Rubia}, J.~{Bond}, L.~R. {Dartnell}, A.~D. {Goldman}, K.~{Lynch}, M.-E. {Naud}, I.~G. {Paulino-Lima}, K.~{Singer}, M.~{Walter-Antonio}, X.~C. {Abrevaya}, R.~{Anderson}, G.~{Arney}, D.~{Atri}, A.~{Az{\'u}a-Bustos}, J.~S. {Bowman}, W.~J. {Brazelton}, G.~A. {Brennecka}, R.~{Carns}, A.~{Chopra}, J.~{Colangelo-Lillis}, C.~J. {Crockett}, J.~{DeMarines}, E.~A. {Frank}, C.~{Frantz}, E.~{de la Fuente}, D.~{Galante}, J.~{Glass}, D.~{Gleeson}, C.~R. {Glein}, C.~{Goldblatt}, R.~{Horak}, L.~{Horodyskyj}, B.~{Ka{\c{c}}ar}, A.~{Kereszturi}, E.~{Knowles}, P.~{Mayeur}, S.~{McGlynn}, Y.~{Miguel}, M.~{Montgomery}, C.~{Neish}, L.~{Noack}, S.~{Rugheimer}, E.~E. {St{\"u}eken}, P.~{Tamez-Hidalgo}, S.~I. {Walker}, and T.~{Wong}.
\newblock {The Astrobiology Primer v2.0}.
\newblock \emph{Astrobiology}, 16\penalty0 (8):\penalty0 561--653, Aug. 2016.
\newblock \doi{10.1089/ast.2015.1460}.

\bibitem[{Donges} et~al.(2017){Donges}, {Lucht}, {M{\"u}ller-Hansen}, and {Steffen}]{DLM17}
J.~F. {Donges}, W.~{Lucht}, F.~{M{\"u}ller-Hansen}, and W.~{Steffen}.
\newblock {The technosphere in Earth System analysis: A coevolutionary perspective}.
\newblock \emph{Anthr. Rev.}, 4\penalty0 (1):\penalty0 23--33, Apr. 2017.
\newblock \doi{10.1177/2053019616676608}.

\bibitem[{Drake} and {Sobel}(1992)]{DS92}
F.~{Drake} and D.~{Sobel}.
\newblock \emph{{Is anyone out there? The scientific search for extraterrestrial intelligence}}.
\newblock New York: Delacorte Press, 1992.

\bibitem[{Drake}(1965)]{Drake1965}
F.~D. {Drake}.
\newblock {The Radio Search for Intelligent Extraterrestrial Life}.
\newblock In G.~{Mamikunian} and M.~H. {Briggs}, editors, \emph{Current Aspects of Exobiology}, pages 323--345. Oxford: Pergamon Press, 1965.
\newblock \doi{10.1016/B978-1-4832-0047-7.50015-0}.

\bibitem[{Driscoll} and {Barnes}(2015)]{DB15}
P.~E. {Driscoll} and R.~{Barnes}.
\newblock {Tidal Heating of Earth-like Exoplanets around M Stars: Thermal, Magnetic, and Orbital Evolutions}.
\newblock \emph{Astrobiology}, 15\penalty0 (9):\penalty0 739--760, Sept. 2015.
\newblock \doi{10.1089/ast.2015.1325}.

\bibitem[{Durand} et~al.(2024){Durand}, {Hofferberth}, and {Schmelzer}]{DHS}
C.~{Durand}, E.~{Hofferberth}, and M.~{Schmelzer}.
\newblock {Planning beyond growth: The case for economic democracy within ecological limits}.
\newblock \emph{J. Clean. Prod.}, 437:\penalty0 140351, Jan. 2024.
\newblock \doi{10.1016/j.jclepro.2023.140351}.

\bibitem[{Dutil} and {Dumas}(2009)]{DD00}
Y.~{Dutil} and S.~{Dumas}.
\newblock {Sustainability: A Tedious Path to Galactic Colonization}.
\newblock In K.~J. {Meech}, J.~V. {Keane}, M.~J. {Mumma}, J.~L. {Siefert}, and D.~J. {Werthimer}, editors, \emph{Bioastronomy 2007: Molecules, Microbes and Extraterrestrial Life}, volume 420 of \emph{Astronomical Society of the Pacific Conference Series}, pages 439--442. San Francisco: Astronomical Society of the Pacific, Dec. 2009.

\bibitem[{Dyson}(1960)]{FJD60}
F.~J. {Dyson}.
\newblock {Search for Artificial Stellar Sources of Infrared Radiation}.
\newblock \emph{Science}, 131\penalty0 (3414):\penalty0 1667--1668, June 1960.
\newblock \doi{10.1126/science.131.3414.1667}.

\bibitem[{Early}(1989)]{JTE89}
J.~T. {Early}.
\newblock {Space-based solar shield to offset greenhouse effect}.
\newblock \emph{J. Br. Interplanet. Soc.}, 42:\penalty0 567--569, Dec. 1989.

\bibitem[{Ehrler} et~al.(2020){Ehrler}, {Alarc{\'o}n-Llad{\'o}}, {Tabernig}, {Veeken}, {Garnett}, and {Polman}]{EAT20}
B.~{Ehrler}, E.~{Alarc{\'o}n-Llad{\'o}}, S.~W. {Tabernig}, T.~{Veeken}, E.~C. {Garnett}, and A.~{Polman}.
\newblock {Photovoltaics Reaching for the Shockley--Queisser Limit}.
\newblock \emph{ACS Energy Lett.}, 5\penalty0 (9):\penalty0 3029--3033, 2020.
\newblock \doi{10.1021/acsenergylett.0c01790}.

\bibitem[{Ekins}(2000)]{PE00}
P.~{Ekins}.
\newblock \emph{{Economic Growth and Environmental Sustainability: The Prospects for Green Growth}}.
\newblock London: Routledge, 2000.

\bibitem[{Fay}(2012)]{MF12}
M.~{Fay}.
\newblock \emph{{Inclusive Green Growth: The Pathway to Sustainable Development}}.
\newblock Washington, D.C.: The World Bank, 2012.

\bibitem[{Finney} and {Jones}(1985)]{FJ85}
B.~R. {Finney} and E.~M. {Jones}, editors.
\newblock \emph{{Interstellar Migration and the Human Experience}}.
\newblock Berkeley: University of California Press, 1985.

\bibitem[{Firth} et~al.(2019){Firth}, {Zhang}, and {Yang}]{FZY19}
A.~{Firth}, B.~{Zhang}, and A.~{Yang}.
\newblock {Quantification of global waste heat and its environmental effects}.
\newblock \emph{Appl. Energy}, 235:\penalty0 1314--1334, Feb. 2019.
\newblock \doi{10.1016/j.apenergy.2018.10.102}.

\bibitem[{Flanner}(2009)]{MGF09}
M.~G. {Flanner}.
\newblock {Integrating anthropogenic heat flux with global climate models}.
\newblock \emph{Geophys. Res. Lett.}, 36\penalty0 (2):\penalty0 L02801, Jan. 2009.
\newblock \doi{10.1029/2008GL036465}.

\bibitem[{Forgan}(2019)]{DF19}
D.~H. {Forgan}.
\newblock \emph{{Solving Fermi's Paradox}}.
\newblock Cambridge: Cambridge University Press, 2019.
\newblock \doi{10.1017/9781316681510}.

\bibitem[{Forman} et~al.(2016){Forman}, {Muritala}, {Pardemann}, and {Meyer}]{FMP16}
C.~{Forman}, I.~K. {Muritala}, R.~{Pardemann}, and B.~{Meyer}.
\newblock {Estimating the global waste heat potential}.
\newblock \emph{Renew. Sustain. Energy Rev.}, 57:\penalty0 1568--1579, 2016.
\newblock \doi{10.1016/j.rser.2015.12.192}.

\bibitem[{Foster}(2022)]{JBF22}
J.~B. {Foster}.
\newblock \emph{{Capitalism in the Anthropocene: Ecological Ruin or Ecological Revolution}}.
\newblock New York: Monthly Review Press, 2022.

\bibitem[{Foster} et~al.(2010){Foster}, {Clark}, and {York}]{FCY10}
J.~B. {Foster}, B.~{Clark}, and R.~{York}.
\newblock \emph{{The Ecological Rift: Capitalism’s War on the Earth}}.
\newblock New York: Monthly Review Press, 2010.

\bibitem[{Frank} et~al.(2017){Frank}, {Kleidon}, and {Alberti}]{FKA17}
A.~{Frank}, A.~{Kleidon}, and M.~{Alberti}.
\newblock {Earth as a Hybrid Planet: The Anthropocene in an Evolutionary Astrobiological Context}.
\newblock \emph{Anthropocene}, 19:\penalty0 13--21, Sept. 2017.
\newblock \doi{10.1016/j.ancene.2017.08.002}.

\bibitem[{Frank} et~al.(2018){Frank}, {Carroll-Nellenback}, {Alberti}, and {Kleidon}]{FCA18}
A.~{Frank}, J.~{Carroll-Nellenback}, M.~{Alberti}, and A.~{Kleidon}.
\newblock {The Anthropocene Generalized: Evolution of Exo-Civilizations and Their Planetary Feedback}.
\newblock \emph{Astrobiology}, 18\penalty0 (5):\penalty0 503--518, May 2018.
\newblock \doi{10.1089/ast.2017.1671}.

\bibitem[{Frank} et~al.(2014){Frank}, {Meyer}, and {Mojzsis}]{FMM14}
E.~A. {Frank}, B.~S. {Meyer}, and S.~J. {Mojzsis}.
\newblock {A radiogenic heating evolution model for cosmochemically Earth-like exoplanets}.
\newblock \emph{Icarus}, 243:\penalty0 274--286, Nov. 2014.
\newblock \doi{10.1016/j.icarus.2014.08.031}.

\bibitem[{Friedman}(2015)]{LF15}
L.~{Friedman}.
\newblock \emph{{Human spaceflight: from Mars to the Stars}}.
\newblock Tucson: University of Arizona Press, 2015.

\bibitem[{Fujii} et~al.(2017){Fujii}, {Del Genio}, and {Amundsen}]{FDA17}
Y.~{Fujii}, A.~D. {Del Genio}, and D.~S. {Amundsen}.
\newblock {NIR-driven Moist Upper Atmospheres of Synchronously Rotating Temperate Terrestrial Exoplanets}.
\newblock \emph{Astrophys. J.}, 848\penalty0 (2):\penalty0 100, Oct. 2017.
\newblock \doi{10.3847/1538-4357/aa8955}.

\bibitem[{Gaidos}(2017)]{EG17}
E.~{Gaidos}.
\newblock {Transit detection of a `starshade' at the inner lagrange point of an exoplanet}.
\newblock \emph{Mon. Not. R. Astron. Soc.}, 469\penalty0 (4):\penalty0 4455--4464, Aug. 2017.
\newblock \doi{10.1093/mnras/stx1078}.

\bibitem[{Galantai}(2005)]{ZG05}
Z.~{Galantai}.
\newblock After kardashev: Farewell to super civilizations.
\newblock \emph{Contact in Context}, 2\penalty0 (2), 2005.

\bibitem[{Gallet} et~al.(2017){Gallet}, {Charbonnel}, {Amard}, {Brun}, {Palacios}, and {Mathis}]{GCA17}
F.~{Gallet}, C.~{Charbonnel}, L.~{Amard}, S.~{Brun}, A.~{Palacios}, and S.~{Mathis}.
\newblock {Impacts of stellar evolution and dynamics on the habitable zone: The role of rotation and magnetic activity}.
\newblock \emph{Astron. Astrophys.}, 597:\penalty0 A14, Jan. 2017.
\newblock \doi{10.1051/0004-6361/201629034}.

\bibitem[{Garofalo} et~al.(2020){Garofalo}, {Bevione}, {Cecchini}, {Mattiussi}, and {Chiolerio}]{GBC20}
E.~{Garofalo}, M.~{Bevione}, L.~{Cecchini}, F.~{Mattiussi}, and A.~{Chiolerio}.
\newblock {Waste Heat to Power: Technologies, Current Applications, and Future Potential}.
\newblock \emph{Energy Technol.}, 8\penalty0 (11):\penalty0 2000413, 2020.
\newblock \doi{10.1002/ente.202000413}.

\bibitem[{Gates}(2021)]{BG21}
B.~{Gates}.
\newblock \emph{{How to Avoid a Climate Disaster: The Solutions We Have and the Breakthroughs We Need}}.
\newblock New York: Alfred A. Knopf, 2021.

\bibitem[{Geffroy} et~al.(2021){Geffroy}, {Lilley}, {Parez}, and {Prasher}]{GLP21}
C.~{Geffroy}, D.~{Lilley}, P.~S. {Parez}, and R.~{Prasher}.
\newblock {Techno-economic analysis of waste-heat conversion}.
\newblock \emph{Joule}, 5\penalty0 (12):\penalty0 3080--3096, 2021.
\newblock \doi{10.1016/j.joule.2021.10.014}.

\bibitem[{Georgescu-Roegen}(1971)]{NGR}
N.~{Georgescu-Roegen}.
\newblock \emph{{The Entropy Law and the Economic Process}}.
\newblock Cambridge: Harvard University Press, 1971.

\bibitem[{Glade} et~al.(2012){Glade}, {Ballet}, and {Bastien}]{GBB12}
N.~{Glade}, P.~{Ballet}, and O.~{Bastien}.
\newblock {A stochastic process approach of the drake equation parameters}.
\newblock \emph{Int. J. Astrobiol.}, 11\penalty0 (2):\penalty0 103--108, Apr. 2012.
\newblock \doi{10.1017/S1473550411000413}.

\bibitem[{Goldblatt} and {Watson}(2012)]{GW12}
C.~{Goldblatt} and A.~J. {Watson}.
\newblock {The runaway greenhouse: implications for future climate change, geoengineering and planetary atmospheres}.
\newblock \emph{Phil. Trans. R. Soc. A}, 370\penalty0 (1974):\penalty0 4197--4216, Sept. 2012.
\newblock \doi{10.1098/rsta.2012.0004}.

\bibitem[{G{\'o}mez-Leal} et~al.(2018){G{\'o}mez-Leal}, {Kaltenegger}, {Lucarini}, and {Lunkeit}]{GKL18}
I.~{G{\'o}mez-Leal}, L.~{Kaltenegger}, V.~{Lucarini}, and F.~{Lunkeit}.
\newblock {Climate Sensitivity to Carbon Dioxide and the Moist Greenhouse Threshold of Earth-like Planets under an Increasing Solar Forcing}.
\newblock \emph{Astrophys. J.}, 869\penalty0 (2):\penalty0 129, Dec. 2018.
\newblock \doi{10.3847/1538-4357/aaea5f}.

\bibitem[{Gong} et~al.(2019){Gong}, {Li}, and {Wasielewski}]{GLW19}
J.~{Gong}, C.~{Li}, and M.~R. {Wasielewski}.
\newblock {Advances in solar energy conversion}.
\newblock \emph{Chem. Soc. Rev.}, 48\penalty0 (7):\penalty0 1862--1864, 2019.
\newblock \doi{10.1039/C9CS90020A}.

\bibitem[{Graus} et~al.(2007){Graus}, {Voogt}, and {Worrell}]{GVW07}
W.~H.~J. {Graus}, M.~{Voogt}, and E.~{Worrell}.
\newblock {International comparison of energy efficiency of fossil power generation}.
\newblock \emph{Energy Policy}, 35\penalty0 (7):\penalty0 3936--3951, 2007.
\newblock \doi{10.1016/j.enpol.2007.01.016}.

\bibitem[{Gray}(2020)]{Gray2020}
R.~H. {Gray}.
\newblock {The Extended Kardashev Scale}.
\newblock \emph{Astron. J.}, 159\penalty0 (5):\penalty0 228, May 2020.
\newblock \doi{10.3847/1538-3881/ab792b}.

\bibitem[{Green} et~al.(2023){Green}, {Dunlop}, {Yoshita}, {Kopidakis}, {Bothe}, {Siefer}, and {Hao}]{GDY23}
M.~A. {Green}, E.~D. {Dunlop}, M.~{Yoshita}, N.~{Kopidakis}, K.~{Bothe}, G.~{Siefer}, and X.~{Hao}.
\newblock {Solar cell efficiency tables (Version 61)}.
\newblock \emph{Prog. Photovolt.}, 32\penalty0 (1):\penalty0 3--13, 2023.
\newblock \doi{10.1002/pip.3646}.

\bibitem[{Gunderson} et~al.(2021){Gunderson}, {Stuart}, and {Petersen}]{GSP21}
R.~{Gunderson}, D.~{Stuart}, and B.~{Petersen}.
\newblock {In search of plan(et) B: Irrational rationality, capitalist realism, and space colonization}.
\newblock \emph{Futures}, 134:\penalty0 102857, 2021.
\newblock \doi{10.1016/j.futures.2021.102857}.

\bibitem[{Haldane}(1905)]{JSH05}
J.~S. {Haldane}.
\newblock {The Influence of High Air Temperatures No. I}.
\newblock \emph{Epidemiol. Infect.}, 5\penalty0 (4):\penalty0 494--513, 1905.
\newblock \doi{10.1017/S0022172400006811}.

\bibitem[{Hallegatte} et~al.(2012){Hallegatte}, {Heal}, {Fay}, and {Treguer}]{HHF12}
S.~{Hallegatte}, G.~{Heal}, M.~{Fay}, and D.~{Treguer}.
\newblock From growth to green growth - a framework.
\newblock Technical report, Cambridge: National Bureau of Economic Research, 2012.

\bibitem[{Haqq-Misra} and {Fauchez}(2022)]{HMF22}
J.~{Haqq-Misra} and T.~J. {Fauchez}.
\newblock {Galactic Settlement of Low-mass Stars as a Resolution to the Fermi Paradox}.
\newblock \emph{Astron. J.}, 164\penalty0 (6):\penalty0 247, Dec. 2022.
\newblock \doi{10.3847/1538-3881/ac9afd}.

\bibitem[{Haqq-Misra} et~al.(2024){Haqq-Misra}, {Profitiliotis}, and {Kopparapu}]{HPK24}
J.~{Haqq-Misra}, G.~{Profitiliotis}, and R.~{Kopparapu}.
\newblock {Projections of Earth's technosphere. I. Scenario modeling, worldbuilding, and overview of remotely detectable technosignatures}.
\newblock \emph{arXiv e-prints}, art. arXiv:2409.00067, Aug. 2024.
\newblock \doi{10.48550/arXiv.2409.00067}.

\bibitem[{Haqq-Misra} and {Baum}(2009)]{HMB09}
J.~D. {Haqq-Misra} and S.~D. {Baum}.
\newblock {The Sustainability Solution To The Fermi Paradox}.
\newblock \emph{J. Br. Interplanet. Soc.}, 62:\penalty0 47--51, Jan. 2009.
\newblock \doi{10.48550/arXiv.0906.0568}.

\bibitem[{Harrison}(2001)]{AAH01}
A.~A. {Harrison}.
\newblock \emph{{Spacefaring: The Human Dimension}}.
\newblock Berkely: University of California Press, 2001.

\bibitem[{Hart}(2005)]{SLH05}
S.~L. {Hart}.
\newblock \emph{{Capitalism At The Crossroads: The Unlimited Business Opportunities In Solving The World's Most Difficult Problems}}.
\newblock Upper Saddle River: Wharton School Publishing, 2005.

\bibitem[{Hartmann}(2016)]{DH16}
D.~L. {Hartmann}.
\newblock \emph{{Global Physical Climatology}}.
\newblock Amsterdam: Elsevier, 2nd edition, 2016.
\newblock \doi{10.1016/C2009-0-00030-0}.

\bibitem[{Hayat} et~al.(2019){Hayat}, {Ali}, {Monyake}, {Alagha}, and {Ahmed}]{HAM19}
M.~B. {Hayat}, D.~{Ali}, K.~C. {Monyake}, L.~{Alagha}, and N.~{Ahmed}.
\newblock {Solar energy-A look into power generation, challenges, and a solar-powered future}.
\newblock \emph{Int. J. Energy Res.}, 43\penalty0 (3):\penalty0 1049--1067, Mar. 2019.
\newblock \doi{10.1002/er.4252}.

\bibitem[{Herrington}(2021)]{GH21}
G.~{Herrington}.
\newblock {Update to limits to growth: Comparing the World3 model with empirical data}.
\newblock \emph{J. Ind. Ecol.}, 25\penalty0 (3):\penalty0 614--626, June 2021.
\newblock \doi{10.1111/jiec.13084}.

\bibitem[{Herrmann-Pillath}(2018)]{CHP18}
C.~{Herrmann-Pillath}.
\newblock {The Case for a New Discipline: Technosphere Science}.
\newblock \emph{Ecol. Econ.}, 149:\penalty0 212--225, July 2018.
\newblock \doi{10.1016/j.ecolecon.2018.03.024}.

\bibitem[{Hickel}(2020)]{JH20}
J.~{Hickel}.
\newblock \emph{{Less is More: How Degrowth Will Save the World}}.
\newblock London: Penguin Random House, 2020.

\bibitem[{Hickel} et~al.(2022){Hickel}, {Kallis}, {Jackson}, {O'Neill}, {Schor}, {Steinberger}, {Victor}, and {{\"U}rge-Vorsatz}]{HKJ22}
J.~{Hickel}, G.~{Kallis}, T.~{Jackson}, D.~W. {O'Neill}, J.~B. {Schor}, J.~K. {Steinberger}, P.~A. {Victor}, and D.~{{\"U}rge-Vorsatz}.
\newblock {Degrowth can work {\textemdash} here's how science can help}.
\newblock \emph{Nature}, 612\penalty0 (7940):\penalty0 400--403, Dec. 2022.
\newblock \doi{10.1038/d41586-022-04412-x}.

\bibitem[Hirsch(1978)]{FH78}
F.~Hirsch.
\newblock \emph{{Social Limits to Growth}}.
\newblock London: Routledge, 2nd edition, 1978.

\bibitem[{Holmes}(1991)]{DLH91}
D.~L. {Holmes}.
\newblock {Archeology in Space: Encountering Alien Trash and Other Remains}.
\newblock In J.~{Heidmann} and M.~J. {Klein}, editors, \emph{Bioastronomy: The Search for Extraterrestial Life --- The Exploration Broadens}, volume 390 of \emph{Lecture Notes in Physics}, pages 327--332. Berlin: Springer, 1991.
\newblock \doi{10.1007/3-540-54752-5_241}.

\bibitem[{Hu} et~al.(2016){Hu}, {Levis}, {Meehl}, {Han}, {Washington}, {Oleson}, {van Ruijven}, {He}, and {Strand}]{HLM16}
A.~{Hu}, S.~{Levis}, G.~A. {Meehl}, W.~{Han}, W.~M. {Washington}, K.~W. {Oleson}, B.~J. {van Ruijven}, M.~{He}, and W.~G. {Strand}.
\newblock {Impact of solar panels on global climate}.
\newblock \emph{Nat. Clim. Change}, 6\penalty0 (3):\penalty0 290--294, Mar. 2016.
\newblock \doi{10.1038/nclimate2843}.

\bibitem[{Ichinose} et~al.(1999){Ichinose}, {Shimodozono}, and {Hanaki}]{ISH99}
T.~{Ichinose}, K.~{Shimodozono}, and K.~{Hanaki}.
\newblock {Impact of anthropogenic heat on urban climate in Tokyo}.
\newblock \emph{Atmos. Environ.}, 33\penalty0 (24):\penalty0 3897--3909, Jan. 1999.
\newblock \doi{10.1016/S1352-2310(99)00132-6}.

\bibitem[{Irvine} et~al.(2016){Irvine}, {Kravitz}, {Lawrence}, and {Muri}]{IKL16}
P.~J. {Irvine}, B.~{Kravitz}, M.~G. {Lawrence}, and H.~{Muri}.
\newblock {An overview of the Earth system science of solar geoengineering}.
\newblock \emph{Wiley Interdiscip. Rev. Clim. Change}, 7\penalty0 (6):\penalty0 815--833, Nov. 2016.
\newblock \doi{10.1002/wcc.423}.

\bibitem[{Ivanov} et~al.(2020){Ivanov}, {Beam{\'\i}n}, {C{\'a}ceres}, and {Minniti}]{IBC20}
V.~D. {Ivanov}, J.~C. {Beam{\'\i}n}, C.~{C{\'a}ceres}, and D.~{Minniti}.
\newblock {A qualitative classification of extraterrestrial civilizations}.
\newblock \emph{Astron. Astrophys.}, 639:\penalty0 A94, July 2020.
\newblock \doi{10.1051/0004-6361/202037597}.

\bibitem[{Jackson} et~al.(2008){Jackson}, {Greenberg}, and {Barnes}]{JGB08}
B.~{Jackson}, R.~{Greenberg}, and R.~{Barnes}.
\newblock {Tidal Heating of Extrasolar Planets}.
\newblock \emph{Astrophys. J.}, 681\penalty0 (2):\penalty0 1631--1638, July 2008.
\newblock \doi{10.1086/587641}.

\bibitem[{Jackson} and {Victor}(2019)]{JV19}
T.~{Jackson} and P.~A. {Victor}.
\newblock {Unraveling the claims for (and against) green growth}.
\newblock \emph{Science}, 366\penalty0 (6468):\penalty0 950--951, Nov. 2019.
\newblock \doi{10.1126/science.aay0749}.

\bibitem[{Jacob}(1999)]{DJJ99}
D.~J. {Jacob}.
\newblock \emph{{Introduction to Atmospheric Chemistry}}.
\newblock Princeton: Princeton University Press, 1999.

\bibitem[{Javaux}(2019)]{EJJ19}
E.~J. {Javaux}.
\newblock {Challenges in evidencing the earliest traces of life}.
\newblock \emph{Nature}, 572\penalty0 (7770):\penalty0 451--460, Aug. 2019.
\newblock \doi{10.1038/s41586-019-1436-4}.

\bibitem[{Jin} et~al.(2005){Jin}, {Dickinson}, and {Zhang}]{JDZ05}
M.~{Jin}, R.~E. {Dickinson}, and D.~{Zhang}.
\newblock {The Footprint of Urban Areas on Global Climate as Characterized by MODIS.}
\newblock \emph{J. Clim.}, 18\penalty0 (10):\penalty0 1551--1565, May 2005.
\newblock \doi{10.1175/JCLI3334.1}.

\bibitem[{Johnson} and {Roy}(2023)]{JR23}
L.~{Johnson} and K.~{Roy}, editors.
\newblock \emph{{Interstellar Travel}}.
\newblock Amsterdam: Elsevier, 2023.
\newblock \doi{10.1016/C2021-0-01046-4}.

\bibitem[{Jones}(1976)]{EMJ76}
E.~M. {Jones}.
\newblock {Colonization of the Galaxy}.
\newblock \emph{Icarus}, 28:\penalty0 421, July 1976.
\newblock \doi{10.1016/0019-1035(76)90156-1}.

\bibitem[{Jones}(1981)]{EMJ81}
E.~M. {Jones}.
\newblock {Discrete calculations of interstellar migration and settlement}.
\newblock \emph{Icarus}, 46\penalty0 (3):\penalty0 328--336, June 1981.
\newblock \doi{10.1016/0019-1035(81)90136-6}.

\bibitem[{Jouhara} et~al.(2018){Jouhara}, {Khordehgah}, {Almahmoud}, {Delpech}, {Chauhan}, and {Tassou}]{JKA18}
H.~{Jouhara}, N.~{Khordehgah}, S.~{Almahmoud}, B.~{Delpech}, A.~{Chauhan}, and S.~A. {Tassou}.
\newblock {Waste heat recovery technologies and applications}.
\newblock \emph{Therm. Sci. Eng. Prog.}, 6:\penalty0 268--289, 2018.
\newblock \doi{10.1016/j.tsep.2018.04.017}.

\bibitem[{Kabir} et~al.(2018){Kabir}, {Kumar}, {Kumar}, {Adelodun}, and {Kim}]{KK18}
E.~{Kabir}, P.~{Kumar}, S.~{Kumar}, A.~A. {Adelodun}, and K.-H. {Kim}.
\newblock {Solar energy: Potential and future prospects}.
\newblock \emph{Renew. Sustain. Energy Rev.}, 82:\penalty0 894--900, 2018.
\newblock \doi{10.1016/j.rser.2017.09.094}.

\bibitem[{Kallis} et~al.(2018){Kallis}, {Kostakis}, {Lange}, {Muraca}, {Paulson}, and {Schmelzer}]{KKL18}
G.~{Kallis}, V.~{Kostakis}, S.~{Lange}, B.~{Muraca}, S.~{Paulson}, and M.~{Schmelzer}.
\newblock {Research On Degrowth}.
\newblock \emph{Annu. Rev. Environ. Resour.}, 43:\penalty0 291--316, 2018.
\newblock \doi{10.1146/annurev-environ-102017-025941}.

\bibitem[{Kallis} et~al.(2020){Kallis}, {Paulson}, {D'Alisa}, and {Demaria}]{KPD20}
G.~{Kallis}, S.~{Paulson}, G.~{D'Alisa}, and F.~{Demaria}.
\newblock \emph{{The Case for Degrowth}}.
\newblock Cambridge: Polity Press, 2020.

\bibitem[{Kardashev}(1964)]{Kardashev1964}
N.~S. {Kardashev}.
\newblock {Transmission of Information by Extraterrestrial Civilizations.}
\newblock \emph{Sov. Astron.}, 8:\penalty0 217, Oct. 1964.

\bibitem[{Kardashev}(1985)]{NSK85}
N.~S. {Kardashev}.
\newblock {On the inevitability and the possible structures of supercivilizations.}
\newblock In M.~D. {Papagiannis}, editor, \emph{The Search for Extraterrestrial Life: Recent Developments}, volume 112, pages 497--504. Dordrecht: Springer, Jan. 1985.

\bibitem[{Kasting} et~al.(1993){Kasting}, {Whitmire}, and {Reynolds}]{KWR93}
J.~F. {Kasting}, D.~P. {Whitmire}, and R.~T. {Reynolds}.
\newblock {Habitable Zones around Main Sequence Stars}.
\newblock \emph{Icarus}, 101\penalty0 (1):\penalty0 108--128, Jan. 1993.
\newblock \doi{10.1006/icar.1993.1010}.

\bibitem[{Keith}(2013)]{DK13}
D.~W. {Keith}.
\newblock \emph{{A Case for Climate Engineering}}.
\newblock Cambridge: The MIT Press, 2013.
\newblock \doi{10.7551/mitpress/9920.001.0001}.

\bibitem[{Kerschner}(2010)]{CK10}
C.~{Kerschner}.
\newblock {Economic de-growth vs. steady-state economy}.
\newblock \emph{J. Clean. Prod.}, 18\penalty0 (6):\penalty0 544--551, Jan. 2010.
\newblock \doi{10.1016/j.jclepro.2009.10.019}.

\bibitem[{Key{\ss}er} and {Lenzen}(2021)]{KL21}
L.~T. {Key{\ss}er} and M.~{Lenzen}.
\newblock {1.5 {\textdegree}C degrowth scenarios suggest the need for new mitigation pathways}.
\newblock \emph{Nat. Commun.}, 12:\penalty0 2676, Jan. 2021.
\newblock \doi{10.1038/s41467-021-22884-9}.

\bibitem[{Khan} and {Santamouris}(2023)]{KS23}
A.~{Khan} and M.~{Santamouris}.
\newblock {On the local warming potential of urban rooftop photovoltaic solar panels in cities}.
\newblock \emph{Sci. Re.}, 13:\penalty0 15623, Sept. 2023.
\newblock \doi{10.1038/s41598-023-40280-9}.

\bibitem[{Khan} and {Arsalan}(2016)]{KA16}
J.~{Khan} and M.~H. {Arsalan}.
\newblock {Solar power technologies for sustainable electricity generation--A review}.
\newblock \emph{Renew. Sustain. Energy Rev.}, 55:\penalty0 414--425, 2016.
\newblock \doi{10.1016/j.rser.2015.10.135}.

\bibitem[{Kim} et~al.(2020{\natexlab{a}}){Kim}, {Lee}, {Jung}, {Shin}, and {Park}]{KLJ20}
J.~Y. {Kim}, J.-W. {Lee}, H.~S. {Jung}, H.~{Shin}, and N.-G. {Park}.
\newblock High-efficiency perovskite solar cells.
\newblock \emph{Chem. Rev.}, 120\penalty0 (15):\penalty0 7867--7918, 2020{\natexlab{a}}.
\newblock \doi{10.1021/acs.chemrev.0c00107}.

\bibitem[{Kim} et~al.(2020{\natexlab{b}}){Kim}, {Lee}, and {Kwak}]{KLK20}
M.~S. {Kim}, J.~H. {Lee}, and M.~K. {Kwak}.
\newblock {Surface Texturing Methods for Solar Cell Efficiency Enhancement}.
\newblock \emph{Int. J. Precis. Eng. Manuf.}, 21\penalty0 (7):\penalty0 1389--1398, 2020{\natexlab{b}}.
\newblock \doi{10.1007/s12541-020-00337-5}.

\bibitem[{Kipping} et~al.(2020){Kipping}, {Frank}, and {Scharf}]{KFS20}
D.~{Kipping}, A.~{Frank}, and C.~{Scharf}.
\newblock {Contact inequality: first contact will likely be with an older civilization}.
\newblock \emph{Int. J. Astrobiol.}, 19\penalty0 (6):\penalty0 430--437, Dec. 2020.
\newblock \doi{10.1017/S1473550420000208}.

\bibitem[{Kirchartz} and {Rau}(2018)]{KR18}
T.~{Kirchartz} and U.~{Rau}.
\newblock {What Makes a Good Solar Cell?}
\newblock \emph{Adv. Energy Mater.}, 8\penalty0 (28):\penalty0 1703385, Oct. 2018.
\newblock \doi{10.1002/aenm.201703385}.

\bibitem[{Kish} and {Quilley}(2017)]{KQ17}
K.~{Kish} and S.~{Quilley}.
\newblock {Wicked Dilemmas of Scale and Complexity in the Politics of Degrowth}.
\newblock \emph{Ecol. Econ.}, 142:\penalty0 306--317, Dec. 2017.
\newblock \doi{10.1016/j.ecolecon.2017.08.008}.

\bibitem[{Kleidon}(2012)]{Kleidon2012}
A.~{Kleidon}.
\newblock {How does the Earth system generate and maintain thermodynamic disequilibrium and what does it imply for the future of the planet?}
\newblock \emph{Phil. Trans. R. Soc. A}, 370\penalty0 (1962):\penalty0 1012--1040, Mar. 2012.
\newblock \doi{10.1098/rsta.2011.0316}.

\bibitem[{Kleidon} et~al.(2023){Kleidon}, {Messori}, {Baidya Roy}, {Didenkulova}, and {Zeng}]{KMB23}
A.~{Kleidon}, G.~{Messori}, S.~{Baidya Roy}, I.~{Didenkulova}, and N.~{Zeng}.
\newblock {Editorial: Global warming is due to an enhanced greenhouse effect, and anthropogenic heat emissions currently play a negligible role at the global scale}.
\newblock \emph{Earth Syst. Dyn.}, 14\penalty0 (1):\penalty0 241--242, Feb. 2023.
\newblock \doi{10.5194/esd-14-241-2023}.

\bibitem[{Kopparapu} et~al.(2021){Kopparapu}, {Arney}, {Haqq-Misra}, {Lustig-Yaeger}, and {Villanueva}]{KAH21}
R.~{Kopparapu}, G.~{Arney}, J.~{Haqq-Misra}, J.~{Lustig-Yaeger}, and G.~{Villanueva}.
\newblock {Nitrogen Dioxide Pollution as a Signature of Extraterrestrial Technology}.
\newblock \emph{Astrophys. J.}, 908\penalty0 (2):\penalty0 164, Feb. 2021.
\newblock \doi{10.3847/1538-4357/abd7f7}.

\bibitem[{Kopparapu} et~al.(2024){Kopparapu}, {Kofman}, {Haqq-Misra}, {Kopparapu}, and {Lingam}]{KKH24}
R.~{Kopparapu}, V.~{Kofman}, J.~{Haqq-Misra}, V.~{Kopparapu}, and M.~{Lingam}.
\newblock {Detectability of Solar Panels as a Technosignature}.
\newblock \emph{Astrophys. J.}, 967\penalty0 (2):\penalty0 119, June 2024.
\newblock \doi{10.3847/1538-4357/ad43d7}.

\bibitem[{Kopparapu} et~al.(2013){Kopparapu}, {Ramirez}, {Kasting}, {Eymet}, {Robinson}, {Mahadevan}, {Terrien}, {Domagal-Goldman}, {Meadows}, and {Deshpande}]{Kopparapu2013}
R.~K. {Kopparapu}, R.~{Ramirez}, J.~F. {Kasting}, V.~{Eymet}, T.~D. {Robinson}, S.~{Mahadevan}, R.~C. {Terrien}, S.~{Domagal-Goldman}, V.~{Meadows}, and R.~{Deshpande}.
\newblock {Habitable Zones around Main-sequence Stars: New Estimates}.
\newblock \emph{Astrophys. J.}, 765\penalty0 (2):\penalty0 131, Mar. 2013.
\newblock \doi{10.1088/0004-637X/765/2/131}.

\bibitem[{Kopparapu} et~al.(2014){Kopparapu}, {Ramirez}, {SchottelKotte}, {Kasting}, {Domagal-Goldman}, and {Eymet}]{KRS14}
R.~K. {Kopparapu}, R.~M. {Ramirez}, J.~{SchottelKotte}, J.~F. {Kasting}, S.~{Domagal-Goldman}, and V.~{Eymet}.
\newblock {Habitable Zones around Main-sequence Stars: Dependence on Planetary Mass}.
\newblock \emph{Astrophys. J. Lett.}, 787\penalty0 (2):\penalty0 L29, June 2014.
\newblock \doi{10.1088/2041-8205/787/2/L29}.

\bibitem[{Kopparapu} et~al.(2016){Kopparapu}, {Wolf}, {Haqq-Misra}, {Yang}, {Kasting}, {Meadows}, {Terrien}, and {Mahadevan}]{KWH16}
R.~K. {Kopparapu}, E.~T. {Wolf}, J.~{Haqq-Misra}, J.~{Yang}, J.~F. {Kasting}, V.~{Meadows}, R.~{Terrien}, and S.~{Mahadevan}.
\newblock {The Inner Edge of the Habitable Zone for Synchronously Rotating Planets around Low-mass Stars Using General Circulation Models}.
\newblock \emph{Astrophys. J.}, 819\penalty0 (1):\penalty0 84, Mar. 2016.
\newblock \doi{10.3847/0004-637X/819/1/84}.

\bibitem[{Kopparapu} et~al.(2017){Kopparapu}, {Wolf}, {Arney}, {Batalha}, {Haqq-Misra}, {Grimm}, and {Heng}]{KWA17}
R.~K. {Kopparapu}, E.~T. {Wolf}, G.~{Arney}, N.~E. {Batalha}, J.~{Haqq-Misra}, S.~L. {Grimm}, and K.~{Heng}.
\newblock {Habitable Moist Atmospheres on Terrestrial Planets near the Inner Edge of the Habitable Zone around M Dwarfs}.
\newblock \emph{Astrophys. J.}, 845\penalty0 (1):\penalty0 5, Aug. 2017.
\newblock \doi{10.3847/1538-4357/aa7cf9}.

\bibitem[{Korpela} et~al.(2015){Korpela}, {Sallmen}, and {Leystra Greene}]{KSL15}
E.~J. {Korpela}, S.~M. {Sallmen}, and D.~{Leystra Greene}.
\newblock {Modeling Indications of Technology in Planetary Transit Light Curves-Dark-side Illumination}.
\newblock \emph{Astrophys. J.}, 809\penalty0 (2):\penalty0 139, Aug. 2015.
\newblock \doi{10.1088/0004-637X/809/2/139}.

\bibitem[{Korycansky}(2004)]{DGK04}
D.~G. {Korycansky}.
\newblock {Astroengineering, or how to save the Earth in only one billion years}.
\newblock In G.~{Garcia-Segura}, G.~{Tenorio-Tagle}, J.~{Franco}, and H.~W. {Yorke}, editors, \emph{Revista Mexicana de Astronomia y Astrofisica Conference Series}, volume~22 of \emph{Revista Mexicana de Astronomia y Astrofisica Conference Series}, pages 117--120, Dec. 2004.

\bibitem[{Korycansky} et~al.(2001){Korycansky}, {Laughlin}, and {Adams}]{KLA01}
D.~G. {Korycansky}, G.~{Laughlin}, and F.~C. {Adams}.
\newblock {Astronomical Engineering: A Strategy For Modifying Planetary Orbits}.
\newblock \emph{Astrophys. Space Sci.}, 275\penalty0 (4):\penalty0 349--366, Mar. 2001.
\newblock \doi{10.1023/A:1002790227314}.

\bibitem[{Kovic}(2021)]{MK21}
M.~{Kovic}.
\newblock {Risks of space colonization}.
\newblock \emph{Futures}, 126:\penalty0 102638, 2021.
\newblock \doi{10.1016/j.futures.2020.102638}.

\bibitem[{Krogstrup} et~al.(2013){Krogstrup}, {J{\o}rgensen}, {Heiss}, {Demichel}, {Holm}, {Aagesen}, {Nygard}, and {Fontcuberta I Morral}]{KJH13}
P.~{Krogstrup}, H.~I. {J{\o}rgensen}, M.~{Heiss}, O.~{Demichel}, J.~V. {Holm}, M.~{Aagesen}, J.~{Nygard}, and A.~{Fontcuberta I Morral}.
\newblock {Single-nanowire solar cells beyond the Shockley-Queisser limit}.
\newblock \emph{Nat. Photonics}, 7\penalty0 (4):\penalty0 306--310, Apr. 2013.
\newblock \doi{10.1038/nphoton.2013.32}.

\bibitem[{Kuhn} and {Berdyugina}(2015)]{KB15}
J.~R. {Kuhn} and S.~V. {Berdyugina}.
\newblock {Global warming as a detectable thermodynamic marker of Earth-like extrasolar civilizations: the case for a telescope like Colossus}.
\newblock \emph{Int. J. Astrobiol.}, 14\penalty0 (3):\penalty0 401--410, July 2015.
\newblock \doi{10.1017/S1473550414000585}.

\bibitem[{Lapenis}(2020)]{AL20}
A.~{Lapenis}.
\newblock {A 50-Year-Old Global Warming Forecast That Still Holds Up}.
\newblock \emph{EOS Transactions}, page 101, Nov. 2020.
\newblock \doi{10.1029/2020EO151822}.

\bibitem[{Latouche}(2009)]{SL09}
S.~{Latouche}.
\newblock \emph{{Farewell to Growth}}.
\newblock Cambridge: Polity Press, 2009.

\bibitem[{Leconte} et~al.(2013){Leconte}, {Forget}, {Charnay}, {Wordsworth}, and {Pottier}]{LFC13}
J.~{Leconte}, F.~{Forget}, B.~{Charnay}, R.~{Wordsworth}, and A.~{Pottier}.
\newblock {Increased insolation threshold for runaway greenhouse processes on Earth-like planets}.
\newblock \emph{Nature}, 504\penalty0 (7479):\penalty0 268--271, Dec. 2013.
\newblock \doi{10.1038/nature12827}.

\bibitem[{Lenton} et~al.(2008){Lenton}, {Held}, {Kriegler}, {Hall}, {Lucht}, {Rahmstorf}, and {Schellnhuber}]{LHK08}
T.~M. {Lenton}, H.~{Held}, E.~{Kriegler}, J.~W. {Hall}, W.~{Lucht}, S.~{Rahmstorf}, and H.~J. {Schellnhuber}.
\newblock {Inaugural Article: Tipping elements in the Earth's climate system}.
\newblock \emph{Proc. Natl. Acad. Sci.}, 105\penalty0 (6):\penalty0 1786--1793, Feb. 2008.
\newblock \doi{10.1073/pnas.0705414105}.

\bibitem[{Lepot}(2020)]{KL20}
K.~{Lepot}.
\newblock {Signatures of early microbial life from the Archean (4 to 2.5 Ga) eon}.
\newblock \emph{Earth Sci. Rev.}, 209:\penalty0 103296, Oct. 2020.
\newblock \doi{10.1016/j.earscirev.2020.103296}.

\bibitem[{Levenson}(2021)]{BPL21}
B.~P. {Levenson}.
\newblock {Habitable zones with an earth climate history model}.
\newblock \emph{Planet. Space Sci.}, 206:\penalty0 105318, Oct. 2021.
\newblock \doi{10.1016/j.pss.2021.105318}.

\bibitem[{Lingam}(2016)]{ML16}
M.~{Lingam}.
\newblock {Interstellar Travel and Galactic Colonization: Insights from Percolation Theory and the Yule Process}.
\newblock \emph{Astrobiology}, 16\penalty0 (6):\penalty0 418--426, June 2016.
\newblock \doi{10.1089/ast.2015.1411}.

\bibitem[{Lingam}(2020)]{ML20}
M.~{Lingam}.
\newblock {Implications of Abiotic Oxygen Buildup for Earth-like Complex Life}.
\newblock \emph{Astron. J.}, 159\penalty0 (4):\penalty0 144, Apr. 2020.
\newblock \doi{10.3847/1538-3881/ab737f}.

\bibitem[{Lingam}(2021)]{Man21}
M.~{Lingam}.
\newblock {A brief history of the term 'habitable zone' in the 19th century}.
\newblock \emph{Int. J. Astrobiol.}, 20\penalty0 (5):\penalty0 332--336, Oct. 2021.
\newblock \doi{10.1017/S1473550421000203}.

\bibitem[{Lingam} and {Balbi}(2024)]{MA24}
M.~{Lingam} and A.~{Balbi}.
\newblock \emph{{From Stars to Life: A Quantitative Approach to Astrobiology}}.
\newblock Cambridge: Cambridge University Press, 2024.
\newblock URL \url{https://cambridge.org/9781009411219}.

\bibitem[{Lingam} and {Loeb}(2017{\natexlab{a}})]{LL17}
M.~{Lingam} and A.~{Loeb}.
\newblock {Natural and artificial spectral edges in exoplanets}.
\newblock \emph{Mon. Not. R. Astron. Soc. Lett.}, 470\penalty0 (1):\penalty0 L82--L86, May 2017{\natexlab{a}}.
\newblock \doi{10.1093/mnrasl/slx084}.

\bibitem[{Lingam} and {Loeb}(2017{\natexlab{b}})]{ML17}
M.~{Lingam} and A.~{Loeb}.
\newblock {Reduced Diversity of Life around Proxima Centauri and TRAPPIST-1}.
\newblock \emph{Astrophys. J. Lett.}, 846\penalty0 (2):\penalty0 L21, Sept. 2017{\natexlab{b}}.
\newblock \doi{10.3847/2041-8213/aa8860}.

\bibitem[{Lingam} and {Loeb}(2019)]{LL19}
M.~{Lingam} and A.~{Loeb}.
\newblock {Colloquium: Physical constraints for the evolution of life on exoplanets}.
\newblock \emph{Rev. Mod. Phys.}, 91\penalty0 (2):\penalty0 021002, Apr. 2019.
\newblock \doi{10.1103/RevModPhys.91.021002}.

\bibitem[{Lingam} and {Loeb}(2020)]{LL20}
M.~{Lingam} and A.~{Loeb}.
\newblock {On the Habitable Lifetime of Terrestrial Worlds with High Radionuclide Abundances}.
\newblock \emph{Astrophys. J. Lett.}, 889\penalty0 (1):\penalty0 L20, Jan. 2020.
\newblock \doi{10.3847/2041-8213/ab68e5}.

\bibitem[{Lingam} and {Loeb}(2021)]{ML21}
M.~{Lingam} and A.~{Loeb}.
\newblock \emph{{Life in the Cosmos: From Biosignatures to Technosignatures}}.
\newblock Cambridge: Harvard University Press, 2021.

\bibitem[{Lingam} et~al.(2023){Lingam}, {Frank}, and {Balbi}]{LFB23}
M.~{Lingam}, A.~{Frank}, and A.~{Balbi}.
\newblock {Planetary Scale Information Transmission in the Biosphere and Technosphere: Limits and Evolution}.
\newblock \emph{Life}, 13\penalty0 (9):\penalty0 1850, Aug. 2023.
\newblock \doi{10.3390/life13091850}.

\bibitem[{Liodakis}(2018)]{GL18}
G.~{Liodakis}.
\newblock {Capital, Economic Growth, and Socio-Ecological Crisis: A Critique of De-Growth}.
\newblock \emph{Int. Crit. Thought}, 8\penalty0 (1):\penalty0 46--65, 2018.
\newblock \doi{10.1080/21598282.2017.1357487}.

\bibitem[{Loeb}(2024)]{AL24}
A.~{Loeb}.
\newblock {Illumination of a Planet by a Black Hole Moon as a Technological Signature}.
\newblock \emph{Res. Notes AAS}, 8\penalty0 (8):\penalty0 200, 2024.
\newblock \doi{10.3847/2515-5172/ad6e7a}.

\bibitem[{Lu} et~al.(2017){Lu}, {Wang}, {Wang}, {Zhang}, {Yu}, and {Qian}]{LWW17}
Y.~{Lu}, H.~{Wang}, Q.~{Wang}, Y.~{Zhang}, Y.~{Yu}, and Y.~{Qian}.
\newblock {Global anthropogenic heat emissions from energy consumption, 1965-2100}.
\newblock \emph{Clim. Change}, 145\penalty0 (3-4):\penalty0 459--468, Dec. 2017.
\newblock \doi{10.1007/s10584-017-2092-z}.

\bibitem[{Maccone}(2012)]{CM12}
C.~{Maccone}.
\newblock {Societal Statistics by virtue of the Statistical Drake Equation}.
\newblock \emph{Acta Astronaut.}, 78:\penalty0 3--15, Sept. 2012.
\newblock \doi{10.1016/j.actaastro.2011.09.014}.

\bibitem[{MacKay}(2009)]{DM09}
D.~J.~C. {MacKay}.
\newblock \emph{{Sustainable Energy -- Without the Hot Air}}.
\newblock Cambridge: IUT, 2009.

\bibitem[{MacMartin} and {Kravitz}(2019)]{MK19}
D.~G. {MacMartin} and B.~{Kravitz}.
\newblock {Mission-driven research for stratospheric aerosol geoengineering}.
\newblock \emph{Proc. Natl. Acad. Sci.}, 116\penalty0 (4):\penalty0 1089--1094, Jan. 2019.
\newblock \doi{10.1073/pnas.1811022116}.

\bibitem[{MacMartin} et~al.(2018){MacMartin}, {Ricke}, and {Keith}]{MRK18}
D.~G. {MacMartin}, K.~L. {Ricke}, and D.~W. {Keith}.
\newblock {Solar geoengineering as part of an overall strategy for meeting the 1.5{\textdegree}C Paris target}.
\newblock \emph{Phil. Trans. R. Soc. A}, 376\penalty0 (2119):\penalty0 20160454, May 2018.
\newblock \doi{10.1098/rsta.2016.0454}.

\bibitem[{Madden} and {Kaltenegger}(2020)]{MK20}
J.~{Madden} and L.~{Kaltenegger}.
\newblock {How surfaces shape the climate of habitable exoplanets}.
\newblock \emph{Mon. Not. R. Astron. Soc.}, 495\penalty0 (1):\penalty0 1--11, June 2020.
\newblock \doi{10.1093/mnras/staa387}.

\bibitem[{Madhusudhan} et~al.(2021){Madhusudhan}, {Piette}, and {Constantinou}]{MPC21}
N.~{Madhusudhan}, A.~A.~A. {Piette}, and S.~{Constantinou}.
\newblock {Habitability and Biosignatures of Hycean Worlds}.
\newblock \emph{Astrophys. J.}, 918\penalty0 (1):\penalty0 1, Sept. 2021.
\newblock \doi{10.3847/1538-4357/abfd9c}.

\bibitem[{Magnan} et~al.(2021){Magnan}, {P{\"o}rtner}, {Duvat}, {Garschagen}, {Guinder}, {Zommers}, {Hoegh-Guldberg}, and {Gattuso}]{MPD21}
A.~K. {Magnan}, H.-O. {P{\"o}rtner}, V.~K.~E. {Duvat}, M.~{Garschagen}, V.~A. {Guinder}, Z.~{Zommers}, O.~{Hoegh-Guldberg}, and J.-P. {Gattuso}.
\newblock {Estimating the global risk of anthropogenic climate change}.
\newblock \emph{Nat. Clim. Change}, 11\penalty0 (10):\penalty0 879--885, Sept. 2021.
\newblock \doi{10.1038/s41558-021-01156-w}.

\bibitem[{Mahendrarajah} et~al.(2023){Mahendrarajah}, {Moody}, {Schrempf}, {Sz{\'a}nth{\'o}}, {Dombrowski}, {Dav{\'\i}n}, {Pisani}, {Donoghue}, {Sz{\"o}ll{\H{o}}si}, {Williams}, and {Spang}]{MMS23}
T.~A. {Mahendrarajah}, E.~R.~R. {Moody}, D.~{Schrempf}, L.~L. {Sz{\'a}nth{\'o}}, N.~{Dombrowski}, A.~A. {Dav{\'\i}n}, D.~{Pisani}, P.~C.~J. {Donoghue}, G.~J. {Sz{\"o}ll{\H{o}}si}, T.~A. {Williams}, and A.~{Spang}.
\newblock {ATP synthase evolution on a cross-braced dated tree of life}.
\newblock \emph{Nat. Commun.}, 14:\penalty0 7456, Nov. 2023.
\newblock \doi{10.1038/s41467-023-42924-w}.

\bibitem[{Malanoski} et~al.(2024){Malanoski}, {Farnsworth}, {Lunt}, {Valdes}, and {Saupe}]{MFL24}
C.~M. {Malanoski}, A.~{Farnsworth}, D.~J. {Lunt}, P.~J. {Valdes}, and E.~E. {Saupe}.
\newblock {Climate change is an important predictor of extinction risk on macroevolutionary timescales}.
\newblock \emph{Science}, 383\penalty0 (6687):\penalty0 1130--1134, Mar. 2024.
\newblock \doi{10.1126/science.adj5763}.

\bibitem[{Malthus}(1798)]{TRM}
T.~R. {Malthus}.
\newblock \emph{{An Essay on the Principle of Population as It Affects the Future Improvement of Society, with Remarks on the Speculations of Mr. Godwin, M. Condorcet, and Other Writers}}.
\newblock London: J. Johnson, 1798.

\bibitem[{Mamikunian} and {Briggs}(1965)]{MB65}
G.~{Mamikunian} and M.~H. {Briggs}.
\newblock \emph{{Current Aspects of Exobiology}}.
\newblock Oxford: Pergamon Press, 1965.

\bibitem[{Markvart}(2022)]{TM22}
T.~{Markvart}.
\newblock {Shockley: Queisser detailed balance limit after 60 years}.
\newblock \emph{Wiley Interdiscip. Rev.: Energy Environ.}, 11\penalty0 (4):\penalty0 e430, July 2022.
\newblock \doi{10.1002/wene.430}.

\bibitem[{Mart{\'\i}nez-Alier} et~al.(2010){Mart{\'\i}nez-Alier}, {Pascual}, {Vivien}, and {Zaccai}]{MPV10}
J.~{Mart{\'\i}nez-Alier}, U.~{Pascual}, F.-D. {Vivien}, and E.~{Zaccai}.
\newblock {Sustainable de-growth: Mapping the context, criticisms and future prospects of an emergent paradigm}.
\newblock \emph{Ecol. Econ.}, 69\penalty0 (9):\penalty0 1741--1747, Jan. 2010.
\newblock \doi{10.1016/j.ecolecon.2010.04.017}.

\bibitem[{Masson} et~al.(2014){Masson}, {Bonhomme}, {Salagnac}, {Briottet}, and {Lemonsu}]{MBS14}
V.~{Masson}, M.~{Bonhomme}, J.-L. {Salagnac}, X.~{Briottet}, and A.~{Lemonsu}.
\newblock {Solar panels reduce both global warming and urban heat island}.
\newblock \emph{Front. Environ. Sci.}, 2:\penalty0 14, 2014.
\newblock \doi{10.3389/fenvs.2014.00014}.

\bibitem[{McInnes}(2002)]{CRM02}
C.~R. {McInnes}.
\newblock {Astronomical Engineering Revisited: Planetary Orbit Modification Using Solar Radiation Pressure}.
\newblock \emph{Astrophys. Space Sci.}, 282\penalty0 (4):\penalty0 765--772, Dec. 2002.
\newblock \doi{10.1023/A:1021178603836}.

\bibitem[{McKay} and {Marinova}(2001)]{MM01}
C.~P. {McKay} and M.~M. {Marinova}.
\newblock {The Physics, Biology, and Environmental Ethics of Making Mars Habitable}.
\newblock \emph{Astrobiology}, 1\penalty0 (1):\penalty0 89--109, Mar. 2001.
\newblock \doi{10.1089/153110701750137477}.

\bibitem[{McPhearson} et~al.(2021){McPhearson}, {Raymond}, {Gulsrud}, {Albert}, {Coles}, {Fagerholm}, {Nagatsu}, {Olafsson}, {Soininen}, and {Vierikko}]{MRG21}
T.~{McPhearson}, C.~M. {Raymond}, N.~{Gulsrud}, C.~{Albert}, N.~{Coles}, N.~{Fagerholm}, M.~{Nagatsu}, A.~S. {Olafsson}, N.~{Soininen}, and K.~{Vierikko}.
\newblock {Radical changes are needed for transformations to a good Anthropocene}.
\newblock \emph{npj Urban Sustainability}, 1\penalty0 (1):\penalty0 5, Feb. 2021.
\newblock \doi{10.1038/s42949-021-00017-x}.

\bibitem[{Meadows} et~al.(1972){Meadows}, {Meadows}, {Randers}, and {Behrens III}]{MMR}
D.~H. {Meadows}, D.~L. {Meadows}, J.~{Randers}, and W.~W. {Behrens III}.
\newblock \emph{{The Limits to Growth}}.
\newblock Washington, D.C.: Potomac Associates, 1972.

\bibitem[{Mieli} et~al.(2023){Mieli}, {Valli}, and {Maccone}]{MVM23}
E.~{Mieli}, A.~M.~F. {Valli}, and C.~{Maccone}.
\newblock {Astrobiology: resolution of the statistical Drake equation by Maccone's lognormal method in 50 steps}.
\newblock \emph{Int. J. Astrobiol.}, 22\penalty0 (4):\penalty0 428--537, Aug. 2023.
\newblock \doi{10.1017/S1473550423000113}.

\bibitem[{Mill}(1848)]{JSM48}
J.~S. {Mill}.
\newblock \emph{Principles of Political Economy, with some of their Applications to Social Philosophy}.
\newblock London: John W. Parker, 1848.

\bibitem[{Millward-Hopkins} et~al.(2020){Millward-Hopkins}, {Steinberger}, {Rao}, and {Oswald}]{MSR20}
J.~{Millward-Hopkins}, J.~K. {Steinberger}, N.~D. {Rao}, and Y.~{Oswald}.
\newblock {Providing decent living with minimum energy: A global scenario}.
\newblock \emph{Glob. Environ. Change}, 65:\penalty0 102168, Nov. 2020.
\newblock \doi{10.1016/j.gloenvcha.2020.102168}.

\bibitem[{Ming} et~al.(2014){Ming}, {de Richter}, {Liu}, and {Caillol}]{MLC14}
T.~{Ming}, R.~{de Richter}, W.~{Liu}, and S.~{Caillol}.
\newblock {Fighting global warming by climate engineering: Is the Earth radiation management and the solar radiation management any option for fighting climate change?}
\newblock \emph{Renew. Sustain. Energy Rev.}, 31:\penalty0 792--834, 2014.
\newblock \doi{10.1016/j.rser.2013.12.032}.

\bibitem[{Moln{\'a}r} et~al.(2020){Moln{\'a}r}, {Kov{\'a}cs}, and {G{\'a}l}]{MKG20}
G.~{Moln{\'a}r}, A.~{Kov{\'a}cs}, and T.~{G{\'a}l}.
\newblock {How does anthropogenic heating affect the thermal environment in a medium-sized Central European city? A case study in Szeged, Hungary}.
\newblock \emph{Urban Clim.}, 34:\penalty0 100673, Dec. 2020.
\newblock \doi{10.1016/j.uclim.2020.100673}.

\bibitem[{Moody} et~al.(2024){Moody}, {Alvarez-Carretero}, {Mahendrarajah}, {Clark}, {Betts}, {Dombrowski}, {Sz{\'a}nth{\'o}}, {Boyle}, {Daines}, {Chen}, {Lane}, {Yang}, {Shields}, {Sz{\"o}ll{\H{o}}si}, {Spang}, {Pisani}, {Williams}, {Lenton}, and {Donoghue}]{MAM24}
E.~R.~R. {Moody}, S.~{Alvarez-Carretero}, T.~{Mahendrarajah}, J.~W. {Clark}, H.~C. {Betts}, N.~{Dombrowski}, L.~L. {Sz{\'a}nth{\'o}}, R.~{Boyle}, S.~{Daines}, X.~{Chen}, N.~{Lane}, Z.~{Yang}, G.~A. {Shields}, G.~J. {Sz{\"o}ll{\H{o}}si}, A.~{Spang}, D.~{Pisani}, T.~A. {Williams}, T.~M. {Lenton}, and P.~C.~J. {Donoghue}.
\newblock {The nature of the last universal common ancestor and its impact on the early Earth system}.
\newblock \emph{Nat. Ecol. Evol.}, 8\penalty0 (9):\penalty0 1654--1666, 2024.
\newblock \doi{10.1038/s41559-024-02461-1}.

\bibitem[{Morgan} et~al.(2022){Morgan}, {Bralower}, {Brugger}, and {W{\"u}nnemann}]{MBB22}
J.~V. {Morgan}, T.~J. {Bralower}, J.~{Brugger}, and K.~{W{\"u}nnemann}.
\newblock {The Chicxulub impact and its environmental consequences}.
\newblock \emph{Nat. Rev. Earth Environ.}, 3\penalty0 (5):\penalty0 338--354, May 2022.
\newblock \doi{10.1038/s43017-022-00283-y}.

\bibitem[{Morton}(2016)]{OM16}
O.~{Morton}.
\newblock \emph{{The Planet Remade: How Geoengineering Could Change the World}}.
\newblock Princeton: Princeton University Press, 2016.

\bibitem[{Mullan} and {Haqq-Misra}(2019)]{MHM19}
B.~{Mullan} and J.~{Haqq-Misra}.
\newblock {Population growth, energy use, and the implications for the search for extraterrestrial intelligence}.
\newblock \emph{Futures}, 106:\penalty0 4--17, Feb. 2019.
\newblock \doi{10.1016/j.futures.2018.06.009}.

\bibitem[{Mun{\'e}var}(2014)]{GM14}
G.~{Mun{\'e}var}.
\newblock {Space exploration and human survival}.
\newblock \emph{Space Policy}, 30\penalty0 (4):\penalty0 197--201, Nov. 2014.
\newblock \doi{10.1016/j.spacepol.2014.10.002}.

\bibitem[{Muri} et~al.(2018){Muri}, {Tjiputra}, {Otter{\aa}}, {Adakudlu}, {Lauvset}, {Grini}, {Schulz}, {Niemeier}, and {Kristj{\'a}nsson}]{MTO18}
H.~{Muri}, J.~{Tjiputra}, O.~H. {Otter{\aa}}, M.~{Adakudlu}, S.~K. {Lauvset}, A.~{Grini}, M.~{Schulz}, U.~{Niemeier}, and J.~E. {Kristj{\'a}nsson}.
\newblock {Climate Response to Aerosol Geoengineering: A Multimethod Comparison}.
\newblock \emph{J. Clim.}, 31\penalty0 (16):\penalty0 6319--6340, 2018.
\newblock \doi{10.1175/JCLI-D-17-0620.1}.

\bibitem[{Naam}(2013)]{RN13}
R.~{Naam}.
\newblock \emph{{The Infinite Resource: The Power of Ideas on a Finite Planet}}.
\newblock Lebanon: University Press of New England, 2013.

\bibitem[{Narasimhan} and {Cui}(2013)]{NC13}
V.~K. {Narasimhan} and Y.~{Cui}.
\newblock {Nanostructures for photon management in solar cells}.
\newblock \emph{Nanophotonics}, 2\penalty0 (3):\penalty0 1, July 2013.
\newblock \doi{10.1515/nanoph-2013-0001}.

\bibitem[{Nelson}(2003)]{JN03}
J.~{Nelson}.
\newblock \emph{{The Physics of Solar Cells}}.
\newblock London: Imperial College Press, 2003.

\bibitem[{Newman} and {Sagan}(1981)]{NS81}
W.~I. {Newman} and C.~{Sagan}.
\newblock {Galactic civilizations: Population dynamics and interstellar diffusion}.
\newblock \emph{Icarus}, 46\penalty0 (3):\penalty0 293--327, June 1981.
\newblock \doi{10.1016/0019-1035(81)90135-4}.

\bibitem[{Nicholson} and {Forgan}(2013)]{NF13}
A.~{Nicholson} and D.~{Forgan}.
\newblock {Slingshot dynamics for self-replicating probes and the effect on exploration timescales}.
\newblock \emph{Int. J. Astrobiol.}, 12\penalty0 (4):\penalty0 337--344, Oct. 2013.
\newblock \doi{10.1017/S1473550413000244}.

\bibitem[{Nickolaenko}(2009)]{AN09}
A.~P. {Nickolaenko}.
\newblock {Concept of planetary thermal balance and global warming}.
\newblock \emph{J. Geophys. Rese. Space Phys.}, 114\penalty0 (A4):\penalty0 A04310, Apr. 2009.
\newblock \doi{10.1029/2008JA013753}.

\bibitem[{Nimmo} et~al.(2020){Nimmo}, {Primack}, {Faber}, {Ramirez-Ruiz}, and {Safarzadeh}]{NPF20}
F.~{Nimmo}, J.~{Primack}, S.~M. {Faber}, E.~{Ramirez-Ruiz}, and M.~{Safarzadeh}.
\newblock {Radiogenic Heating and Its Influence on Rocky Planet Dynamos and Habitability}.
\newblock \emph{Astrophys. J. Lett.}, 903\penalty0 (2):\penalty0 L37, Nov. 2020.
\newblock \doi{10.3847/2041-8213/abc251}.

\bibitem[{Niven}(1970)]{LN70}
L.~{Niven}.
\newblock \emph{{Ringworld}}.
\newblock New York: Ballantine Books, 1970.

\bibitem[{Obama}(2017)]{BO17}
B.~{Obama}.
\newblock {The irreversible momentum of clean energy}.
\newblock \emph{Science}, 355\penalty0 (6321):\penalty0 126--129, Jan. 2017.
\newblock \doi{10.1126/science.aam6284}.

\bibitem[{Obara} and {Tanaka}(2021)]{OT21}
S.~{Obara} and R.~{Tanaka}.
\newblock {Waste heat recovery system for nuclear power plants using the gas hydrate heat cycle}.
\newblock \emph{Appl. Energy}, 292:\penalty0 116667, June 2021.
\newblock \doi{10.1016/j.apenergy.2021.116667}.

\bibitem[{Odum}(2007)]{HTO07}
H.~T. {Odum}.
\newblock \emph{{Environment, Power, and Society for the Twenty-First Century: The Hierarchy of Energy}}.
\newblock New York: Columbia University Press, 2nd edition, 2007.

\bibitem[{Ohashi} et~al.(2007){Ohashi}, {Genchi}, {Kondo}, {Kikegawa}, {Yoshikado}, and {Hirano}]{OGK07}
Y.~{Ohashi}, Y.~{Genchi}, H.~{Kondo}, Y.~{Kikegawa}, H.~{Yoshikado}, and Y.~{Hirano}.
\newblock {Influence of Air-Conditioning Waste Heat on Air Temperature in Tokyo during Summer: Numerical Experiments Using an Urban Canopy Model Coupled with a Building Energy Model}.
\newblock \emph{J. Appl. Meteorol. Climatol.}, 46\penalty0 (1):\penalty0 66, Jan. 2007.
\newblock \doi{10.1175/JAM2441.1}.

\bibitem[{O'Neill} et~al.(2018){O'Neill}, {Fanning}, {Lamb}, and {Steinberger}]{OFL18}
D.~W. {O'Neill}, A.~L. {Fanning}, W.~F. {Lamb}, and J.~K. {Steinberger}.
\newblock {A good life for all within planetary boundaries}.
\newblock \emph{Nat. Sustain.}, 1\penalty0 (2):\penalty0 88--95, Feb. 2018.
\newblock \doi{10.1038/s41893-018-0021-4}.

\bibitem[{O'Neill}(1977)]{ONeill1977}
G.~K. {O'Neill}.
\newblock \emph{{The High Frontier: Human Colonies in Space}}.
\newblock New York: William Morrow, 1977.

\bibitem[{Opatrn{\'y}} et~al.(2017){Opatrn{\'y}}, {Richterek}, and {Bakala}]{Opatrny2016}
T.~{Opatrn{\'y}}, L.~{Richterek}, and P.~{Bakala}.
\newblock {Life under a black sun}.
\newblock \emph{Am. J. Phys.}, 85\penalty0 (1):\penalty0 14--22, Jan. 2017.
\newblock \doi{10.1119/1.4966905}.

\bibitem[{Pamplany} et~al.(2020){Pamplany}, {Gordijn}, and {Brereton}]{PGB20}
A.~{Pamplany}, B.~{Gordijn}, and P.~{Brereton}.
\newblock {The Ethics of Geoengineering: A Literature Review}.
\newblock \emph{Sci. Eng. Ethics}, 26:\penalty0 3069--3119, 2020.
\newblock \doi{10.1007/s11948-020-00258-6}.

\bibitem[{Pearson} et~al.(2006){Pearson}, {Oldson}, and {Levin}]{POL06}
J.~{Pearson}, J.~{Oldson}, and E.~{Levin}.
\newblock {Earth rings for planetary environment control}.
\newblock \emph{Acta Astronaut.}, 58\penalty0 (1):\penalty0 44--57, Jan. 2006.
\newblock \doi{10.1016/j.actaastro.2005.03.071}.

\bibitem[{Persson} et~al.(2022){Persson}, {Carney Almroth}, {Collins}, {Cornell}, {de Wit}, {Diamond}, {Fantke}, {Hassell{\"o}v}, {MacLeod}, {Ryberg}, {S{\o}gaard J{\o}rgensen}, {Villarrubia-G{\'o}mez}, {Wang}, and {Hauschild}]{PCC22}
L.~{Persson}, B.~M. {Carney Almroth}, C.~D. {Collins}, S.~{Cornell}, C.~A. {de Wit}, M.~L. {Diamond}, P.~{Fantke}, M.~{Hassell{\"o}v}, M.~{MacLeod}, M.~W. {Ryberg}, P.~{S{\o}gaard J{\o}rgensen}, P.~{Villarrubia-G{\'o}mez}, Z.~{Wang}, and M.~Z. {Hauschild}.
\newblock {Outside the Safe Operating Space of the Planetary Boundary for Novel Entities}.
\newblock \emph{Environ. Sci. Technol.}, 56\penalty0 (3):\penalty0 1510--1521, Feb. 2022.
\newblock \doi{10.1021/acs.est.1c04158}.

\bibitem[{Pierrehumbert} and {Gaidos}(2011)]{PG11}
R.~{Pierrehumbert} and E.~{Gaidos}.
\newblock {Hydrogen Greenhouse Planets Beyond the Habitable Zone}.
\newblock \emph{Astrophys. J. Lett.}, 734\penalty0 (1):\penalty0 L13, June 2011.
\newblock \doi{10.1088/2041-8205/734/1/L13}.

\bibitem[{Polewsky} et~al.(2024){Polewsky}, {Hankammer}, {Kleer}, and {Antons}]{PHK}
M.~{Polewsky}, S.~{Hankammer}, R.~{Kleer}, and D.~{Antons}.
\newblock {Degrowth vs. Green Growth. A computational review and interdisciplinary research agenda}.
\newblock \emph{Ecol. Econ.}, 217:\penalty0 108067, Mar. 2024.
\newblock \doi{10.1016/j.ecolecon.2023.108067}.

\bibitem[{Popp} et~al.(2016){Popp}, {Schmidt}, and {Marotzke}]{PSM16}
M.~{Popp}, H.~{Schmidt}, and J.~{Marotzke}.
\newblock {Transition to a Moist Greenhouse with CO$_{2}$ and solar forcing}.
\newblock \emph{Nat. Commun.}, 7:\penalty0 10627, Feb. 2016.
\newblock \doi{10.1038/ncomms10627}.

\bibitem[{Quilley}(2013)]{SQ13}
S.~{Quilley}.
\newblock {De-Growth Is Not a Liberal Agenda: Relocalisation and the Limits to Low Energy Cosmopolitanism}.
\newblock \emph{Environ. Values}, 22\penalty0 (2):\penalty0 261--285, 2013.
\newblock \doi{10.3197/096327113X13581561725310}.

\bibitem[{Raj} et~al.(2020){Raj}, {Paul}, {Chakraborty}, and {Kuttippurath}]{RPC20}
S.~{Raj}, S.~K. {Paul}, A.~{Chakraborty}, and J.~{Kuttippurath}.
\newblock {Anthropogenic forcing exacerbating the urban heat islands in India}.
\newblock \emph{J. Environ. Manag.}, 257:\penalty0 110006, Mar. 2020.
\newblock \doi{10.1016/j.jenvman.2019.110006}.

\bibitem[{Ramirez} et~al.(2019){Ramirez}, {Abbot}, {Fujii}, {Hamano}, {Kite}, {Levi}, {Lingam}, {Lueftinger}, {Robinson}, {Rushby}, {Schaefer}, {Tasker}, {Vladilo}, and {Wordsworth}]{RAF19}
R.~{Ramirez}, D.~S. {Abbot}, Y.~{Fujii}, K.~{Hamano}, E.~{Kite}, A.~{Levi}, M.~{Lingam}, T.~{Lueftinger}, T.~D. {Robinson}, A.~{Rushby}, L.~{Schaefer}, E.~{Tasker}, G.~{Vladilo}, and R.~D. {Wordsworth}.
\newblock {Habitable zone predictions and how to test them}.
\newblock \emph{Bull. Am. Astron. Soc.}, 51\penalty0 (3):\penalty0 31, May 2019.
\newblock \doi{10.48550/arXiv.1903.03706}.

\bibitem[{Ramirez}(2018)]{RMR18}
R.~M. {Ramirez}.
\newblock {A More Comprehensive Habitable Zone for Finding Life on Other Planets}.
\newblock \emph{Geosciences}, 8\penalty0 (8):\penalty0 280, July 2018.
\newblock \doi{10.3390/geosciences8080280}.

\bibitem[{Ramirez}(2020)]{RMR20}
R.~M. {Ramirez}.
\newblock {The effect of high nitrogen pressures on the habitable zone and an appraisal of greenhouse states}.
\newblock \emph{Mon. Not. R. Astron. Soc.}, 494\penalty0 (1):\penalty0 259--270, May 2020.
\newblock \doi{10.1093/mnras/staa603}.

\bibitem[{Randers}(2012)]{JR12}
J.~{Randers}.
\newblock \emph{{2052: A Global Forecast for the Next Forty Years}}.
\newblock White River Junction: Chelsea Green Publishing, 2012.

\bibitem[{Rasch} et~al.(2008){Rasch}, {Tilmes}, {Turco}, {Robock}, {Oman}, {Chen}, {Stenchikov}, and {Garcia}]{RTT08}
P.~J. {Rasch}, S.~{Tilmes}, R.~P. {Turco}, A.~{Robock}, L.~{Oman}, C.-C.~J. {Chen}, G.~L. {Stenchikov}, and R.~R. {Garcia}.
\newblock {An overview of geoengineering of climate using stratospheric sulphate aerosols}.
\newblock \emph{Phil. Trans. R. Soc. A}, 366\penalty0 (1882):\penalty0 4007--4037, Nov. 2008.
\newblock \doi{10.1098/rsta.2008.0131}.

\bibitem[{Raut} et~al.(2011){Raut}, {Ganesh}, {Nair}, and {Ramakrishna}]{RGN11}
H.~K. {Raut}, V.~A. {Ganesh}, A.~S. {Nair}, and S.~{Ramakrishna}.
\newblock {Anti-reflective coatings: A critical, in-depth review}.
\newblock \emph{Energy Environ. Sci.}, 4\penalty0 (10):\penalty0 3779--3804, 2011.
\newblock \doi{10.1039/C1EE01297E}.

\bibitem[{Raworth}(2017)]{KR17}
K.~{Raworth}.
\newblock \emph{{Doughnut Economics: Seven Ways to Think Like a 21st-Century Economist}}.
\newblock White River Junction: Chelsea Green Publishing, 2017.

\bibitem[{Reichl}(1998)]{LR98}
L.~E. {Reichl}.
\newblock \emph{{A Modern Course in Statistical Physics}}.
\newblock Weinheim: Wiley-VCH, 2nd edition, 1998.

\bibitem[{Richardson} et~al.(2023){Richardson}, {Steffen}, {Lucht}, {Bendtsen}, {Cornell}, {Donges}, {Dr{\"u}ke}, {Fetzer}, {Bala}, {von Bloh}, {Feulner}, {Fiedler}, {Gerten}, {Gleeson}, {Hofmann}, {Huiskamp}, {Kummu}, {Mohan}, {Nogu{\'e}s-Bravo}, {Petri}, {Porkka}, {Rahmstorf}, {Schaphoff}, {Thonicke}, {Tobian}, {Virkki}, {Wang-Erlandsson}, {Weber}, and {Rockstr{\"o}m}]{RSL23}
K.~{Richardson}, W.~{Steffen}, W.~{Lucht}, J.~{Bendtsen}, S.~E. {Cornell}, J.~F. {Donges}, M.~{Dr{\"u}ke}, I.~{Fetzer}, G.~{Bala}, W.~{von Bloh}, G.~{Feulner}, S.~{Fiedler}, D.~{Gerten}, T.~{Gleeson}, M.~{Hofmann}, W.~{Huiskamp}, M.~{Kummu}, C.~{Mohan}, D.~{Nogu{\'e}s-Bravo}, S.~{Petri}, M.~{Porkka}, S.~{Rahmstorf}, S.~{Schaphoff}, K.~{Thonicke}, A.~{Tobian}, V.~{Virkki}, L.~{Wang-Erlandsson}, L.~{Weber}, and J.~{Rockstr{\"o}m}.
\newblock {Earth beyond six of nine planetary boundaries}.
\newblock \emph{Sci. Adv.}, 9\penalty0 (37):\penalty0 eadh2458, Sept. 2023.
\newblock \doi{10.1126/sciadv.adh2458}.

\bibitem[{Ricke} et~al.(2023){Ricke}, {Wan}, {Saenger}, and {Lutsko}]{RWS23}
K.~{Ricke}, J.~S. {Wan}, M.~{Saenger}, and N.~J. {Lutsko}.
\newblock {Hydrological Consequences of Solar Geoengineering}.
\newblock \emph{Annu. Rev. Earth Planet. Sci.}, 51:\penalty0 447--470, May 2023.
\newblock \doi{10.1146/annurev-earth-031920-083456}.

\bibitem[{Ritchie}(2024)]{HR24}
H.~{Ritchie}.
\newblock \emph{Not the End of the World: How We Can Be the First Generation to Build a Sustainable Planet}.
\newblock New York: Little, Brown Spark, 2024.

\bibitem[{Ritchie} et~al.(2024){Ritchie}, {Rosado}, and {Roser}]{OWID20}
H.~{Ritchie}, P.~{Rosado}, and M.~{Roser}.
\newblock {Energy Production and Consumption}.
\newblock \url{https://ourworldindata.org/energy-production-consumption}, 2024.
\newblock Accessed: 2024-05-08.

\bibitem[{Ritchie} et~al.(2021){Ritchie}, {Clarke}, {Cox}, and {Huntingford}]{RCC21}
P.~D.~L. {Ritchie}, J.~J. {Clarke}, P.~M. {Cox}, and C.~{Huntingford}.
\newblock {Overshooting tipping point thresholds in a changing climate}.
\newblock \emph{Nature}, 592\penalty0 (7855):\penalty0 517--523, Apr. 2021.
\newblock \doi{10.1038/s41586-021-03263-2}.

\bibitem[{Rockstr{\"o}m} et~al.(2009){Rockstr{\"o}m}, {Steffen}, {Noone}, {Persson}, {Chapin}, {Lambin}, {Lenton}, {Scheffer}, {Folke}, {Schellnhuber}, {Nykvist}, {de Wit}, {Hughes}, {van der Leeuw}, {Rodhe}, {S{\"o}rlin}, {Snyder}, {Costanza}, {Svedin}, {Falkenmark}, {Karlberg}, {Corell}, {Fabry}, {Hansen}, {Walker}, {Liverman}, {Richardson}, {Crutzen}, and {Foley}]{RSN09}
J.~{Rockstr{\"o}m}, W.~{Steffen}, K.~{Noone}, {\r{A}}.~{Persson}, F.~S. {Chapin}, E.~F. {Lambin}, T.~M. {Lenton}, M.~{Scheffer}, C.~{Folke}, H.~J. {Schellnhuber}, B.~{Nykvist}, C.~A. {de Wit}, T.~{Hughes}, S.~{van der Leeuw}, H.~{Rodhe}, S.~{S{\"o}rlin}, P.~K. {Snyder}, R.~{Costanza}, U.~{Svedin}, M.~{Falkenmark}, L.~{Karlberg}, R.~W. {Corell}, V.~J. {Fabry}, J.~{Hansen}, B.~{Walker}, D.~{Liverman}, K.~{Richardson}, P.~{Crutzen}, and J.~A. {Foley}.
\newblock {A safe operating space for humanity}.
\newblock \emph{Nature}, 461\penalty0 (7263):\penalty0 472--475, Sept. 2009.
\newblock \doi{10.1038/461472a}.

\bibitem[{Rockstr{\"o}m} et~al.(2023){Rockstr{\"o}m}, {Gupta}, {Qin}, {Lade}, {Abrams}, {Andersen}, {Armstrong McKay}, {Bai}, {Bala}, {Bunn}, {Ciobanu}, {DeClerck}, {Ebi}, {Gifford}, {Gordon}, {Hasan}, {Kanie}, {Lenton}, {Loriani}, {Liverman}, {Mohamed}, {Nakicenovic}, {Obura}, {Ospina}, {Prodani}, {Rammelt}, {Sakschewski}, {Scholtens}, {Stewart-Koster}, {Tharammal}, {van Vuuren}, {Verburg}, {Winkelmann}, {Zimm}, {Bennett}, {Bringezu}, {Broadgate}, {Green}, {Huang}, {Jacobson}, {Ndehedehe}, {Pedde}, {Rocha}, {Scheffer}, {Schulte-Uebbing}, {de Vries}, {Xiao}, {Xu}, {Xu}, {Zafra-Calvo}, and {Zhang}]{RGQ23}
J.~{Rockstr{\"o}m}, J.~{Gupta}, D.~{Qin}, S.~J. {Lade}, J.~F. {Abrams}, L.~S. {Andersen}, D.~I. {Armstrong McKay}, X.~{Bai}, G.~{Bala}, S.~E. {Bunn}, D.~{Ciobanu}, F.~{DeClerck}, K.~{Ebi}, L.~{Gifford}, C.~{Gordon}, S.~{Hasan}, N.~{Kanie}, T.~M. {Lenton}, S.~{Loriani}, D.~M. {Liverman}, A.~{Mohamed}, N.~{Nakicenovic}, D.~{Obura}, D.~{Ospina}, K.~{Prodani}, C.~{Rammelt}, B.~{Sakschewski}, J.~{Scholtens}, B.~{Stewart-Koster}, T.~{Tharammal}, D.~{van Vuuren}, P.~H. {Verburg}, R.~{Winkelmann}, C.~{Zimm}, E.~M. {Bennett}, S.~{Bringezu}, W.~{Broadgate}, P.~A. {Green}, L.~{Huang}, L.~{Jacobson}, C.~{Ndehedehe}, S.~{Pedde}, J.~{Rocha}, M.~{Scheffer}, L.~{Schulte-Uebbing}, W.~{de Vries}, C.~{Xiao}, C.~{Xu}, X.~{Xu}, N.~{Zafra-Calvo}, and X.~{Zhang}.
\newblock {Safe and just Earth system boundaries}.
\newblock \emph{Nature}, 619\penalty0 (7968):\penalty0 102--111, July 2023.
\newblock \doi{10.1038/s41586-023-06083-8}.

\bibitem[{Rockstr{\"o}m} et~al.(2024){Rockstr{\"o}m}, {Donges}, {Fetzer}, {Martin}, {Wang-Erlandsson}, and {Richardson}]{RDF24}
J.~{Rockstr{\"o}m}, J.~F. {Donges}, I.~{Fetzer}, M.~A. {Martin}, L.~{Wang-Erlandsson}, and K.~{Richardson}.
\newblock {Planetary Boundaries guide humanity’s future on Earth}.
\newblock \emph{Nat. Rev. Earth Environ.}, 5\penalty0 (11):\penalty0 773--788, 2024.
\newblock \doi{10.1038/s43017-024-00597-z}.

\bibitem[{Romm}(2022)]{JJR22}
J.~J. {Romm}.
\newblock \emph{{Climate Change: What Everyone Needs to Know}}.
\newblock Oxford: Oxford University Press, 3rd edition, 2022.
\newblock \doi{10.1093/wentk/9780197647127.001.0001}.

\bibitem[{Rovetto}(2013)]{RJR13}
R.~J. {Rovetto}.
\newblock {The essential role of human spaceflight}.
\newblock \emph{Space Policy}, 29\penalty0 (4):\penalty0 225--228, Nov. 2013.
\newblock \doi{10.1016/j.spacepol.2013.08.001}.

\bibitem[{Roy}(2022)]{KR22}
K.~{Roy}.
\newblock {The solar shield concept: Current status and future possibilities}.
\newblock \emph{Acta Astronaut.}, 197:\penalty0 368--374, Aug. 2022.
\newblock \doi{10.1016/j.actaastro.2022.02.022}.

\bibitem[{R{\"u}hle}(2016)]{SR16}
S.~{R{\"u}hle}.
\newblock {Tabulated values of the Shockley-Queisser limit for single junction solar cells}.
\newblock \emph{Solar Energy}, 130:\penalty0 139--147, June 2016.
\newblock \doi{10.1016/j.solener.2016.02.015}.

\bibitem[{Rushby} et~al.(2013){Rushby}, {Claire}, {Osborn}, and {Watson}]{Rushby2013}
A.~J. {Rushby}, M.~W. {Claire}, H.~{Osborn}, and A.~J. {Watson}.
\newblock {Habitable Zone Lifetimes of Exoplanets around Main Sequence Stars}.
\newblock \emph{Astrobiology}, 13\penalty0 (9):\penalty0 833--849, Sept. 2013.
\newblock \doi{10.1089/ast.2012.0938}.

\bibitem[{Rushby} et~al.(2019){Rushby}, {Shields}, and {Joshi}]{RSJ19}
A.~J. {Rushby}, A.~L. {Shields}, and M.~{Joshi}.
\newblock {The Effect of Land Fraction and Host Star Spectral Energy Distribution on the Planetary Albedo of Terrestrial Worlds}.
\newblock \emph{Astrophys. J.}, 887\penalty0 (1):\penalty0 29, Dec. 2019.
\newblock \doi{10.3847/1538-4357/ab4da6}.

\bibitem[{Sagan}(1963)]{CS63}
C.~{Sagan}.
\newblock {Direct contact among galactic civilizations by relativistic interstellar spaceflight}.
\newblock \emph{Planet. Space Sci.}, 11\penalty0 (5):\penalty0 485--498, May 1963.
\newblock \doi{10.1016/0032-0633(63)90072-2}.

\bibitem[{Sagan}(1973)]{CS73}
C.~{Sagan}.
\newblock \emph{{The Cosmic Connection: An Extraterrestrial Perspective}}.
\newblock New York: Doubleday, 1973.

\bibitem[{Sagan}(1994)]{CS94}
C.~{Sagan}.
\newblock \emph{{Pale Blue Dot: A Vision of the Human Future in Space}}.
\newblock New York: Random House, 1994.

\bibitem[{Sailor} et~al.(2021){Sailor}, {Anand}, and {King}]{SAK21}
D.~J. {Sailor}, J.~{Anand}, and R.~R. {King}.
\newblock {Photovoltaics in the built environment: A critical review}.
\newblock \emph{Energy Build.}, 253:\penalty0 111479, Dec. 2021.
\newblock \doi{10.1016/j.enbuild.2021.111479}.

\bibitem[{Saito}(2022)]{KS22}
K.~{Saito}.
\newblock \emph{{Marx in the Anthropocene: Towards the Idea of Degrowth Communism}}.
\newblock Cambridge: Cambridge University Press, 2022.

\bibitem[{Salamanca} et~al.(2014){Salamanca}, {Georgescu}, {Mahalov}, {Moustaoui}, and {Wang}]{SGM14}
F.~{Salamanca}, M.~{Georgescu}, A.~{Mahalov}, M.~{Moustaoui}, and M.~{Wang}.
\newblock {Anthropogenic heating of the urban environment due to air conditioning}.
\newblock \emph{J. Geophys. Res. Atmos.}, 119\penalty0 (10):\penalty0 5949--5965, May 2014.
\newblock \doi{10.1002/2013JD021225}.

\bibitem[{Sark{\i}n} et~al.(2020){Sark{\i}n}, {Ekren}, and {Sa{\u{g}}lam}]{SES20}
A.~S. {Sark{\i}n}, N.~{Ekren}, and {\c{S}}.~{Sa{\u{g}}lam}.
\newblock {A review of anti-reflection and self-cleaning coatings on photovoltaic panels}.
\newblock \emph{Sol. Energy}, 199:\penalty0 63--73, Mar. 2020.
\newblock \doi{10.1016/j.solener.2020.01.084}.

\bibitem[{Savitch} et~al.(2021){Savitch}, {Frank}, {Carroll-Nellenback}, {Haqq-Misra}, {Kleidon}, and {Alberti}]{SFC21}
E.~{Savitch}, A.~{Frank}, J.~{Carroll-Nellenback}, J.~{Haqq-Misra}, A.~{Kleidon}, and M.~{Alberti}.
\newblock {Triggering a Climate Change Dominated ``Anthropocene'': Is It Common among Exocivilizations?}
\newblock \emph{Astron. J.}, 162\penalty0 (5):\penalty0 196, Nov. 2021.
\newblock \doi{10.3847/1538-3881/ac1a71}.

\bibitem[{Schmelzer}(2015)]{MS15}
M.~{Schmelzer}.
\newblock {The growth paradigm: History, hegemony, and the contested making of economic growthmanship}.
\newblock \emph{Ecol. Econ.}, 118:\penalty0 262--271, Oct. 2015.
\newblock \doi{10.1016/j.ecolecon.2015.07.029}.

\bibitem[{Schmelzer} et~al.(2022){Schmelzer}, {Vetter}, and {Vansintjan}]{SVV22}
M.~{Schmelzer}, A.~{Vetter}, and A.~{Vansintjan}.
\newblock \emph{{The Future is Degrowth: A Guide to a World Beyond Capitalism}}.
\newblock London: Verso Books, 2022.

\bibitem[{Schramski} et~al.(2015){Schramski}, {Gattie}, and {Brown}]{SGB15}
J.~R. {Schramski}, D.~K. {Gattie}, and J.~H. {Brown}.
\newblock {Human domination of the biosphere: Rapid discharge of the earth-space battery foretells the future of humankind}.
\newblock \emph{Proc. Natl. Acad. Sci.}, 112\penalty0 (31):\penalty0 9511--9517, Aug. 2015.
\newblock \doi{10.1073/pnas.1508353112}.

\bibitem[{Schroeder}(2020)]{DVS20}
D.~V. {Schroeder}.
\newblock \emph{{An Introduction to Thermal Physics}}.
\newblock Oxford: Oxford University Press, 2020.

\bibitem[{Schwartz} et~al.(2021){Schwartz}, {Wells-Jensen}, {Traphagan}, {Weibel}, and {Smith}]{SWT21}
J.~{Schwartz}, S.~{Wells-Jensen}, J.~{Traphagan}, D.~{Weibel}, and K.~{Smith}.
\newblock {What do we need to ask before settling space?}
\newblock \emph{J. Br. Interplanet. Soc.}, 74:\penalty0 140--149, Apr. 2021.

\bibitem[{Schwartz} and {Milligan}(2016)]{SM16}
J.~S.~J. {Schwartz} and T.~{Milligan}, editors.
\newblock \emph{{The Ethics of Space Exploration}}.
\newblock Space and Society. Cham: Springer, 2016.
\newblock \doi{10.1007/978-3-319-39827-3}.

\bibitem[{Schwartzman}(2012)]{DS12}
D.~{Schwartzman}.
\newblock {A Critique of Degrowth and its Politics}.
\newblock \emph{Capital. Nat. Social.}, 23\penalty0 (1):\penalty0 119--125, 2012.
\newblock \doi{10.1080/10455752.2011.648848}.

\bibitem[{Schwieterman} et~al.(2019){Schwieterman}, {Reinhard}, {Olson}, {Harman}, and {Lyons}]{SRO19}
E.~W. {Schwieterman}, C.~T. {Reinhard}, S.~L. {Olson}, C.~E. {Harman}, and T.~W. {Lyons}.
\newblock {A Limited Habitable Zone for Complex Life}.
\newblock \emph{Astrophys. J.}, 878\penalty0 (1):\penalty0 19, June 2019.
\newblock \doi{10.3847/1538-4357/ab1d52}.

\bibitem[{Shanmugam} et~al.(2020){Shanmugam}, {Pugazhendhi}, {Madurai Elavarasan}, {Kasiviswanathan}, and {Das}]{SPM20}
N.~{Shanmugam}, R.~{Pugazhendhi}, R.~{Madurai Elavarasan}, P.~{Kasiviswanathan}, and N.~{Das}.
\newblock {Anti-Reflective Coating Materials: A Holistic Review from PV Perspective}.
\newblock \emph{Energies}, 13\penalty0 (10):\penalty0 2631, 2020.
\newblock \doi{10.3390/en13102631}.

\bibitem[{Shelhamer}(2017)]{MS17}
M.~{Shelhamer}.
\newblock {Why send humans into space? Science and non-science motivations for human space flight}.
\newblock \emph{Space Policy}, 42:\penalty0 37--40, Nov. 2017.
\newblock \doi{10.1016/j.spacepol.2017.10.001}.

\bibitem[{Shermer}(2002)]{MS02}
M.~{Shermer}.
\newblock {Why ET Hasn't Called}.
\newblock \emph{Sci. Am.}, 287\penalty0 (2):\penalty0 33--33, Aug. 2002.
\newblock \doi{10.1038/scientificamerican0802-33}.

\bibitem[{Sherwood} and {Huber}(2010)]{SH10}
S.~C. {Sherwood} and M.~{Huber}.
\newblock {An adaptability limit to climate change due to heat stress}.
\newblock \emph{Proc. Natl. Acad. Sci.}, 107\penalty0 (21):\penalty0 9552--9555, May 2010.
\newblock \doi{10.1073/pnas.0913352107}.

\bibitem[{Shields} et~al.(2016){Shields}, {Ballard}, and {Johnson}]{SBJ16}
A.~L. {Shields}, S.~{Ballard}, and J.~A. {Johnson}.
\newblock {The habitability of planets orbiting M-dwarf stars}.
\newblock \emph{Phys. Rep.}, 663:\penalty0 1, Dec. 2016.
\newblock \doi{10.1016/j.physrep.2016.10.003}.

\bibitem[{Shklovskii} and {Sagan}(1966)]{Sagan1966}
I.~S. {Shklovskii} and C.~{Sagan}.
\newblock \emph{{Intelligent life in the universe}}.
\newblock Holden-Day: San Francisco, 1966.

\bibitem[{Shockley} and {Queisser}(1961)]{SQ61}
W.~{Shockley} and H.~J. {Queisser}.
\newblock {Detailed Balance Limit of Efficiency of p-n Junction Solar Cells}.
\newblock \emph{J. Appl. Phys.}, 32\penalty0 (3):\penalty0 510--519, Mar. 1961.
\newblock \doi{10.1063/1.1736034}.

\bibitem[{Shostak}(2011)]{SS11}
S.~{Shostak}.
\newblock {L: How Long Do They Last?}
\newblock In H.~P. {Shuch}, editor, \emph{Searching for Extraterrestrial Intelligence}, pages 451--466. Berlin: Springer, 2011.
\newblock \doi{10.1007/978-3-642-13196-7_23}.

\bibitem[{Slobodian}(2015)]{RES15}
R.~E. {Slobodian}.
\newblock {Selling space colonization and immortality: A psychosocial, anthropological critique of the rush to colonize Mars}.
\newblock \emph{Acta Astronaut.}, 113:\penalty0 89--104, Aug. 2015.
\newblock \doi{10.1016/j.actaastro.2015.03.027}.

\bibitem[{Smil}(2000)]{VS00}
V.~{Smil}.
\newblock {Perils of Long-Range Energy Forecasting: Reflections on Looking Far Ahead}.
\newblock \emph{Technol. Forecast. Soc. Change}, 65\penalty0 (3):\penalty0 251--264, 2000.
\newblock \doi{10.1016/S0040-1625(99)00097-9}.

\bibitem[{Smil}(2008)]{VS08}
V.~{Smil}.
\newblock \emph{{Energy in Nature and Society: General Energetics of Complex Systems}}.
\newblock Cambridge: The MIT Press, 2008.

\bibitem[{Smil}(2015)]{VS15}
V.~{Smil}.
\newblock \emph{{Power Density: A Key to Understanding Energy Sources and Uses}}.
\newblock Cambridge: The MIT Press, 2015.

\bibitem[{Smil}(2017)]{Smil2017}
V.~{Smil}.
\newblock \emph{{Energy and Civilization: A History}}.
\newblock Cambridge: The MIT Press, 2017.

\bibitem[{Smil}(2019)]{VS19}
V.~{Smil}.
\newblock \emph{{Growth: From Microorganisms to Megacities}}.
\newblock Cambridge: The MIT Press, 2019.

\bibitem[{Smil}(2020)]{VS20}
V.~{Smil}.
\newblock \emph{{Numbers Don't Lie: 71 Things You Need to Know About the World}}.
\newblock New York: Penguin, 2020.

\bibitem[{Socas-Navarro} et~al.(2021){Socas-Navarro}, {Haqq-Misra}, {Wright}, {Kopparapu}, {Benford}, {Davis}, and {TechnoClimes 2020 workshop participants}]{SHW21}
H.~{Socas-Navarro}, J.~{Haqq-Misra}, J.~T. {Wright}, R.~{Kopparapu}, J.~{Benford}, R.~{Davis}, and {TechnoClimes 2020 workshop participants}.
\newblock {Concepts for future missions to search for technosignatures}.
\newblock \emph{Acta Astronaut.}, 182:\penalty0 446--453, May 2021.
\newblock \doi{10.1016/j.actaastro.2021.02.029}.

\bibitem[{Song} et~al.(2021){Song}, {Kemp}, {Tian}, {Chu}, {Song}, and {Dai}]{SKT21}
H.~{Song}, D.~B. {Kemp}, L.~{Tian}, D.~{Chu}, H.~{Song}, and X.~{Dai}.
\newblock {Thresholds of temperature change for mass extinctions}.
\newblock \emph{Nat. Commun.}, 12:\penalty0 4694, Aug. 2021.
\newblock \doi{10.1038/s41467-021-25019-2}.

\bibitem[{Spash}(2020)]{CLS20}
C.~L. {Spash}.
\newblock {A tale of three paradigms: Realising the revolutionary potential of ecological economics}.
\newblock \emph{Ecol. Econ.}, 169:\penalty0 106518, Mar. 2020.
\newblock \doi{10.1016/j.ecolecon.2019.106518}.

\bibitem[{Stapledon}(1937)]{Stap37}
O.~{Stapledon}.
\newblock \emph{Star Maker}.
\newblock London: Methuen, 1937.

\bibitem[{Steffen} et~al.(2015){Steffen}, {Richardson}, {Rockstr{\"o}m}, {Cornell}, {Fetzer}, {Bennett}, {Biggs}, {Carpenter}, {De Vries}, {De Wit}, {Folke}, {Gerten}, {Heinke}, {Mace}, {Persson}, {Ramanathan}, and {Reyers}]{SRR15}
W.~{Steffen}, K.~{Richardson}, J.~{Rockstr{\"o}m}, S.~E. {Cornell}, I.~{Fetzer}, E.~M. {Bennett}, R.~{Biggs}, S.~R. {Carpenter}, W.~{De Vries}, C.~A. {De Wit}, C.~{Folke}, D.~{Gerten}, J.~{Heinke}, G.~M. {Mace}, L.~M. {Persson}, V.~{Ramanathan}, and B.~{Reyers}.
\newblock {Planetary boundaries: Guiding human development on a changing planet}.
\newblock \emph{science}, 347\penalty0 (6223):\penalty0 1259855, 2015.
\newblock \doi{10.1126/science.1259855}.

\bibitem[{Steiglechner}(2018)]{Steiglechner2021}
P.~{Steiglechner}.
\newblock {Estimating global warming from anthropogenic heat emissions}.
\newblock Master's thesis, Universit{\"a}t Potsdam, 2018.
\newblock URL \url{10.25932/publishup-49886}.

\bibitem[{Steiglechner} et~al.(2019){Steiglechner}, {Martin}, and {Feulner}]{SMF19}
P.~{Steiglechner}, M.~{Martin}, and G.~{Feulner}.
\newblock {Estimating global warming from anthropogenic heat emissions}.
\newblock In \emph{EGU General Assembly Conference Abstracts}, EGU General Assembly Conference Abstracts, page 502, Apr. 2019.

\bibitem[{Stephens} et~al.(2015){Stephens}, {O'Brien}, {Webster}, {Pilewski}, {Kato}, and {Li}]{SOW15}
G.~L. {Stephens}, D.~{O'Brien}, P.~J. {Webster}, P.~{Pilewski}, S.~{Kato}, and J.-l. {Li}.
\newblock {The albedo of Earth}.
\newblock \emph{Rev. Geophys.}, 53\penalty0 (1):\penalty0 141--163, Mar. 2015.
\newblock \doi{10.1002/2014RG000449}.

\bibitem[{Stewart} and {Kennedy}(2017)]{SK17}
I.~D. {Stewart} and C.~A. {Kennedy}.
\newblock {Metabolic heat production by human and animal populations in cities}.
\newblock \emph{Int. J. Biometeorol.}, 61\penalty0 (7):\penalty0 1159--1171, July 2017.
\newblock \doi{10.1007/s00484-016-1296-7}.

\bibitem[{Strona} and {Bradshaw}(2018)]{SB18}
G.~{Strona} and C.~J.~A. {Bradshaw}.
\newblock {Co-extinctions annihilate planetary life during extreme environmental change}.
\newblock \emph{Sci. Rep.}, 8:\penalty0 16724, Nov. 2018.
\newblock \doi{10.1038/s41598-018-35068-1}.

\bibitem[{Struchtrup}(2024)]{HS24}
H.~{Struchtrup}.
\newblock \emph{{Thermodynamics and Energy Conversion}}.
\newblock Cham: Springer, 2nd edition, 2024.
\newblock \doi{10.1007/978-3-031-60556-7}.

\bibitem[{Strunz} and {Bartkowski}(2018)]{SSB18}
S.~{Strunz} and B.~{Bartkowski}.
\newblock {Degrowth, the project of modernity, and liberal democracy}.
\newblock \emph{J. Clean. Prod.}, 196:\penalty0 1158--1168, Sept. 2018.
\newblock \doi{10.1016/j.jclepro.2018.06.148}.

\bibitem[{Summers}(1971)]{CMS71}
C.~M. {Summers}.
\newblock {The Conversion of Energy}.
\newblock \emph{Sci. Am.}, 225\penalty0 (3):\penalty0 148--160, Sept. 1971.
\newblock \doi{10.1038/scientificamerican0971-148}.

\bibitem[{Sun} et~al.(2022){Sun}, {Zou}, {Qin}, {Zhang}, and {Wu}]{SZQ22}
C.~{Sun}, Y.~{Zou}, C.~{Qin}, B.~{Zhang}, and X.~{Wu}.
\newblock {Temperature effect of photovoltaic cells: a review}.
\newblock \emph{Adv. Compos. Hybrid Mater.}, 5\penalty0 (4):\penalty0 2675--2699, 2022.
\newblock \doi{10.1007/s42114-022-00533-z}.

\bibitem[{Sun} et~al.(2018){Sun}, {Wang}, and {Chen}]{SWC18}
R.~{Sun}, Y.~{Wang}, and L.~{Chen}.
\newblock {A distributed model for quantifying temporal-spatial patterns of anthropogenic heat based on energy consumption}.
\newblock \emph{J. Clean. Prod.}, 170:\penalty0 601--609, Jan. 2018.
\newblock \doi{10.1016/j.jclepro.2017.09.153}.

\bibitem[{Szocik}(2019)]{KS19}
K.~{Szocik}.
\newblock {Should and could humans go to Mars? Yes, but not now and not in the near future}.
\newblock \emph{Futures}, 105:\penalty0 54--66, 2019.
\newblock \doi{10.1016/j.futures.2018.08.004}.

\bibitem[{Taleb}(2012)]{NNT12}
N.~N. {Taleb}.
\newblock \emph{{Antifragile: Things That Gain From Disorder}}.
\newblock New York: Random House, 2012.

\bibitem[{Tarter}(2001)]{JT01}
J.~{Tarter}.
\newblock {The Search for Extraterrestrial Intelligence (SETI)}.
\newblock \emph{Annu. Rev. Astron. Astrophys.}, 39:\penalty0 511--548, Jan. 2001.
\newblock \doi{10.1146/annurev.astro.39.1.511}.

\bibitem[{Tarter} et~al.(2018){Tarter}, {Rummel}, {Siemion}, {Rees}, {Maccone}, and {Hellbourg}]{TRS18}
J.~{Tarter}, J.~{Rummel}, A.~{Siemion}, M.~{Rees}, C.~{Maccone}, and G.~{Hellbourg}.
\newblock {Three Versions of the Third Law: Technosignatures and Astrobiology}.
\newblock Technical report, National Academies of Sciences, Engineering, and Medicine, 2018.
\newblock URL \url{https://zenodo.org/records/2539473}.

\bibitem[{The Borexino Collaboration}(2020)]{AAA20}
{The Borexino Collaboration}.
\newblock {Comprehensive geoneutrino analysis with Borexino}.
\newblock \emph{Phys. Rev. D}, 101\penalty0 (1):\penalty0 012009, Jan. 2020.
\newblock \doi{10.1103/PhysRevD.101.012009}.

\bibitem[{Trainer}(2020)]{TT20}
T.~{Trainer}.
\newblock {De-growth: Some suggestions from the Simpler Way perspective}.
\newblock \emph{Ecol. Econ.}, 167:\penalty0 106436, Jan. 2020.
\newblock \doi{10.1016/j.ecolecon.2019.106436}.

\bibitem[{Traphagan}(2019)]{JT19}
J.~W. {Traphagan}.
\newblock {Which humanity would space colonization save?}
\newblock \emph{Futures}, 110:\penalty0 47--49, 2019.
\newblock \doi{10.1016/j.futures.2019.02.016}.

\bibitem[{Trenberth} et~al.(2009){Trenberth}, {Fasullo}, and {Kiehl}]{TFK09}
K.~E. {Trenberth}, J.~T. {Fasullo}, and J.~{Kiehl}.
\newblock {Earth's Global Energy Budget}.
\newblock \emph{Bull. Am. Meteorol. Soc.}, 90\penalty0 (3):\penalty0 311--324, 2009.
\newblock \doi{10.1175/2008BAMS2634.1}.

\bibitem[{Tuchow} and {Wright}(2023)]{TW23}
N.~W. {Tuchow} and J.~T. {Wright}.
\newblock {The Abundance of Belatedly Habitable Planets and Ambiguities in Definitions of the Continuously Habitable Zone}.
\newblock \emph{Astrophys. J.}, 944\penalty0 (1):\penalty0 71, Feb. 2023.
\newblock \doi{10.3847/1538-4357/acb054}.

\bibitem[{Tziperman}(2022)]{ET22}
E.~{Tziperman}.
\newblock \emph{{Global Warming Science: A Quantitative Introduction to Climate Change and Its Consequences}}.
\newblock Princeton: Princeton University Press, 2022.

\bibitem[{Vakoch} and {Dowd}(2015)]{Vakoch2015}
D.~A. {Vakoch} and M.~F. {Dowd}.
\newblock \emph{{The Drake Equation}}.
\newblock Cambridge: Cambridge University Press, 2015.

\bibitem[{Valentine}(2012)]{DV12}
D.~{Valentine}.
\newblock {Exit Strategy: Profit, Cosmology, and the Future of Humans in Space}.
\newblock \emph{Anthropol. Q.}, 85\penalty0 (4):\penalty0 1045--1067, 2012.
\newblock \doi{10.1353/anq.2012.0073}.

\bibitem[{Vallis}(2012)]{GVV12}
G.~K. {Vallis}.
\newblock \emph{{Climate and the Oceans}}.
\newblock Princeton: Princeton University Press, 2012.

\bibitem[{van den Bergh}(2011)]{VDB11}
J.~C.~J.~M. {van den Bergh}.
\newblock {Environment versus growth {\textemdash} A criticism of ``degrowth'' and a plea for ``a-growth''}.
\newblock \emph{Ecol. Econ.}, 70\penalty0 (5):\penalty0 881--890, Jan. 2011.
\newblock \doi{10.1016/j.ecolecon.2010.09.035}.

\bibitem[{van den Bergh}(2017)]{VDB17}
J.~C.~J.~M. {van den Bergh}.
\newblock {A third option for climate policy within potential limits to growth}.
\newblock \emph{Nat. Clim. Change}, 7\penalty0 (2):\penalty0 107--112, Feb. 2017.
\newblock \doi{10.1038/nclimate3113}.

\bibitem[{Vaughan} and {Lenton}(2011)]{VL11}
N.~E. {Vaughan} and T.~M. {Lenton}.
\newblock {A review of climate geoengineering proposals}.
\newblock \emph{Climatic change}, 109\penalty0 (3):\penalty0 745--790, 2011.
\newblock \doi{10.1007/s10584-011-0027-7}.

\bibitem[{von Hoerner}(1975)]{SVH75}
S.~{von Hoerner}.
\newblock {Population explosion and interstellar expansion}.
\newblock \emph{J. Br. Interplanet. Soc.}, 28:\penalty0 691--712, Jan. 1975.

\bibitem[{Ware} et~al.(2022){Ware}, {Young}, {Truitt}, and {Spacek}]{WYT22}
A.~{Ware}, P.~{Young}, A.~{Truitt}, and A.~{Spacek}.
\newblock {Continuous Habitable Zones: Using Bayesian Methods to Prioritize Characterization of Potentially Habitable Worlds}.
\newblock \emph{Astrophys. J.}, 929\penalty0 (2):\penalty0 143, Apr. 2022.
\newblock \doi{10.3847/1538-4357/ac5c4e}.

\bibitem[{Webb}(2015)]{Webb2015}
S.~{Webb}.
\newblock \emph{If the Universe Is Teeming with Aliens ... Where is Everybody?}
\newblock Cham: Springer International Publishing, 2nd edition, 2015.
\newblock \doi{10.1007/978-3-319-13236-5}.

\bibitem[{Weinstein} et~al.(2015){Weinstein}, {Loomis}, {Bhatia}, {Bierman}, {Wang}, and {Chen}]{WLB15}
L.~A. {Weinstein}, J.~{Loomis}, B.~{Bhatia}, D.~M. {Bierman}, E.~N. {Wang}, and G.~{Chen}.
\newblock {Concentrating Solar Power}.
\newblock \emph{Chem. Rev.}, 115\penalty0 (23):\penalty0 12797--12838, 2015.
\newblock \doi{10.1021/acs.chemrev.5b00397}.

\bibitem[{Westall} et~al.(2023){Westall}, {Brack}, {Fair{\'e}n}, and {Schulte}]{WBF23}
F.~{Westall}, A.~{Brack}, A.~G. {Fair{\'e}n}, and M.~D. {Schulte}.
\newblock {Setting the geological scene for the origin of life and continuing open questions about its emergence}.
\newblock \emph{Front. Astron. Space Sci.}, 9:\penalty0 1095701, Jan. 2023.
\newblock \doi{10.3389/fspas.2022.1095701}.

\bibitem[{Wiedmann} et~al.(2020){Wiedmann}, {Lenzen}, {Key{\ss}er}, and {Steinberger}]{WLK20}
T.~{Wiedmann}, M.~{Lenzen}, L.~T. {Key{\ss}er}, and J.~K. {Steinberger}.
\newblock {Scientists' warning on affluence}.
\newblock \emph{Nat. Commun.}, 11:\penalty0 3107, June 2020.
\newblock \doi{10.1038/s41467-020-16941-y}.

\bibitem[{Williams}(2010)]{LW10}
L.~{Williams}.
\newblock {Irrational Dreams of Space Colonization}.
\newblock \emph{Peace Review}, 22\penalty0 (1):\penalty0 4--8, 2010.
\newblock \doi{10.1080/10402650903539828}.

\bibitem[{Wolf} and {Toon}(2015)]{WT15}
E.~T. {Wolf} and O.~B. {Toon}.
\newblock {The evolution of habitable climates under the brightening Sun}.
\newblock \emph{J. Geophys. Res. Atmospheres}, 120\penalty0 (12):\penalty0 5775--5794, June 2015.
\newblock \doi{10.1002/2015JD023302}.

\bibitem[{Wong}(2017)]{KVW17}
K.~V. {Wong}.
\newblock {Anthropogenic Heat Generation and Heat Exhaust to the Ultimate Sink}.
\newblock \emph{J. Energy Resour. Technol.}, 139\penalty0 (3):\penalty0 034701, 2017.
\newblock \doi{10.1115/1.4034859}.

\bibitem[{Wong} and {Bartlett}(2022)]{WB22}
M.~L. {Wong} and S.~{Bartlett}.
\newblock {Asymptotic burnout and homeostatic awakening: a possible solution to the Fermi paradox?}
\newblock \emph{J. R. Soc. Interface}, 19\penalty0 (190):\penalty0 20220029, 2022.
\newblock \doi{10.1098/rsif.2022.0029}.

\bibitem[{Wright} et~al.(2022){Wright}, {Haqq-Misra}, {Frank}, {Kopparapu}, {Lingam}, and {Sheikh}]{WHF22}
J.~T. {Wright}, J.~{Haqq-Misra}, A.~{Frank}, R.~{Kopparapu}, M.~{Lingam}, and S.~Z. {Sheikh}.
\newblock {The Case for Technosignatures: Why They May Be Abundant, Long-lived, Highly Detectable, and Unambiguous}.
\newblock \emph{Astrophys. J. Lett.}, 927\penalty0 (2):\penalty0 L30, Mar. 2022.
\newblock \doi{10.3847/2041-8213/ac5824}.

\bibitem[{Wunderling} et~al.(2024){Wunderling}, {von der Heydt}, {Aksenov}, {Barker}, {Bastiaansen}, {Brovkin}, {Brunetti}, {Couplet}, {Kleinen}, {Lear}, {Lohmann}, {Roman-Cuesta}, {Sinet}, {Swingedouw}, {Winkelmann}, {Anand}, {Barichivich}, {Bathiany}, {Baudena}, {Bruun}, {Chiessi}, {Coxall}, {Docquier}, {Donges}, {Falkena}, {Klose}, {Obura}, {Rocha}, {Rynders}, {Steinert}, and {Willeit}]{WVA24}
N.~{Wunderling}, A.~S. {von der Heydt}, Y.~{Aksenov}, S.~{Barker}, R.~{Bastiaansen}, V.~{Brovkin}, M.~{Brunetti}, V.~{Couplet}, T.~{Kleinen}, C.~H. {Lear}, J.~{Lohmann}, R.~M. {Roman-Cuesta}, S.~{Sinet}, D.~{Swingedouw}, R.~{Winkelmann}, P.~{Anand}, J.~{Barichivich}, S.~{Bathiany}, M.~{Baudena}, J.~T. {Bruun}, C.~M. {Chiessi}, H.~K. {Coxall}, D.~{Docquier}, J.~F. {Donges}, S.~K.~J. {Falkena}, A.~K. {Klose}, D.~{Obura}, J.~{Rocha}, S.~{Rynders}, N.~J. {Steinert}, and M.~{Willeit}.
\newblock {Climate Tipping Point Interactions And Cascades: A Review}.
\newblock \emph{Earth Syst. Dyn.}, 15:\penalty0 41--74, Jan. 2024.
\newblock \doi{10.5194/esd-15-41-2024}.

\bibitem[{Xu} et~al.(2024){Xu}, {Li}, {Qin}, and {Bach}]{XLQ24}
Z.~{Xu}, Y.~{Li}, Y.~{Qin}, and E.~{Bach}.
\newblock {A global assessment of the effects of solar farms on albedo, vegetation, and land surface temperature using remote sensing}.
\newblock \emph{Sol. Energy}, 268:\penalty0 112198, Jan. 2024.
\newblock \doi{10.1016/j.solener.2023.112198}.

\bibitem[{Yang} et~al.(2014){Yang}, {Bou{\'e}}, {Fabrycky}, and {Abbot}]{YBF14}
J.~{Yang}, G.~{Bou{\'e}}, D.~C. {Fabrycky}, and D.~S. {Abbot}.
\newblock {Strong Dependence of the Inner Edge of the Habitable Zone on Planetary Rotation Rate}.
\newblock \emph{Astrophys. J. Lett.}, 787\penalty0 (1):\penalty0 L2, May 2014.
\newblock \doi{10.1088/2041-8205/787/1/L2}.

\bibitem[{Zalasiewicz} et~al.(2017){Zalasiewicz}, {Williams}, {Waters}, {Barnosky}, {Palmesino}, {R{\"o}nnskog}, {Edgeworth}, {Neal}, {Cearreta}, {Ellis}, {Grinevald}, {Haff}, {Ivar do Sul}, {Jeandel}, {Leinfelder}, {McNeill}, {Odada}, {Oreskes}, {Price}, {Revkin}, {Steffen}, {Summerhayes}, {Vidas}, {Wing}, and {Wolfe}]{ZWW17}
J.~{Zalasiewicz}, M.~{Williams}, C.~N. {Waters}, A.~D. {Barnosky}, J.~{Palmesino}, A.-S. {R{\"o}nnskog}, M.~{Edgeworth}, C.~{Neal}, A.~{Cearreta}, E.~C. {Ellis}, J.~{Grinevald}, P.~{Haff}, J.~A. {Ivar do Sul}, C.~{Jeandel}, R.~{Leinfelder}, J.~R. {McNeill}, E.~{Odada}, N.~{Oreskes}, S.~J. {Price}, A.~{Revkin}, W.~{Steffen}, C.~{Summerhayes}, D.~{Vidas}, S.~{Wing}, and A.~P. {Wolfe}.
\newblock {Scale and diversity of the physical technosphere: A geological perspective}.
\newblock \emph{Anthr. Rev.}, 4\penalty0 (1):\penalty0 9--22, Apr. 2017.
\newblock \doi{10.1177/2053019616677743}.

\bibitem[{Zarrouk} and {Moon}(2014)]{ZM14}
S.~J. {Zarrouk} and H.~{Moon}.
\newblock {Efficiency of geothermal power plants: A worldwide review}.
\newblock \emph{Geothermics}, 51:\penalty0 142--153, July 2014.
\newblock \doi{10.1016/j.geothermics.2013.11.001}.

\bibitem[{Zhang} et~al.(2013){Zhang}, {Cai}, and {Hu}]{ZCH13}
G.~J. {Zhang}, M.~{Cai}, and A.~{Hu}.
\newblock {Energy consumption and the unexplained winter warming over northern Asia and North America}.
\newblock \emph{Nat. Clim. Change}, 3\penalty0 (5):\penalty0 466--470, May 2013.
\newblock \doi{10.1038/nclimate1803}.

\bibitem[{Zhang} and {Xu}(2020)]{ZX20}
X.~{Zhang} and M.~{Xu}.
\newblock {Assessing the Effects of Photovoltaic Powerplants on Surface Temperature Using Remote Sensing Techniques}.
\newblock \emph{Remote Sens.}, 12\penalty0 (11):\penalty0 1825, June 2020.
\newblock \doi{10.3390/rs12111825}.

\bibitem[{Ziar} et~al.(2019){Ziar}, {S{\"o}nmez}, {Isabella}, and {Zeman}]{ZSI19}
H.~{Ziar}, F.~F. {S{\"o}nmez}, O.~{Isabella}, and M.~{Zeman}.
\newblock {A comprehensive albedo model for solar energy applications: Geometric spectral albedo}.
\newblock \emph{Appl. Energy}, 255:\penalty0 113867, Dec. 2019.
\newblock \doi{10.1016/j.apenergy.2019.113867}.

\bibitem[{Zotin} et~al.(2001){Zotin}, {Lamprecht}, and {Zotin}]{ZLZ01}
A.~A. {Zotin}, I.~{Lamprecht}, and A.~I. {Zotin}.
\newblock {Bioenergetic Progress and Heat Barriers}.
\newblock \emph{J. Non Equilib. Thermodyn.}, 26\penalty0 (2):\penalty0 191--202, July 2001.
\newblock \doi{10.1515/JNETDY.2001.013}.

\bibitem[{Zsom}(2015)]{AZ15}
A.~{Zsom}.
\newblock {A Population-based Habitable Zone Perspective}.
\newblock \emph{Astrophys. J.}, 813\penalty0 (1):\penalty0 9, Nov. 2015.
\newblock \doi{10.1088/0004-637X/813/1/9}.

\bibitem[{Zubrin}(2000)]{RZ00}
R.~{Zubrin}.
\newblock \emph{{Entering Space: Creating a Spacefaring Civilization}}.
\newblock New York: Penguin, 2000.

\bibitem[{Zubrin}(2019)]{RZ19}
R.~{Zubrin}.
\newblock \emph{{The Case for Space: How the Revolution in Spaceflight Opens Up a Future of Limitless Possibility}}.
\newblock Buffalo: Prometheus Books, 2019.

\bibitem[{Zubrin}(2021)]{RZ21}
R.~{Zubrin}.
\newblock \emph{{The Case for Mars: The Plan to Settle the Red Planet and Why We Must}}.
\newblock New York: The Free Press, 3rd edition, 2021.

\end{thebibliography}

\end{document}